\documentclass[12pt]{article}
\usepackage[english]{babel}
\usepackage{hyperref}
\usepackage{amsmath}
\usepackage{amsfonts}
\usepackage{amssymb}
\begin{document}

\title{
{\bf{}BRST approach to Lagrangian formulation for mixed-symmetry
fermionic higher-spin fields}}

\author{\textsc{P.Yu. Moshin}${}^{a,b}$\thanks{E-mail: moshin@dfn.if.usp.br
\hspace{2cm} ${}^{\dagger}$E-mail: reshet@tspu.edu.ru} and
\textsc{A.A. Reshetnyak}$^{c\dagger}$
\\$^{a}$Instituto de
F\'{\i}sica, Universidade de S\~{a}oPaulo, \\Caixa Postal
66318-CEP, 05315-970 S\~{a}o Paulo, S.P., Brazil
\\ $^{b}$Tomsk State Pedagogical University, 634041 Tomsk, Russia\\ $^{c}$Laboratory of Computer-Aided Design of Materials,
 Institute of \\ Strength Physics and
Materials Science, 634055 Tomsk, Russia}
\date{}

\maketitle \thispagestyle{empty}

\begin{abstract}
We construct a Lagrangian description of irreducible half-integer
higher-spin representations of the Poincare group with the
corresponding Young tableaux having two rows, on a basis of the
BRST approach. Starting with a description of fermionic
higher-spin fields in a flat space of any dimension in terms of an
auxiliary Fock space, we realize a conversion of the initial
operator constraint system (constructed with respect to the relations
extracting irreducible Poincare-group representations) into
a first-class constraint system. For this purpose, we find
 auxiliary representations of the constraint subsuperalgebra
containing the subsystem of second-class constraints  in terms of
Verma modules. We propose a universal procedure of constructing
gauge-invariant Lagrangians with reducible gauge symmetries
describing the dynamics of both massless and massive fermionic
fields of any spin. No off-shell constraints for the fields and
gauge parameters are used from the very beginning. It is shown
that the space of BRST cohomologies with a vanishing ghost number
is determined only by the constraints corresponding to an
irreducible Poincare-group representation. To illustrate the
general construction, we obtain a Lagrangian description of
fermionic fields with generalized spin (3/2,1/2) and (3/2,3/2) on
a flat background containing the complete set of auxiliary fields
and gauge symmetries.

\end{abstract}

\section{Introduction}

The study of  various aspects of higher-spin (HS) field theory
has attracted a considerable attention for a long time due to
the hope of discovering new possible approaches to the unification
of the fundamental interactions. Higher-spin field theory is closely
related to superstring theory, which operates with an infinite
tower of bosonic and fermionic higher-spin fields. The problem of
a covariant Lagrangian description  of fields with an arbitrary
spin propagating on flat \cite{flatin}--\cite{Bonelli2} and
(A)dS \cite{AdSin}--\cite{AlkalaevVasiliev} backgrounds as well as
the problem of constructing an interacting higher-spin field
theory are in the permanent focus of research (for reviews and
more references, see, e.g., \cite{reviews}). One of the attractive
features of investigating higher-spin gauge theories in AdS
spaces is due to a possible relation of this study to the
tensionless limit of superstring theory on the $AdS_5 \times S_5$
Ramond--Ramond background \cite{Heslop,Bonelli1} and the conformal
$\mathcal{N} = 4$ SYM theory  in the context of the AdS/CFT
correspondence \cite{Maldacena}.

At present, the dynamics of totally symmetric higher-spin fields
presents the most developed direction in the variety of
unitary representations of the Poincare and AdS algebras
\cite{Singh,Fronsdal,AdSin,Vasiliev ads,Deser}.
To a great extent, this is caused by the fact that
in a $4d$ space-time there is no place for mixed-symmetry
irreducible representations with the exception of dual
theories\footnote{For a detailed discussion of dual theories in
various dimensions, see \cite{Bekaert,reviews,Hull}.}.
In higher space-time dimensions, there appear
mixed-symmetry representations determined by more than one
spin-like parameters, and the problem of their field-\-theoretic
description is not so well-developed as for totally symmetric
irreps.
 Starting from the papers of Fierz--Pauli and
Singh--Hagen \cite{flatin,Singh} for higher-spin field
theories in the Minkowski space, it has been known that all such
theories include, together with  the basic fields of a given spin,
also some auxiliary fields of lower spins, necessary to provide a
compatibility of the Lagrangian equations of motion with the
relations that determine irreducible representations of the
Poincare group. Attempts to construct Lagrangian descriptions of
free and interacting higher-spin field theories have resulted in
consistency problems, which are not completely  resolved until now.

  The present work is devoted to the construction of
gauge-invariant Lagrangians for both massless and massive
mixed-symmetry spin-tensor fields of rank $n_1 + n_2 + ... + n_k$,
with any integer numbers $n_1 \geq n_2 \geq ... \geq n_k \geq 1$
for $k =2$ in a $d$-dimensional Minkowski space, the fields
being elements of Poincare-group irreps with a Young tableaux having two
rows. In the case of the Minkowski space, several approaches have been
proposed to study mixed-symmetry higher-spin fields
\cite{Labastida,Pashnev1,Bekaert,Medeiros}. Our approach is based
on the BFV--BRST construction \cite{BFV}, see also the reviews \cite{bf,Henneaux},
which was initially developed for a Hamiltonian quantization of
dynamical systems subject to first-class constraints. Following a
tradition accepted in string theory and higher-spin field theory,
we further refer to this method as the BRST method, and to the
corresponding BFV charge, as the BRST operator. The application of
the BRST construction to higher-spin field theory consists of
three steps. First, the conditions that determine the representations
with a given spin are regarded as a system of first- and
second-class operator constraints in an auxiliary Fock space.
Second, the system of the initial constraints is converted, with
a preservation of the initial algebraic structure, into a system of
first-class constraints alone in an enlarged Fock space (see
\cite{conversion} for the development of conversion methods),
with respect to which one constructs the BRST charge.
 Third, the
Lagrangian for a higher-spin field is constructed in terms of the
BRST charge in such a way that the corresponding equations of
motion reproduce the initial constraints. We emphasize that this
approach automatically implies a gauge-invariant Lagrangian
description reflecting the general fact of BV--BFV duality
\cite{BV-BFV,Lyakh-Shar}, realized in order to reproduce a
Lagrangian action or a probability amplitude   by means of a
Hamiltonian object.

The construction of the flat dynamics of mixed-symmetry gauge fields
has been examined in \cite{Curtright,Ouvry,Labastida,Pashnev1,BurdikPashnev,Bekaert,Medeiros},
including the construction of Lagrangians in the BRST approach
for massless bosonic higher-spin fields with two rows of the Young
tableaux \cite{BurdikPashnev}, and recently also for
interacting bosonic HS fields \cite{Tsulaia} and for those of lower
spins \cite{Bizdadea} on the basis of the BV cohomological
deformation theory \cite{BarnichHenneaux1}. Lagrangian descriptions
of massless mixed-symmetry fermionic and
bosonic higher-spin fields in the (A)dS spaces have been suggested
within a ``frame-like'' approach in \cite{AlkalaevVasiliev},
whereas for massive fields of lower superspins in the flat and
(A)dS spaces they have been examined in \cite{Zinoviev}. To be
complete, note that for free totally symmetric higher-spin fields
of integer spins the BRST approach has been used to derive
Lagrangians in the flat space \cite{Pashnev1,0505092} and in the
(A)dS space \cite{symint-ads}. The corresponding programme of a
Lagrangian description of fermionic HS fields has been realized
in the flat space \cite{symferm-flat} and in the (A)dS space
\cite{symferm-ads}.

In this paper, we construct a gauge-invariant Lagrangian
description of fermionic HS fields in Minkowski space of any
dimension, corresponding to a unitary irreducible Poincare-group
representation with a Young tableaux having two rows of length
$n_1, n_2$ ($n_1 \geq n_2$).

The paper is organized as follows. In Section~\ref{Symmalgebra},
we formulate a closed Lie superalgebra of operators, based on the
constraints in an auxiliary Fock space that determines an
irreducible representation of the Poincare group with a
generalized spin $\mathbf{s} = (n_1 + 1/2, n_2+1/2)$. In
Section~\ref{Vermamodule}, we construct a Verma module, being an
auxiliary representation for a rank-2 subsuperalgebra  of the
superalgebra of the initial constraints corresponding to the
subsystem of second-class constraints. This representation is then
realized in terms of new (additional) creation and annihilation
operators in Fock space. Note that a similar construction for
bosonic HS fields in a flat space has been presented in
\cite{0001195}. In Section~\ref{conversionBRST}, we carry out a
conversion of the initial system of first- and second-class
constraints into a system of first-class constraints in the space
being the tensor product of the initial and new Fock spaces. Next,
we construct a BRST operator for the converted constraint
superalgebra. The construction of an action and of a sequence of
reducible gauge transformations describing the propagation of a
mixed-symmetry fermionic field of an arbitrary spin is realized in
Section~\ref{LagrFormulation}. We demonstrate that the Lagrangian
description of a massive half-integer mixed-symmetry HS field
in a $d$-dimensional Minkowski space can be deduced not only by
using the same algorithm, with allowance for the presence of 4 additional
second-class constraints instead of the respective first-class ones
with first-order derivatives, but also by using dimensional reduction
for a massless HS field theory of the same type in a $(d+1)$-dimensional
flat space. In Section~\ref{Proof}, we sketch a proof of the fact that
the resulting action reproduces the correct conditions for a field that
determine an irreducible representation of the Poincare group with
a fixed $\mathbf{s}=(n_1+1/2,n_2+1/2)$  spin. We illustrate the general
formalism by a construction of gauge-invariant Lagrangian actions
for massless and massive spin-$(1+1/2,1/2)$ and
spin-$(1+1/2,1+1/2)$ fields in Section~\ref{Examples}. In
Conclusion, we summarize the results of this work and outline some
open problems.

In addition to the conventions of
\cite{BurdikPashnev,symferm-flat,0001195},
 we use the notation $\varepsilon(A)$, $gh(A)$ for
the respective values of Grassmann parity and ghost number of a
quantity $A$, and denote by $[A,\,B\}$ the supercommutator of
quantities $A, B$, which in the case of definite values of Grassmann parity
is given by $[A\,,B\}$ = $AB - (-1)^{\varepsilon(A)\varepsilon(B)}BA$.

\section{Half-integer HS Symmetry Algebra in Flat
Space-time }\label{Symmalgebra}

In general, a massless half-integer irreducible representation of
the Poincare group in a $d$-dimensional Minkowski space is
described by a spin-tensor field
$\Phi_{\mu_1\ldots\mu_{n_1},\nu_1\ldots\nu_{n_2},...,\rho_1\ldots\rho_{n_k}}(x)$,
with the Dirac index being suppressed, of rank $n_1 + n_2 + ... +
n_k$ and generalized spin $\mathbf{s} = (n_1 +1/2,
n_2+1/2,
 ... , n_k+1/2)$, which corresponds to a Young tableaux with $k$
rows of length $n_1, n_2,  ..., n_k$, respectively, \textbf{and $k \leq
[d/2]$.} This field is symmetric with respect to the
permutations of each type of indices $\mu_i$,
 $i=1,...,k$.

In this paper, we restrict ourselves to the fields characterized
by a Young tableaux with $k=2$ rows:
\begin{equation}\label{Young k2}
\begin{array}{|c|c|c c c|c|c|c|c|c| c| c|}\hline
  \!\mu_1 \!&\! \mu_2\! & \cdot \ & \cdot \ & \cdot \ & \cdot\ & \cdot\ & \cdot\  & \cdot\ &
  \cdot\
  & \cdot\    &\!\! \mu_{n_1}\!\! \\
   \hline
    \! \nu_1\! &\! \nu_2\! & \cdot\
   & \cdot\ & \cdot & \cdot & \cdot & \cdot & \cdot & \!\!\nu_{n_2}\!\!   \\
  \cline{1-10}
\end{array}\ .
\end{equation}
The field $\Phi_{(\mu)_{n_1},(\nu)_{n_2}}(x) \equiv
\Phi_{\mu_1\ldots\mu_{n_1},\nu_1\ldots\nu_{n_2}}(x)$, as an element
of a Poincare-group irrep, obeys the mass-shell and
$\gamma$-traceless conditions for each type of
indices\footnotemark \footnotetext{Throughout the paper, we use
the mostly minus signature $\eta_{\mu\nu} = diag (+, -,...,-)$,
$\mu, \nu = 0,1,...,d-1$, and the Dirac matrices satisfy the
relations $\{\gamma^{\mu}, \gamma^{\nu}\} = 2\eta^{\mu\nu}.$}
\begin{eqnarray}
\label{Eq-0} &&
\imath\gamma^{\mu}\partial_{\mu}\Phi_{(\mu)_{n_1},(\nu)_{n_2}}(x)
=0\,,
\\
&& \gamma^{\mu_1} \Phi_{\mu_1\mu_2\ldots\mu_{n_1},(\nu)_{n_2}}(x)
=0\,,
 \label{Eq-1}\\
 &&
\gamma^{\nu_1} \Phi_{(\mu)_{n_1},\nu_1\nu_2 ... \nu_{n_2}}(x)
=0\,.
 \label{Eq-2}
\end{eqnarray}
The correspondence with a given Young tableaux implies that after
the symmetrization of all the vector indices of the first row with any
vector index of the second row the field
$\Phi_{(\mu)_{n_1},(\nu)_{n_2}}(x)$ becomes equal to zero:
\begin{equation}\label{Eq-3}
    \Phi_{\{(\mu)_{n_1},\nu_1\}\nu_2...\nu_{n_2}}(x) \equiv \sum^{n_1}_{i=1}
\Phi_{\mu_1...\mu_{i-1}\nu_1\mu_{i+1}...\mu_{n_1},\mu_i\nu_2...\nu_{n_2}}(x)
+ \Phi_{(\mu)_{n_1},(\nu)_{n_2}}(x) = 0\,,
\end{equation}
where in the case $i=1$ it is implied that
$\Phi_{\mu_{0}\nu_1\mu_{2}...\mu_{n_1},\mu_1\nu_2...\nu_{n_2}}(x)
\equiv \Phi_{\nu_1\mu_{2}...\mu_{n_1},\mu_1\nu_2...\nu_{n_2}}(x)$.

In order to describe all the irreducible representations
simultaneously, it is convenient to introduce an auxiliary Fock
space $\mathcal{H}$ generated by creation and annihilation
operators $a^{i+}_{\mu}, a^{j}_{\mu}$ with additional internal
indices, $i,j=1,2$,
\begin{eqnarray}\label{comrels}
[a^i_\mu, a_\nu^{j+}]=-\eta_{\mu\nu}\delta^{ij}\,, \qquad
\delta^{ij} = diag(1,1)\,.
\end{eqnarray}
The general state (a Dirac-like spinor) of the Fock space has the
form
\begin{eqnarray}
\label{PhysState} |\Phi\rangle &=&
\sum_{n_1=0}^{\infty}\sum_{n_2=0}^{n_1}\Phi_{(\mu)_{n_1},(\nu)_{n_2}}(x)\,
a^{+\mu_1}_1\ldots\,a^{+\mu_{n_1}}_1a^{+\nu_1}_2\ldots\,a^{+\nu_{n_2}}_2|0\rangle
,
\end{eqnarray}
providing the symmetry property of
$\Phi_{(\mu)_{n_1},(\nu)_{n_2}}(x)$ under the permutation of
indices of the same type.
 We refer to the vector (\ref{PhysState}) as the basic vector.

Because of the property of translational invariance of the vacuum,
$\partial_\mu |0\rangle = 0$, the conditions
(\ref{Eq-0})--(\ref{Eq-2}) can be equivalently expressed in terms
of the bosonic operators
\begin{eqnarray}
  && \tilde{t}_0 =  i\gamma^{\mu}\partial_\mu\,, \qquad \tilde{t}^i= \gamma^{\mu}a^i_\mu\,,
  \label{tildet0} \\
   &&  t = a^{1+}_\mu a^{2{}\mu} \label{t}
\end{eqnarray}
as follows:
\begin{equation}\label{t0t1t}
    \tilde{t}_0|\Phi\rangle =
\tilde{t}^i|\Phi\rangle = t|\Phi\rangle =  0.
\end{equation}
Thus, the constraints (\ref{t0t1t}) with each component
$\Phi_{(\mu)_{n_1},(\nu)_{n_2}}(x)$ of the vector
(\ref{PhysState}) subject to (\ref{Eq-0})--(\ref{Eq-2}) describe a
field of spin $(n_1+1/2,n_2+1/2)$.

Because of the fermionic nature of equations
(\ref{Eq-0})--(\ref{Eq-2}) with respect to the standard
Lorentz-like Grassmann parity, and due to the bosonic nature of
the primary constraint operators $\tilde{t}_0, \tilde{t}^i$,
$\varepsilon(\tilde{t}_0) = \varepsilon(\tilde{t}^i)= 0$, in order
to equivalently transform  these operators into fermionic ones, we
now introduce  a set of $d+1$ Grassmann-odd gamma-matrix-like
objects $\tilde{\gamma}^\mu$, $\tilde{\gamma}$,  subject to the
conditions
\begin{eqnarray}
\{\tilde{\gamma}^\mu,\tilde{\gamma}^\nu\} = 2\eta^{\mu\nu}, \qquad
\{\tilde{\gamma}^\mu,\tilde{\gamma}\}=0, \qquad
\tilde{\gamma}^2=-1, \label{tgammas}
\end{eqnarray}
and related  to the conventional gamma-matrices as
follows:\footnote{For more details, see \cite{symferm-flat}.
The quantities $\tilde{\gamma}^\mu, \tilde{\gamma}$
may be viewed as intermediate objects as compared
to ${\gamma}^\mu$. Indeed, the final Lagrangian description
in terms of ghost-independent or spin-tensor forms depends
only on the standard Grassmann-even matrices ${\gamma}^\mu$,
and does not depend on $\tilde{\gamma}^\mu, \tilde{\gamma}$,
because the latter enter the Lagrangian only in even degrees,
and also due to the homogeneity of the reducible gauge
transformations w.r.t. $\tilde{\gamma}$, as shown by
the examples in Section~\ref{Examples}.}
\begin{eqnarray}\label{gammas}
\gamma^{\mu} = \tilde{\gamma}^{\mu} \tilde{\gamma}.
\end{eqnarray}

We can now define Grassmann-odd constraints,
\begin{eqnarray}
\label{t0ti} {t}_0 = -\imath\tilde{\gamma}^\mu \partial_\mu\,,
\qquad {t}^i =  \tilde{\gamma}^\mu a^i_\mu ,
\end{eqnarray}
related to the operators (\ref{tildet0}) as follows:
\begin{eqnarray}
\left(t_0, t^i\right) = \tilde{\gamma}\left(-\tilde{t}_0,
\tilde{t}^i\right) .
\end{eqnarray}

We next define an odd scalar product:
\begin{eqnarray}
\label{sproduct} \langle\tilde{\Psi}|\Phi\rangle & =  &
\int d^dx
         \sum_{n_1, k_1, n_2, k_2 =0}^{\infty}
         \langle 0|a_1^{\rho_1}\ldots\,a_1^{\rho_{k_1}} a_2^{\sigma_1}\ldots\,a_2^{\sigma_{k_2}}
         \Psi^+_{(\rho)_{k_1},(\sigma)_{k_2}}(x)
         \tilde{\gamma}_0
         \Phi_{(\mu)_{n_1},(\nu)_{n_2}}(x) \times \nonumber \\
         &{}& \qquad \qquad a_1^{+\mu_1}\ldots\,a_1^{+\mu_{n_1}}
         a_2^{+\nu_1}\ldots\,a_2^{+\nu_{n_2}}
         |0\rangle .
\end{eqnarray}

The operators $t_0, t^i , t$ in (\ref{t}) and (\ref{t0ti}), with
$t^{i+} = \tilde{\gamma}^\mu a^{i+}_\mu $ and $t^+ = a^{2+}_\mu
a^{1\mu}$ being Hermitian conjugate, respectively, to $t^i, t$
with reference to the scalar product (\ref{sproduct}), generate an
operator Lie superalgebra composed of the operators
\begin{align}
& {t}_0 = -\imath\tilde{\gamma}^\mu \partial_\mu \,, \label{t0} &&
{}
\\
& t^i=\tilde{\gamma}^\mu a^i_\mu\,, \label{ti} &&
t^{i+}=\tilde{\gamma}^\mu a_\mu^{i+}
\,,
\\
& t = a_\mu^{1+}a^{2{}\mu} \,, \label{tt+} && t^+ =
a_\mu^{2+}a^{1{}\mu} \,, \\
 & l^i=- ia^i_\mu \partial^\mu \,, \label{li} &&
l^{i+} = - i a^{i+}_\mu \partial^\mu \,,
\\
& l^{ij}={\textstyle\frac{1}{2}}\,a^{i{}\mu} a^j_{\mu}\,,
\label{lij} && l^{ij+} =
{\textstyle\frac{1}{2}}\,a^{i{}\mu+}a^{j{}+}_\mu\footnotemark \,,
\\
& l_0 = \partial^\mu\partial_\mu\,, && g^i_0=-a^{i+}_\mu
a^{i{}\mu}+{\textstyle\frac{d}{2}} \label{g0} \,,
\end{align}
\footnotetext{For the operators $l^{12}, l^{12+}$ in (\ref{lij}),
we have used a definition slightly different from that of
\cite{0001195}, where $(\hat{l}^{12}, \hat{l}^{12+})$ = $2(l^{12},
l^{12+})$.}which is invariant under Hermitian conjugation.

The operators (\ref{t0})--(\ref{g0}) form a superalgebra given by
Table~\ref{table in}, with an omission of the Poincare-group
Casimir operator $l_0$ being the central charge of this algebra,
\begin{table}[t]
{\small
\begin{eqnarray*}
\begin{array}{||c||c|c|c|c|c|c|c|c|c||c||}\hline\hline\vphantom{\biggm|}\hspace{-0.3em}
[\;\downarrow\;,\to\}\hspace{-0.4em}& {t}_0&t^i&t^{i{}+}& t & t^+
& l^i &l^{i{}+} & l^{ij}
&l^{ij{}+} &g^i_0\\
\hline\hline\vphantom{\biggm|} {t}_0
   &-2{l}_0& 2l^i & 2l^{i{}+} & 0 & 0 & 0 & 0
    & 0 & 0&0\\
\hline\vphantom{\biggm|} t^k
   & 2l^k&4l^{ki}& A^{ki} &-t^2\delta^{k1}& -t^1\delta^{k2} &
   0 & -t_0\delta^{ki} & 0 & B^{k,ij} & t^i\delta^{ki} \\
\hline\vphantom{\biggm|} t^{k{}+}
   & 2l^{k{}+} & A^{ik} &4l^{ki+} & t^{1+}\delta^{k2} & t^{2+}\delta^{k1} &
   t_0\delta^{ki}
    & 0 & C^{k,ij} & 0 & - t^{i+}\delta^{ki} \\
\hline\vphantom{\biggm|} t
   &0& t^2\delta^{i1}  & \hspace{-0.2em}- t^{1+}\delta^{i2} \hspace{-0.2em}
    & 0 & \hspace{-0.2em}g_0^1 - g_0^2 \hspace{-0.2em}&
   l^2\delta^{i1} &\hspace{-0.2em} - l^{1+}\delta^{i2} \hspace{-0.2em}&
   D^{ij} & E^{ij} & F^i \\
\hline\vphantom{\biggm|} t^+
   & 0 & t^1\delta^{i2} & \hspace{-0.2em}- t^{2+}\delta^{i1} \hspace{-0.2em}
   & \hspace{-0.2em} g_0^2 - g_0^1 \hspace{-0.2em}
    & 0 & l^1\delta^{i2} & \hspace{-0.2em}-l^{2+}\delta^{i1}\hspace{-0.2em}
     & G^{ij} & H^{ij} & I^i \\
\hline\vphantom{\biggm|} l^k
   &0&0& -t_0\delta^{ik} & - l^2\delta^{k1}  & -l^1\delta^{k2} & 0 & l_0\delta^{ik}
    & 0 & J^{k,ij} & l^i\delta^{ik} \\
\hline\vphantom{\biggm|} l^{k+} & 0
   & t_0\delta^{ik} & 0 & l^{1+}\delta^{k2} & l^{2+}\delta^{k1} & -l_0\delta^{ik}
   & 0 & K^{k,ij} & 0 & -l^{i+}\delta^{ik}  \\
\hline\vphantom{\biggm|} l^{kl}
   & 0 & 0 &-C^{i,kl}& -D^{kl} & -G^{kl} & 0 & - K^{i,kl} & 0 & L^{kl,ij} &
   l^{i\{k}\delta^{l\}i} \\
\hline\vphantom{\biggm|} l^{kl+}
   & 0 &\hspace{-0.2em} -B^{i,kl}\hspace{-0.2em} & 0 & -E^{kl} & -H^{kl} &
   -J^{i,kl} & 0 & - L^{ij,kl} & 0 &
   \hspace{-0.3em} -l^{i\{k+}\delta^{l\}i} \hspace{-0.3em} \\
\hline\hline\vphantom{\biggm|} g_0^k
   & 0 & -t_k\delta^{ik}& t^{k+}\delta^{ik} & -F^k &-I^k & -l^{k}\delta^{ik}
   & l^{k+}\delta^{ik} & \hspace{-0.3em}-l^{k\{i}\delta^{j\}k}\hspace{-0.3em}
    & \hspace{-0.3em} l^{k\{i+}\delta^{j\}k}
   \hspace{-0.3em} & 0 \\
   \hline\hline
\end{array}
\end{eqnarray*}}
\caption{The superalgebra of the initial operators}\label{table
in}
\end{table}
where the quantities $A^{ik}$, $B^{k,ij}$, $C^{k,ij}$, $D^{ij}$,
$E^{ij}$, $F^i$, $I^i$, $G^{ij}$, $H^{ij}$, $J^{k,ij}$,
$K^{k,ij}$, $L^{kl,ij}$ are defined by the relations
\begin{align}
&  A^{ik} = -2g_0^i\delta^{ik}+2t\delta^{i2}\delta^{k1} +
2t^+\delta^{i1}\delta^{k2}\,,\label{Akj}&& {}
\\
   & B^{k,ij} = - \textstyle\frac{1}{2}t^{\{i+}\delta^{j\}k}\,,\label{Bkij} &&
   C^{k,ij} =  \textstyle\frac{1}{2}t^{\{i}\delta^{j\}k} \,,\\
 & D^{ij} = l^{\{i 2}\delta^{j\}1}\,,\label{Dij} && E^{ij} = -l^{1\{i+}\delta^{j\}2} \,,\\
 & F^i = t(\delta^{i2}-\delta^{i1})  \,,\label{Fi}&&  I^i = t^+(\delta^{i1}-\delta^{i2})\,,\\
 &  G^{ij} = l^{1\{i}\delta^{j\}2} \,, \label{Gij} && H^{ij} = -l^{\{i 2+}\delta^{j\}1}  \,,\\
 &  J^{k,ij} = - \textstyle\frac{1}{2}l^{\{i+}\delta^{j\}k} \,, \label{Jkij}  &&
 K^{k,ij} =  \textstyle\frac{1}{2}l^{\{i}\delta^{j\}k}  \,,
\end{align}
\vspace{-4ex}
\begin{eqnarray}
  L^{kl,ij} &=&   \textstyle\frac{1}{4}\Bigl\{\delta^{ik}
\delta^{lj}\Bigl[2g_0^k\delta^{kl} + g_0^k + g_0^l\Bigr]  \nonumber \\
&& -
\delta^{ik}\Bigl[t\Bigl(\delta^{l2}(\delta^{j1}+\delta^{k1}\delta^{kj})+
\delta^{k2}\delta^{j1}\delta^{lk}\Bigr)
   + t^+\Bigl(\delta^{l1}(\delta^{j2}+\delta^{k2}\delta^{kj})+
\delta^{k1}\delta^{j2}\delta^{lk}\Bigr)\Bigr] \nonumber \\
&& -
\delta^{lj}\Bigl[t\Bigl(\delta^{k2}(\delta^{i1}+\delta^{l1}\delta^{li})+
\delta^{l2}\delta^{i1}\delta^{kl}\Bigr)
    + t^+\Bigl(\delta^{k1}(\delta^{i2}+\delta^{l2}\delta^{li})+
\delta^{l1}\delta^{i2}\delta^{lk}\Bigr)\Bigr]\Bigr\}
 \,.\label{Lklij}
\end{eqnarray}
We call this algebra the half-integer higher-spin symmetry algebra
in Minkowski space with a Young tableaux having two rows.

From the viewpoint of constraint system theory, the above
superalgebra is a system of constraints, except for the operators
$g_0^k$, being non-degenerate in $\mathcal{H}$. These operators, as
follows from Table~\ref{table in},  determine an invertible
operator supermatrix of commutators for the subsystem of
second-class constraints, $\{t_k, t_k^+, l_{ij}, l_{ij}^+, t,
t^+\}$, with the other constraints, ${t_0, l_0, l_i, l_i^+}$, being
first-class ones.

A conversion of this constraint system $\{o_I\}$, including the
operators $g_0^k$, into a first-class constraint system $\{O_I\}$
by means of an additive composition of  $o_I$ with certain operators
$o'_I$ depending on some new creation and annihilation operators, $o_I
\to O_I = o_I + o'_I$, can be effectively realized only for the
 subsuperalgebra of the entire symmetry superalgebra that
contains the subsystem of second class-constraints and $g_0^k$.
The only requirement, as shown in \cite{0001195}, is that each of
the Hermitian operators $g_0^k$ should have a linear dependence
on an arbitrary parameter $h^k$, whose values are to be determined
later.

\section{Auxiliary Representation for the  Superalgebra with \\ Second-class
Constraints}\label{Vermamodule}
\setcounter{equation}{0}

In this section, we describe the method of Verma module
construction for the Lie superalgebra with second-class
constraints alone. Having denoted $\{o_a\} = \{t_k, t^{+}_k,
l_{ij}, l^+_{ij}, g_0^k\}$, $o_a \in \{o_I\}$, as the basis
elements of the above superalgebra, and using the requirements that
${o}_a$, $o'_a$ must supercommute, $\{{o}_a, o'_a] = 0$, and that
the converted constraints must be in involution, $\{O_a, O_b] \sim
O_c$, we find that the superalgebra of the additional parts $o'_a$
is uniquely determined by the same algebraic conditions as those
for the initial constraints. In this case, it is unnecessary to
convert the subsystem of the initial first-class constraints not
entering $\{o_a\}$, and therefore they remain intact.

Following \cite{symferm-flat} and the general method of Verma
module construction for mixed-symmetry  integer-spin HS fields
\cite{0001195}, let us denote $E^{\alpha} \equiv (t^k; l^{ij},t) =
(E^{\alpha_0}; E^{\alpha_1}),$ (${\alpha}_0 > 0, {\alpha}_1 > 0$)
for $i\leq j$, and define
\begin{equation}\label{convconstr}
 \mathcal{H}^i = g_0^i +  g_0^{\prime i}\,, \quad g_0^{\prime i} = h^i + ... \,,  \qquad
 \mathcal{E}^{\alpha} =  {E}^{\alpha} + {E}^{\prime\alpha}(h)\,,
 \quad \alpha_0=1,2\,,\ \alpha_1=1,2,3,4\,.
\end{equation}
The quantities $g_0^i, E^{\alpha_1}, E^{- \alpha_1}$ are the
Cartan generators, positive and negative root vectors, except for
$\alpha_1 = 2$ (see footnote \thefootnote) of the subalgebra
$so(3,2)$ in the superalgebra of second-class constraints, and the
odd generators  $E^{\alpha_0}, E^{- \alpha_0}$ supplement the
basis $so(3,2)$ up to that of the above superalgebra. The
quantities $g_0^{\prime i}, E^{\prime\alpha}, E^{- \prime\alpha}$
and $\mathcal{H}^i, \mathcal{E}^{\alpha}, \mathcal{E}^{- \alpha}$
have the same identification respectively for the additional and
enlarged operators of the symmetry superalgebra.

Consider the highest-weight representation  of the superalgebra of
the additional parts with the highest-weight vector $|0 \rangle_V$
annihilated by the positive roots and being the proper vector of
the Cartan generators:
\begin{equation}\label{eigenvector}
 E^{\prime\alpha}|0\rangle_V = 0\,, \alpha >0\,, \qquad g_0^{\prime i}|0\rangle_V =
 h^i |0\rangle_V\,.
\end{equation}

Following the Poincare--Birkhoff--Witt  theorem, the basis space
of this representation, called the Verma module in the mathematical
literature \cite{Dixmier}, is given by the vectors
\begin{equation}
\label{basisV} \left|\vec{n}_k^0, \vec{n}_{ij}, n\rangle_V \right.
 = \bigl(E^{\prime - \alpha^0_{1}}\bigr){}^{n^0_1}\bigl(E^{\prime -
\alpha^0_{2}}\bigr){}^{n^0_2}\bigl(E^{\prime -
\alpha^1_{1}}\bigr){}^{n_{11}}\bigl(E^{\prime -
\alpha^2_{1}}\bigr){}^{n_{12}}\bigl(E^{\prime -
\alpha^3_{1}}\bigr){}^{n_{22}}\bigl(E^{\prime -
\alpha^4_{1}}\bigr){}^{n}|0\rangle_V\,,
\end{equation}
where $\vec{n}_k^0 = ({n}_1^0, {n}_2^0)$, $\vec{n}_{ij} =
({n}_{11}, {n}_{12}, {n}_{22})$, $n_1^0, n_2^0 = 0, 1$, $n_{ij}
\in \mathbb{N}_0$. Note that the restriction for the values of
$\vec{n}_k^0$ in (\ref{basisV}) is due to the identities
\begin{equation}
\label{identities} \{E^{\prime - \alpha^0_{i}}, E^{\prime -
\alpha^0_{j}}] = 4(E^{\prime - \alpha^1_{1}}, E^{\prime -
\alpha^2_{1}}, E^{\prime - \alpha^3_{1}}) = 4(l^{\prime+}_{11},
l^{\prime+}_{12}, l^{\prime+}_{22})\,, \qquad i,j=1,2, i\leq j\,.
\end{equation}

Using the commutation relations of the superalgebra given by
Table~\ref{table in} and the formula for the product of graded
operators,
\begin{eqnarray}\label{product}
&&    AB^n =
\sum^{n}_{k=0}(-1)^{\varepsilon(A)\varepsilon(B)(n-k)}C^{(s)}{
}^n_k B^{n-k}\mathrm{ad}^k_B{}A\,, \  n\geq 0\,,s = \varepsilon(B)\,,  \nonumber\\
&& \mathrm{ad}^k_B{}A=
\mathrm{ad}_B\left(\mathrm{ad}^{k-1}_B{}A\right), \
\mathrm{ad}_B{}A = \{A,B]\,,
\end{eqnarray}
we can calculate the explicit form of the Verma module. In
(\ref{product}), we have introduced generalized coefficients for a
number of graded combinations, $C^{(s)}{ }^n_k$, that coincide with
the standard ones only for the bosonic operator $B$:
${C^{(0)}{}^n_k} = C^n_k = \frac{n!}{k!(n-k)!}$. These
coefficients are defined recursively, by the relations
\begin{eqnarray}\label{combination}
&&    C^{(s)}{}^{n+1}_{k} = (-1)^{s(n+k+1)}C^{(s)}{ }^n_{k-1} +
C^{(s)}{}^n_k \,, \qquad
 n, k \geq 0\,,\\
&&  C^{(s)}{}^{n}_{0} = C^{(s)}{}^{n}_{n}=1\,, \qquad
C^{(s)}{}^{n}_{k}=0\,, \ n<k
\end{eqnarray}
and possess the properties $C^{(s)}{}^n_k=C^{(s)}{}^n_{n-k}$. The
corresponding values of $C^{(1)}{}^n_k$ are defined, for $n\geq k$,
by the formulae
\begin{equation}\label{expressions}
C^{(1)}{}^n_k =
\sum^{n-k+1}_{i_k=1}\sum^{n-i_k-k+2}_{i_{k-1}=1}\ldots \sum^{n-
\sum^{k}_{j=3}i_j-1}_{i_2=1} \sum^{n-\sum^{k}_{j=2}i_j}_{i_1=1}
(-1)^{k(n+1) +
\sum\limits^{[(k+1)/2]}_{j=1}\left(i_{2j-1}+1\right)},
\end{equation}
which follow by induction, and in which $[a]$ stands for the
integer part of the number $a$. For our purposes, due to $n^0_k =
0,1$ in (\ref{basisV}), (\ref{product}), it is sufficient to know
that $C^{(1)}{}^0_0 = C^{(1)}{}^1_0 = 1$ and $C^{(1)}{}^{n^0_k}_1
= n^0_k$.

Then, following \cite{Burdik} and making use of the mapping
\begin{equation}\label{map}
    \left|\vec{n}_k^0, \vec{n}_{ij}, n\rangle_V \right.
    \leftrightarrow \left|\vec{n}_k^0, \vec{n}_{ij}, n\rangle \right.
 = \bigl(f^+_{1}\bigr){}^{n^0_1}\bigl(f^+_{2}\bigr){}^{n^0_2}\bigl(b^{+}_{11}\bigr){}^{
 n_{11}}\bigl(b^{+}_{12}\bigr){}^{n_{12}}\bigl(b^{+}_{22}\bigr){}^{n_{22}}
 \bigl(b^{+}\bigr){}^{n}|0\rangle\,,
\end{equation}
where $\left|\vec{n}_k^0, \vec{n}_{ij}, n\rangle\right.$, for
${n}_k^0 = 0,1$, $n_{ij} \in \mathbb{N}_0$, are the basis vectors
of a Fock space $\mathcal{H}'$ generated by new fermionic,
$f^+_{k}, f_{k}$, $k=0,1$, and bosonic, $b^{+}_{ij}, b^+, b_{ij},
b$, $i,j=1,2, i\leq j$, creation and annihilation operators with
the standard (only nonvanishing) commutation relations
\begin{equation}\label{commrelations}
 \{f_k\,, f^+_l\} = \delta_{kl}\,,\qquad [b_{ij}, b^+_{lk}] =
 \delta_{il}\delta_{jk}\,, \ i\leq j, k\leq l\,,  \qquad [b\,,b^+]
 =1\,,
\end{equation}
we can represent the Verma module as polynomials in the creation
operators of the Fock space $\mathcal{H}'$.

First, we find the action of the negative root operators
$E^{\prime-\alpha}$ on the basis vectors. After a simple
calculation, one obtains
\begin{eqnarray}\label{t'+i}
 t^{\prime i +} \left|\vec{n}_k^0, \vec{n}_{lm}, n\rangle_V
 \right. & = &\textstyle \delta_{i1}\left(1+\left[\frac{n_1^0+1}{2}\right]\right)
 \left|{n}_1^0 + 1{} mod{} 2, n_2^0,n_{11} + \left[\frac{n_1^0+1}{2}\right]
 , {n}_{12}, {n}_{22}, n\right\rangle_V
 \\
 && +\textstyle \delta_{i2}(-1)^{n_1^0}\Bigl\{\hspace{-0.2em}\left(1+\hspace{-0.2em}
 \left[\frac{n_2^0+1}{2}\hspace{-0.2em}\right]\hspace{-0.2em}\right)
 \hspace{-0.2em}\left|{n}_1^0, n_2^0  + 1{} mod{} 2, n_{11}
 , {n}_{12}, {n}_{22}+ \hspace{-0.2em}\left[\frac{n_2^0+1}{2}\hspace{-0.2em}\right], n
 \hspace{-0.2em}\right\rangle_V
 \nonumber \\
 &&
  - 4n_1^0\left|{n}_1^0-1, n_2^0, n_{11}
 , {n}_{12}+1, {n}_{22}, n\rangle_V \right.\Bigr\}\,,
 \nonumber \\
l^{\prime+}_{ij}\left|\vec{n}_k^0, \vec{n}_{lm}, n\rangle_V
 \right. & = & \left|\vec{n}_k^0, \vec{n}_{lm}+ \delta_{il}\delta_{jm}, n\rangle_V
 \right. \,,
 \label{l'+ij} \\
 t^{\prime+}  \left|\vec{n}_k^0, \vec{n}_{lm}, n\rangle_V
 \right. & = & \left|\vec{n}_k^0, \vec{n}_{lm}, n+1\rangle_V \right.  -
 2n_{11}\left|\vec{n}_k^0, {n}_{11}-1, n_{12}+1, n_{22}, n\rangle_V
 \right. \\
 &&  -{n}_1^0\textstyle\left(1+\left[\frac{n_2^0+1}{2}\right]\right)
\left|{n}_1^0-1, n_2^0  + 1{} mod{} 2, n_{11}
 , {n}_{12}, {n}_{22}+ \left[\frac{n_2^0+1}{2}\right], n \right\rangle_V
 \nonumber \\
  && - n_{12}\left|\vec{n}_k^0, {n}_{11}, n_{12}-1, n_{22} +1, n\rangle_V
 \right.\,,
 \label{t'+}
 \nonumber \\
 g_0^{\prime i} \left|\vec{n}_k^0, \vec{n}_{lm}, n\rangle_V
 \right. & = & \left(n_k^0\delta_{ik} + \sum_{l\leq m}n_{lm}\left(\delta^{il}+
 \delta^{im}\right) + n\left(\delta^{i2}-\delta^{i1}\right) + h^i\right)
 \left|\vec{n}_k^0, \vec{n}_{lm}, n\rangle_V
 \right.\,.\label{g'0i}
\end{eqnarray}
Second, for the positive root operators $E^{\prime\alpha}$ we find
\begin{eqnarray}
   \label{t'1}
 t^{\prime 1 } \left|\vec{n}_k^0, \vec{n}_{lm}, n\rangle_V
 \right. &=&
-2{n}_1^0(2n_{11}+ n_{12}  - n +h^1)
   \left|{n}_1^0-1, {n}_2^0, \vec{n}_{lm},
     n \rangle_V\right. \\
&&
 \textstyle -(-1)^{n_1^0 } \hspace{-0.2em}\left(1+
 \hspace{-0.2em}\left[\frac{n_1^0+1}{2}
\hspace{-0.2em}\right]\hspace{-0.2em}\right)n_{11}
 \hspace{-0.2em}\left|{n}_1^0 + 1{} mod{} 2, n_2^0,n_{11}-1 +
 \hspace{-0.2em}\left[\frac{n_1^0+1}{2}
\hspace{-0.2em}\right]
 , {n}_{12}, {n}_{22}, n\hspace{-0.2em}\right\rangle_V \nonumber\\
   && \textstyle
    - (-1)^{n_1^0 + n_2^0}\frac{n_{12}}{2}\hspace{-0.2em}\left(1+\hspace{-0.2em}
    \left[\frac{n_2^0 + 1}{2}
 \hspace{-0.2em}\right]\hspace{-0.2em}\right)\hspace{-0.2em}
 \left|{n}_1^0, n_2^0 + 1 {} mod{} 2, n_{11}
 , {n}_{12}-1, {n}_{22} + \hspace{-0.2em}\left[\frac{n_2^0+1}{2}\hspace{-0.2em}
 \right], n\hspace{-0.2em}\right\rangle_V   \nonumber \\
   &&
+  2(-1)^{n_1^0} n_2^0 \Bigl(-n_{12} \left|{n}_1^0, {n}_2^0 -1,
n_{11}, n_{12}-1, {n}_{22} +
   1,  n \rangle_V\right. \nonumber\\
  &&  + \left|{n}_1^0, {n}_2^0 -1,
\vec{n}_{lm},
 n+1\rangle_V\right.\Bigr),  \nonumber
   \end{eqnarray}
   \vspace{-3ex}
   \begin{eqnarray}
   \label{t'2}
 t^{\prime 2 } \left|\vec{n}_k^0, \vec{n}_{lm}, n\rangle_V
 \right. & = &
   (-1)^{n_1^0}\Bigl\{
- 2 n_2^0 (2n_{22}-n_1^0+n+h^2)\left|{n}_1^0, n_2^0 - 1,
\vec{n}_{lm}, n\right\rangle_V \\
&&
   -
   \textstyle\frac{n_{12}}{2}\left(1+\left[\frac{n_1^0 + 1}{2}
 \right]\right)\left|{n}_1^0 + 1 {} mod{} 2, n_2^0, n_{11} + \left[\frac{n_1^0+1}{2}\right]
 , {n}_{12}-1, {n}_{22}, n\right\rangle_V     \nonumber\\
   && - (-1)^{n_2^0} n_{22} \textstyle\hspace{-0.2em}\left(1+\hspace{-0.2em}
   \left[\frac{n_2^0 + 1}{2}\hspace{-0.2em}
 \right]\hspace{-0.2em}\right) \hspace{-0.2em}\left|{n}_1^0, n_2^0 + 1 {} mod{} 2, n_{11},
 {n}_{12}, {n}_{22}-1 + \hspace{-0.2em}\left[\frac{n_2^0 + 1}{2}
 \hspace{-0.2em}\right], n\hspace{-0.2em}\right\rangle_V  \nonumber   \\
   &&
   - 2 n_1^0 \Bigl(n(h^1-h^2 -n +1)\left|{n}_1^0-1, n_2^0, {\vec{n}}_{lm},
    n-1\right\rangle_V \nonumber \\
   && - 2{n}_{22}\left|{n}_1^0-1, n_2^0, {n}_{11}, {n}_{12}+1, n_{22}-1,
    n\right\rangle_V
    \nonumber \\
&&
      -n_{12}\left|{n}_1^0-1, n_2^0, {n}_{11}+1, {n}_{12}-1, n_{22},
    n\right\rangle_V  \Bigr)
 \Bigr\}\,,  \nonumber
 \end{eqnarray}
   \vspace{-3ex}
   \begin{eqnarray}
  \label{l'11}
  l^{\prime 11}
\hspace{-0.2em}\left|\vec{n}_k^0, \vec{n}_{lm}, n\rangle_V
 \right. \hspace{-0.25em}&\hspace{-0.25em}=\hspace{-0.25em}&
 \hspace{-0.25em}
   {n}_{11}({n}_{11}+{n}_{12}+n_1^0- n -1 + h^1)
   \left|\vec{n}_k^0, {n}_{11} -1, {n}_{12}, {n}_{22}, n\rangle_V
 \right. \\ &&\hspace{-0.25em}
 - \textstyle\frac{{n}_{12}}{2}\left|\vec{n}_k^0, {n}_{11}, {n}_{12} -1, {n}_{22}, n+1\rangle_V
 \right. + \textstyle\frac{{n}_{12}({n}_{12}-1)}{4}
 \left|\vec{n}_k^0, {n}_{11}, {n}_{12} -2, {n}_{22}+1, n\rangle_V
 \right.
 \nonumber \\
 \hspace{-0.25em}&\hspace{-0.25em}- &\hspace{-0.25em}
    n_1^0 \textstyle\Bigl\{2 n_2^0\Bigl( \left|\overrightarrow{n_k^0-1}, \vec{n}_{lm},  n +
1\rangle_V\right. - n_{12}\left|\overrightarrow{n_k^0-1},
{n}_{11}, {n}_{12} -1, {n}_{22}+1, n\rangle_V
 \right.\Bigr)
 \nonumber \\
\hspace{-0.25em}&\hspace{-0.25em}& \hspace{-0.25em}\textstyle
-\frac{(-1)^{n_2^0}}{2}n_{12}\hspace{-0.25em}\left(1+\hspace{-0.25em}
 \left[\frac{n_2^0 + 1}{2}\hspace{-0.25em}
 \right]\hspace{-0.2em}\right)\hspace{-0.25em}\left|{n}_1^0 -1, n_2^0+ 1 {} mod{} 2, n_{11}
 , {n}_{12}-1, {n}_{22} + \hspace{-0.25em}\left[\frac{n_2^0+1}{2}\hspace{-0.25em}\right], n
 \hspace{-0.25em}\right\rangle_V\Bigr\},
 \nonumber
   \end{eqnarray}
   \vspace{-3ex}
   \begin{eqnarray}
   \label{l'12}
  l^{\prime 12}
\left|\vec{n}_k^0, \vec{n}_{lm}, n\rangle_V
 \right. &=&  \textstyle \frac{{n}_{12}}{4}\hspace{-0.15em}\Bigl(\hspace{-0.15em}2{n}_{11}+{n}_{12}+2{n}_{22}+
 \hspace{-0.15em}
 \displaystyle\sum_{k}\hspace{-0.1em}(n_k^0 +h^k)-1\hspace{-0.15em}\Bigr)\hspace{-0.2em}\left|\vec{n}_k^0,
 {n}_{11} , {n}_{12}\hspace{-0.1em}-\hspace{-0.1em}1, {n}_{22}, n\hspace{-0.1em}\rangle_V
 \right.
 \\
&& + \textstyle\frac{1}{2} n n_{11}(h^2-h^1 + n -
1)\left|\vec{n}_k^0,
 {n}_{11}-1, {n}_{12}, {n}_{22}, n-1\rangle_V \right.
 \nonumber \\
 && \textstyle  + n_{11}n_{22}\left|\vec{n}_k^0,
 {n}_{11}-1 , {n}_{12}+1, {n}_{22} - 1, n\rangle_V
 \right.
\nonumber \\
&&  \textstyle - \frac{n_{22}}{2}\left|\vec{n}_k^0,
 {n}_{11}, {n}_{12}, {n}_{22} - 1, n+1\rangle_V
 \right. \nonumber \\
 &+&   \textstyle
 \frac{n^0_1}{2}\Bigl\{2{n_2^0}
(n+2n_{22}+h^2)\left|\overrightarrow{{n}_k^0-1},
 \vec{n}_{lm}, n\rangle_V \right.+ \textstyle(-1)^{n^0_2}\left(1+
 \left[\frac{n_2^0 + 1}{2}
\right]\right) \times
 \nonumber \\
 && \times \textstyle
 n_{22}\left|{n}_1^0 -1, n_2^0+ 1 {} mod{} 2, n_{11}
 , {n}_{12}, {n}_{22}-1 + \left[\frac{n_2^0+1}{2}
\right], n\right\rangle_V\Bigr\}
 \nonumber \\
 &+& \textstyle \frac{n^0_2}{2}
 n_{11}\hspace{-0.2em}\left(1+\hspace{-0.2em}\left[\frac{n_1^0 + 1}{2}
 \hspace{-0.2em}\right]\hspace{-0.2em}\right)
 \hspace{-0.2em}\left|{n}_1^0 + 1 {} mod{} 2, n_2^0-1, n_{11}-1 +
 \hspace{-0.2em}\left[\frac{n_1^0+1}{2}\hspace{-0.2em}\right]
 , {n}_{12}, {n}_{22}, n\hspace{-0.2em}\right\rangle_V,
 \nonumber
    \end{eqnarray}
   \vspace{-3ex}
   \begin{eqnarray}
\label{l'22}
  l^{\prime 22}
\left|\vec{n}_k^0, \vec{n}_{lm}, n\rangle_V
 \right. &=&  n_{22}(n_{12} + n + n_2^0 + n_{22} -1 + h^2)
 \left|\vec{n}_k^0, {n}_{11}, {n}_{12}, {n}_{22}-1, n\rangle_V
 \right. \\
  && + \textstyle \frac{n_{12}n}{2}(n-1 + h^2 - h^1)\left|\vec{n}_k^0, {n}_{11}, {n}_{12}-1, {n}_{22},
 n-1\rangle_V\,,
 \right.
\nonumber \\
 && + \textstyle \frac{n_{12}(n_{12}-1)}{4}\left|\vec{n}_k^0, {n}_{11}+1, {n}_{12}-2, {n}_{22},
 n\rangle_V
 \right.
\nonumber \\
& +& \textstyle
\frac{n_2^0n_{12}}{2}\hspace{-0.2em}\left(1+\hspace{-0.2em}\left[\frac{n_1^0
+ 1}{2} \hspace{-0.2em} \right]
\hspace{-0.2em}\right)\hspace{-0.2em}\left|{n}_1^0 + 1 {} mod{} 2,
n_2^0-1, n_{11} +
\hspace{-0.2em}\left[\frac{n_1^0+1}{2}\hspace{-0.2em}\right]
 , {n}_{12}-1, {n}_{22}, n\hspace{-0.2em}\right\rangle_V,
 \nonumber \\
 \label{t'}
  t^{\prime}
\left|\vec{n}_k^0, \vec{n}_{lm}, n\rangle_V
 \right. &=& n(h^1-h^2-n+1)\left|\vec{n}_k^0, \vec{n}_{lm}, n-1\rangle_V
 \right.  \\
  && -n_{12}\hspace{-0.2em}\left|\vec{n}_k^0, {n}_{11}+1, {n}_{12}-1, {n}_{22}, n\rangle_V
 \right. - 2n_{22}\hspace{-0.2em}\left|\vec{n}_k^0, {n}_{11}, {n}_{12}+1, {n}_{22}-1,
 n\rangle_V
 \right.
 \nonumber\\
 &-&  n_2^0\textstyle \left(1+\left[\frac{n_1^0 +
1}{2}
 \right]\right)\left|{n}_1^0 + 1 {} mod{} 2, n_2^0-1, n_{11} + \left[\frac{n_1^0+1}{2}\right]
 , {n}_{12}, {n}_{22}, n\right\rangle_V.
 \nonumber
\end{eqnarray}
Using expressions (\ref{t'+i})--(\ref{t'}) and the mapping
(\ref{map}), we  reconstruct the action of the operators
$E^{\prime \alpha}$, $E^{\prime -\alpha}$, $g_0^{\prime i}$ in the
Fock space $\mathcal{H}'$, namely,
\begin{eqnarray}\label{t'+iF}
 t^{\prime i +} & = & f^+_i + 2b_{ii}^+f_i
 +4\delta_{i2}b_{12}^+f_1
  \,,
 \\
 t^{\prime+}   & = & b^+ - 2 b_{12}^+b_{11}-b_{22}^+b_{12} -
 f_2^+f_1 + 2 b_{22}^+ f_1f_2
   \,, \label{t'+F}\\
   g_0^{\prime i}& = & f_i^+f_i + \sum_{l\leq m}
 b_{lm}^+b_{lm}(\delta^{il}+\delta^{im}) + b^+b(\delta^{i2}-
 \delta^{i1}) +h^i
 \,,\label{g'0iF} \\
 l^{\prime+}_{ij} & = & b_{ij}^+\,.
 \label{l'+ijF}
\end{eqnarray}
\vspace{-3ex}
\begin{eqnarray}
  t^{\prime 1} &=& - f_1^+b_{11} - \textstyle\frac{1}{2}f_2^+b_{12} -
  2\left(b_{11}^+b_{11}
  +b_{12}^+b_{12}-b^+b + h^1\right)f_1+ \left(2b^+ - b_{22}^+b_{12}\right)f_2 \,, \label{t'1F}\\
t^{\prime 2} &=& - f_2^+b_{22} - \textstyle\frac{1}{2}f_1^+b_{12}
 - 2\left(b_{22}^+b_{22}
  -f_1^+f_1 + b^+b + h^2\right)f_2  \label{t'2F}\\
   && + 2\left(h^1-h^2- b^+b\right)bf_1 - \left(b_{11}^+b_{12}+ 4b_{12}^+b_{22}\right)f_1 \,,
   \nonumber\\
l^{\prime 11} &=& \left(b_{11}^+b_{11}+b_{12}^+b_{12}-b^+b +
f^+_1f_1 + h^1\right)b_{11} - \textstyle\frac{1}{2}b^+b_{12} +
\textstyle\frac{1}{4}b^+_{22}b^2_{12}
\label{l'11F}\\
   && + \textstyle\frac{1}{2}(f_2^++2b_{22}^+f_2)f_1b_{12} -2 b^+f_2f_1\,,
   \nonumber
 \end{eqnarray}
\vspace{-3ex}
\begin{eqnarray}
  l^{\prime 12} &=& \textstyle\frac{1}{4}\Bigl[\displaystyle\sum_{i}\left(2b_{ii}^+b_{ii} + f_i^+f_i + h^i\right)
  + b_{12}^+b_{12}\Bigr]b_{12}  + \textstyle\frac{1}{2}\left(b^+b + h^2 -h^1\right)b b_{11}
  \label{l'12F} \\
   && +\left(b_{12}^+b_{11} - \textstyle\frac{1}{2}b^+\right)b_{22}
   + \textstyle\frac{1}{2}f_2^+f_1b_{22} + \textstyle\frac{1}{2}f_1^+f_2b_{11} \nonumber\\
  && +
   \left(b^+b + b_{11}^+b_{11} + b_{22}^+b_{22}+h^2\right)f_2f_1  \,, \nonumber\\
  l^{\prime 22} &=& \left(b_{22}^+b_{22}+b_{12}^+b_{12}+b^+b +
f^+_2f_2 + h^2\right)b_{22} + \textstyle\frac{1}{2}\left(b^+b +
h^2 -h^1\right)b_{12}b \label{l'22F}\\
   &&  + \textstyle\frac{1}{4}b_{11}^+b_{12}^2
+\textstyle\frac{1}{2}f_1^+f_2b_{12}
 + b_{11}^+b_{12}f_2f_1\,,  \nonumber \\
t^{\prime } &=& \left(h^1-h^2 - b^+b\right)b - b_{11}^+b_{12} -2b_{12}^+b_{22} - f_1^+f_2 -2 b_{11}^+f_2f_1
 \label{t'F}\,.
\end{eqnarray}

Note that the additional parts $E^{\prime\alpha}, E^{-
\prime\alpha}$ do not obey the usual properties
\begin{equation}
 \left(E^{\prime\alpha}\right)^+\neq E^{- \prime\alpha}\,,
\label{hermcong}
\end{equation}
if one should use the standard rules of Hermitian conjugation for
the new creation and annihilation operators,
\begin{equation}
(b_{ij})^+=b_{ij}^+, \qquad (b)^+=b^+, \qquad (f_i)^+=f_i^+.
\end{equation}
To restore the proper Hermitian conjugation properties for the
additional parts, we change the scalar product in the Fock space
$\mathcal{H}'$ as follows:
\begin{eqnarray}
\langle\tilde{\Psi}_1|\Psi_2\rangle_{\mathrm{new}} =
\langle\tilde{\Psi}_1|K'|\Psi_2\rangle\,, \label{newsprod}
\end{eqnarray}
for any vectors $|\Psi_1\rangle, |\Psi_2\rangle$ with some, yet
unknown, operator $K'$. This operator is determined by the
condition that all the operators of the algebra  must have the
proper Hermitian properties with respect to the new scalar
product:
\begin{align}
& \langle\tilde{\Psi}_1|K'E^{- \prime\alpha}|\Psi_2\rangle =
\langle\tilde{\Psi}_2|K'E^{\prime\alpha}|\Psi_1\rangle^* ,  &&
\langle\tilde{\Psi}_1|K'g_0^{\prime i}|\Psi_2\rangle =
\langle\tilde{\Psi}_2|K'g_0^{\prime i}|\Psi_1\rangle^*.
\end{align}
These relations permit one to determine the operator $K'$,
Hermitian with respect to the usual scalar product
$\langle\, |\, \rangle$, as follows:
\begin{eqnarray}
\label{explicit K} K'=Z^+Z, \qquad
Z=\sum_{(\vec{n}_{lm},n)=(\vec{0},0)}^{\infty}\sum_{\vec{n}^0_k=(0,0)}^{{(1,1)}}
\left|\vec{n}^0_k,\vec{n}_{lm},n\rangle_V\right.\frac{1}{(\vec{n}_{lm})!n!}\langle
0|b^{n}b_{11}^{n_{11}}b_{12}^{n_{12}}b_{22}^{n_{22}}f_1^{n_1^0}f_2^{n_2^0},
\end{eqnarray}
where $(\vec{n}_{lm})! ={n}_{11}!{n}_{12}!{n}_{22}! $. One can
show by direct calculation  that the following relation holds true:
${}_V\left\langle\vec{n}^{\prime{}0}_k,\vec{n}'_{lm},n'\right.
\left|\vec{n}^0_k,\vec{n}_{lm},n\rangle_V\right.\sim
\delta^{n^0_1+2n_{11}+n_{12}-n}_{n^{\prime
{}0}_1+2n'_{11}+n'_{12}-n'}\delta^{n^0_2+n_{12}+2n_{22}+n}_{n^{\prime
{}0}_2+n'_{12}+2n'_{22}+n'}$.
 For low pairs of
numbers $(n^0_1+ 2n_{11}+n_{12} - n, n^0_2 +n_{12}+2n_{22}+ n)$,
with $n, n_{ij}$ being the numbers of ``particles'' associated
with $b^+, b_{ij}^+$ for $i\leq j$ (where $b^+$
reduces the spin number $s_1$ by one unit and increases the spin number
$s_2$ by one unit simultaneously) and $n^0_k$ being the number of
``particles'' associated with $f_k^+$, the operator $K'$ reads
\begin{eqnarray}
K' &=& |0\rangle\langle0| + (h^1-h^2)b^+|0\rangle\langle0|b
 -2h^i f^+_i|0\rangle\langle0|f_i + 2f_2^+|0\rangle\langle0|(h^1-h^2) bf_1 \nonumber \\
 && + f^+_1b^+|0\rangle\langle0|\Bigl(2 bf_1(h^2-h^1)(h^1-1) +
2 f_2(h^1-h^2)\Bigr)  \nonumber \\
&& + b_{12}^+|0\rangle\langle0|\Bigl( \frac{1}{4}b_{12} (h^1+h^2)
+ f_2f_1 h^2 +  \frac{1}{2}
 b_{11}b (h^2-h^1)\Bigr) + h^ib_{ii}^+|0\rangle\langle 0|b_{ii} \nonumber \\
  && + b_{11}^+b^+|0\rangle\langle0|\Bigl(b_{11}b (h^1-h^2)(h^1-1)
  + \frac{1}{2}b_{12} (h^2-h^1) + 2f_2f_1 (h^2-h^1)
 \Bigr) \nonumber \\
 && + f^+_1f^+_2|0\rangle\langle0|\Bigl(4 f_2f_1 (h^2h^1 + h^2 -
 h^1) + b_{12} h^2 + 2 b_{11}b (h^2-h^1)\Bigr)+
 \ldots \,.\label{K}
\end{eqnarray}
This expression for the operator $K'$ will be used later in
constructing the examples of Section~\ref{Examples}.

Thus,  we have constructed  the additional parts $o'_a$,
(\ref{t'+iF})--(\ref{t'F}), for the constraints $o_a$. In the next
section, we determine the algebra of the extended constraints and
find the BRST operator corresponding to this algebra.

\section{The Converted Superalgebra and the BFV--BRST \\ Operator}\label{conversionBRST}
\setcounter{equation}{0}

The superalgebra of the converted operators $O_I$,
\begin{equation}\label{OI}
    O_I = (O_a, O_p), \qquad O_a = o_a + o'_a, \qquad O_p =
    o_p\,,\
    o_p \in \{t_0, l_0, l_i, l^+_i\}\,,
\end{equation}
has the same form as the superalgebra of the initial operators
$o_I$, and therefore it is determined by the  relations of
Table~\ref{table in} under the replacement $o_I \leftrightarrow
O_I$. Despite the fact that the operators $\mathcal{H}^i$ do not
belong to the constraint system, and in order to provide a
Lorentz-covariant description of BRST cohomology  spaces, we do
not impose the restrictions $\mathcal{H}^i|\chi\rangle_{def} = 0$
on the vector $|\chi\rangle_{def}$, being the vector $|\Phi\rangle$
(\ref{PhysState}) enlarged into the tensor product of the Fock spaces
$\mathcal{H}_{def} = \mathcal{H}\otimes \mathcal{H}'$,
\begin{eqnarray}
\label{defState} |\chi\rangle_{def} =
\sum_{k_a}\left(f_i^+\right)^{k_i}\left(b_{lm}^+\right)^{k_{lm}}\left(b^+\right)^k
a^{+\mu_1}_1\ldots\,a^{+\mu_{k_{10}}}_1a^{+\nu_1}_2\ldots\,a^{+\nu_{k_{20}}}_2
\chi^{k_1{}k_2{}k_{11}{}k_{12}{}k_{22}{}k}_{(\mu)_{k_{10}},(\nu)_{k_{20}}}(x)\,
|0\rangle \,,
\end{eqnarray}
and include $\mathcal{H}^i$ into the converted first-class
constraint system, with respect to which we construct the BRST
operator $Q'$. The sum in (\ref{defState}) is taken  over $k_{i0},
k_{lm}, k$, running from 0 to infinity, and over $k_i$, running
from 0 to 1 \textbf{for $i =0,1$, $l,m =1,2$, $l\leq m$}. Having
constructed $Q'$, we extract from it the operators
$\mathcal{H}^i$, enlarged by means of the ghost variables
$\mathcal{C}, \mathcal{P}$ up to new operators $\sigma^i$,
$\sigma^i = \left(\mathcal{H}^i - h^i +
O\left(\mathcal{C}\mathcal{P}\right)\right)$, which will be used
to describe, by virtue of the equations $(\sigma^i +
h^i)|\chi\rangle = 0$, the direct sum of the Fock subspaces
$\mathcal{H}_{(n_1,n_2)}$ of a definite generalized spin
$\mathbf{s}=(n_1+\frac{1}{2},n_2+\frac{1}{2})$ in the enlarged
Hilbert space $\mathcal{H}_{tot} = \mathcal{H} \otimes
\mathcal{H}'\otimes \mathcal{H}_{gh}$ for $|\chi\rangle \in
\mathcal{H}_{tot}$. In this case, the remaining operator $Q$,
independent of the ghost variables $\eta^i_{\mathcal{H}},
\mathcal{P}^i_{\mathcal{H}}$ associated with $\mathcal{H}^i$, in
$Q'$ = $Q + O\left(\eta^i_{\mathcal{H}},
\mathcal{P}^i_{\mathcal{H}}\right)$, is  covariant and nilpotent
in each space
 $\mathcal{H}_{(n_1,n_2)}$ for the converted constraint system $O_I$
 without $\mathcal{H}^i$. Then, substituting instead of the
 parameters  $-h^i$ the operators $\sigma^i$, we obtain a nilpotent
 BRST operator  in the complete space $\mathcal{H}_{tot}$ without
 $\eta^i_{\mathcal{H}},  \mathcal{P}^i_{\mathcal{H}}$, which
 encodes the superalgebra of the converted constraints  $\{O_I\}\setminus \{\mathcal{H}^i\}$
 for fermionic HS fields with two rows of the Young tableaux.

 The construction of a nilpotent fermionic BRST operator for a
 Lie superalgebra is based on a principle similar to those
 developed  in \cite{symferm-flat,symferm-ads}: see the
 general analysis of the
 BFV quantization in the reviews \cite{BFV,bf,Henneaux}. Following the
 prescription of \cite{bf}, the
BRST operator constructed on a basis of the superalgebra presented
in Table~\ref{table in} can be found in an exact form, with the
use of the $(\mathcal{C} \mathcal{P})$-ordering of the ghost
coordinate $\mathcal{C}^I$ and momenta $\mathcal{P}_I$ operators,
as follows:
\begin{equation}\label{generalQ'}
    Q'  = {O}_I\mathcal{C}^I + \frac{1}{2}
    \mathcal{C}^I\mathcal{C}^Jf^K_{JI}\mathcal{P}_K (-1)^{\varepsilon({O}_K) + \varepsilon({O}_I)}
\end{equation}
with the constants $f^K_{IJ}$ written in a compact $x$-local
representation, $\{O_I\,, O_J] = f^K_{IJ}O_K$, and, according to
Table~\ref{table in}, $Q'$ has the form
\begin{eqnarray}
\label{Q'} {Q}' \hspace{-0.4em} &=&\hspace{-0.4em} q_0T_0+q_i^+T^i
+ T_i^+q^i +\eta_0L_0+\eta_i^+L^i+ L_i^+\eta^i +\eta_{lm}^+L^{lm}
+L_{lm}^+\eta^{lm} + \eta^+T + T^+\eta+
\eta^i_{\mathcal{H}}\mathcal{H}_i \nonumber
\\\hspace{-0.4em}&&
{}\hspace{-0.4em} +i(\eta_i^+q^i-\eta^iq_i^+)p_0 +
(\eta^i_{\mathcal{H}}q_i+\eta_{ii}q_i^+)p_i^+ +
(\eta^i_{\mathcal{H}}q_i^++\eta_{ii}^+q^i)p_i \nonumber
\\
\hspace{-0.4em}&&{}\hspace{-0.4em}
-\imath(q_0^2-\eta_i^+\eta^i){\cal{}P}_0 - \imath(2q^iq_i^+ -
\eta_{ii}^+\eta^{ii}){\cal{}P}^i_{\mathcal{H}} + (
\eta^i_{\mathcal{H}}\eta_i^++\eta_{ii}^+\eta^i
-2q_0q_i^+){\cal{}P}^i \nonumber
\\
\hspace{-0.4em}&&{}\hspace{-0.4em} +(\eta_i\eta^i_{\mathcal{H}}
+\eta_i^+\eta^{ii}-2q_0q^i){\cal{}P}_i^+ +2(\eta^i_{\mathcal{H}}
\eta_{ii}^+-q_i^{+2}){\cal{}P}_{ii} +2(\eta_{ii}
\eta^i_{\mathcal{H}}
-q_i^2){\cal{}P}_{ii}^+ \nonumber\\
\hspace{-0.4em} &-& \hspace{-0.4em} 2
\left[\textstyle\frac{1}{2}(\eta^1_{\mathcal{H}}+\eta^2_{\mathcal{H}})\eta_{12}
-\eta^+\eta_{22} -\eta\eta_{11} +2 q_1q_2\right]\mathcal{P}_{12}^+
+ \eta^+\eta_{12}\mathcal{P}_{11}^+ +
\eta\eta_{12}\mathcal{P}_{22}^+ \nonumber
\\
\hspace{-0.4em}&& \hspace{-0.4em}+2
\left[\textstyle\frac{1}{2}(\eta^1_{\mathcal{H}}+\eta^2_{\mathcal{H}})\eta_{12}^+
-\eta^+\eta^+_{11} -\eta\eta_{22}^+ -2
q_1^+q_2^+\right]\mathcal{P}_{12} -\eta\eta_{12}^+\mathcal{P}_{11}
- \eta^+\eta_{12}^+\mathcal{P}_{22} \nonumber
\\
\hspace{-0.4em}&&{}\hspace{-0.4em}- \imath
\eta\eta^+\left({\cal{}P}^1_{\mathcal{H}}-{\cal{}P}^2_{\mathcal{H}}\right)
+ \textstyle
\frac{\imath}{4}\eta_{12}^+\eta_{12}\left({\cal{}P}^1_{\mathcal{H}}+
{\cal{}P}^2_{\mathcal{H}}\right)\nonumber\\
 \phantom{{Q}' \hspace{-3em} =}
\hspace{-0.4em}&&{}\hspace{-0.4em}+
\left[\textstyle\frac{1}{2}\eta_{12}^+\eta_{11}+
\textstyle\frac{1}{2}\eta_{22}^+\eta_{12} - 2q_1q_2^+ +
(\eta^2_{\mathcal{H}}-\eta^1_{\mathcal{H}})\eta^+\right]\mathcal{P}
\nonumber
\\
\hspace{-0.4em}&&{} \hspace{-0.4em} +
\left[\textstyle\frac{1}{2}\eta_{11}^+\eta_{12}+
\textstyle\frac{1}{2}\eta_{12}^+\eta_{22} - 2q_2q_1^+ -
(\eta^2_{\mathcal{H}}-\eta^1_{\mathcal{H}})\eta\right]\mathcal{P}^+
\nonumber
\\
\hspace{-0.4em}&&{}\hspace{-0.4em}+
\left[\textstyle\frac{1}{2}q_1\eta_{12}^+ - \eta^+q_1^+\right]p_2
+ \left[\textstyle\frac{1}{2}q_1^+\eta_{12} - \eta q_1\right]p_2^+
+ \left[\textstyle\frac{1}{2}q_2\eta_{12}^+ - \eta
q_2^+\right]p_1\nonumber
\\
\hspace{-0.4em}&& \hspace{-0.4em}+
\left[\textstyle\frac{1}{2}q_2^+\eta_{12} - \eta^+ q_2\right]p_1^+
+ \left[\textstyle\frac{1}{2}\eta_{12}^+\eta_2 - \eta
\eta_2^+\right]\mathcal{P}_1  +
\left[\textstyle\frac{1}{2}\eta_2^+\eta_{12} + \eta^+
\eta_2\right]\mathcal{P}_1^+\nonumber
\\
\hspace{-0.4em}&&{}\hspace{-0.4em}
 + \left[\textstyle\frac{1}{2}\eta_{12}^+\eta_1 - \eta^+
\eta_1^+\right]\mathcal{P}_2  +
\left[\textstyle\frac{1}{2}\eta_1^+\eta_{12} - \eta_1
\eta\right]\mathcal{P}_2^+ .
\end{eqnarray}
Here, we imply summation over the repeated index $i$,  and the
raising (lowering) of the indices $i,j$ in quantities $f^{ij}$ is
made by the two-dimensional Euclidian metric tensor $g^{ij}$
($g_{ij}$), $g^{ij} = diag(1, 1)$. The quantities
 $q_0$, $q_i$, $q_i^+$ and  $\eta_0$, $\eta_i^+$, $\eta_i$,
$\eta_{lm}^+$, $\eta_{lm}$, $\eta$, $\eta^+$,
$\eta^i_{\mathcal{H}}$ are, respectively, bosonic and fermionic
ghost ``coordinates'' corresponding to their canonically conjugate
ghost ``momenta'' $p_0$, $p_i^+$, $p_i$, ${\cal{}P}_0$,
${\cal{}P}_i$, ${\cal{}P}_i^+$, ${\cal{}P}_{lm}$,
${\cal{}P}_{lm}^+$, $\mathcal{P}^+$, $\mathcal{P}$,
${\cal{}P}^i_{\mathcal{H}}$ for $i,l,m =1,2$, $l\leq m$.  They
form a set of Wick ghost pairs, $\bigl\{\left(q_i, p_i^+\right)$,
$\left(p_i, q_i^+\right)$, $\left({\eta}_i,
\mathcal{P}_i^+\right)$, $\left(\mathcal{P}_i, {\eta}_i^+\right)$,
$\left({\eta}_{lm}, \mathcal{P}_{lm}^+\right)$,
$\left(\mathcal{P}_{lm}, {\eta}_{lm}^+\right)$, $\left({\eta},
\mathcal{P}^+\right)$, $\left(\mathcal{P}, {\eta}^+\right)
\bigr\}$, and a set of zero-mode pairs, $\bigl\{\left(q_0,
p_0\right)$, $\left({\eta}_0, \mathcal{P}_0\right)$,
$\left({\eta}^i_{\mathcal{H}}, \mathcal{P}^i_{\mathcal{H}}\right)
\bigr\}$. Following \cite{bf}, they obey the nonvanishing
(anti)commutation relations
\begin{align}\label{ghosts}
& \{\eta,{\cal{}P}^+\}= \{{\cal{}P}, \eta^+\}=
\{\eta_i,{\cal{}P}_i^+\}= \{{\cal{}P}_i, \eta_i^+\}=1\,, &&
\{\eta_{lm},{\cal{}P}_{lm}^+\}=
\{{\cal{}P}_{lm}, \eta_{lm}^+\} =1\,, \nonumber \\
& [q_i, p_i^+] = [p_i, q_i^+] = 1\,,  &&  [q_0, p_0] =
\{\eta_0,{\cal{}P}_0\}= \{\eta^i_{\mathcal{H}},
{\cal{}P}^i_{\mathcal{H}}\}
 = \imath\,;
\end{align}
they also possess the  standard  ghost number distribution,
$gh(\mathcal{C}^i)$ = $ - gh(\mathcal{P}_i)$ = $1$, providing the
property  $gh(\tilde{Q}')$ = $1$, and have the Hermitian
conjugation properties of zero-mode pairs,\footnote{In terms of
the redefinition $\left(p_i, {\cal{}P}_0, {\cal{}P}^i_{\mathcal{H}}
\right) \mapsto \imath \left(p_i, {\cal{}P}_0,
{\cal{}P}^i_{\mathcal{H}} \right)$, the BRST operator (\ref{Q'})
and relations (\ref{ghosts}) are written in the notation of
\cite{symferm-flat, symferm-ads}.}
\begin{eqnarray}\label{Hermnull}
 \left(q_0, \eta_0, \eta^i_{\mathcal{H}}, p_0, {\cal{}P}_0,
{\cal{}P}^i_{\mathcal{H}} \right)^+ & = & \left(q_0, \eta_0,
\eta^i_{\mathcal{H}}, p_0, - {\cal{}P}_0,
-{\cal{}P}^i_{\mathcal{H}}\right).
\end{eqnarray}
The property of the BRST operator to be Hermitian is defined by
the rule
\begin{eqnarray}\label{HermQ}
  Q^{\prime +}K = K Q'\,,
  \end{eqnarray}
and is calculated with respect to the scalar product $\langle \ |\
\rangle$ in $\mathcal{H}_{tot}$ with the measure $d^dx$, which, in
its turn, is constructed as the direct product of the scalar
products in $\mathcal{H}, \mathcal{H}'$ and $\mathcal{H}_{gh}$.
The operator $K$ in (\ref{HermQ}) is the tensor product of the
operator $K'$ in $\mathcal{H}'$ and the unit operators in
$\mathcal{H}$, $\mathcal{H}_{gh}$
\begin{eqnarray} \label{tK}
  K &=&  \hat{1} \otimes K' \otimes \hat{1}_{gh}\,.
\end{eqnarray}

 Thus, we have constructed a Hermitian BRST
operator for the entire superalgebra of $O_I$. In the next
section, this operator will be used to construct a Lagrangian
action for fermionic HS fields of spin $(s_1, s_2)$ in a flat
space.

\section{Construction of Lagrangian Actions}\label{LagrFormulation}
\setcounter{equation}{0}

The construction of Lagrangians for fermionic  higher-spin fields
in a $d$-dimensional Minkowski  space can be developed by partially
following the algorithm of \cite{symferm-flat}, which is a
particular case of our construction, corresponding to $n_2 = 0$.
As a first step, we extract the dependence of the BRST operator
$Q'$ (\ref{Q'}) on the ghosts $\eta^i_{\mathcal{H}},
{\cal{}P}^i_{\mathcal{H}}$, so as to obtain the BRST operator $Q$
only for the system of converted first-class constraints
$\{O_I\} \setminus \{\mathcal{H}^i\}$:
\begin{eqnarray} \label{Q'decomp}
Q'&=&Q+\eta^i_{\mathcal{H}}(\sigma^i+h^i)+\mathcal{A}^i
\mathcal{P}^i_{\mathcal{H}}\,,
\end{eqnarray}
where
\begin{eqnarray}
\label{Q} Q &=& q_0T_0+q_i^+T^i + T_i^+q^i +\eta_0L_0+\eta_i^+L^i+
L_i^+\eta^i +\eta_{lm}^+L^{lm} +L_{lm}^+\eta^{lm} + \eta^+T +
T^+\eta \nonumber
\\&&
{} +i(\eta_i^+q_i-\eta_iq_i^+)p_0 +\eta_{ii}q_i^+p_i^+
+\eta_{ii}^+q_ip_i -i(q_0^2-\eta_i^+\eta_i){\cal{}P}_0
+(\eta_{ii}^+\eta_i-2q_0q_i^+){\cal{}P}_i \nonumber
\\
&&{} +(\eta_i^+\eta_{ii}-2q_0q_i){\cal{}P}_i^+
-2q_i^{+2}{\cal{}P}_{ii}
-2q_i^2{\cal{}P}_{ii}^+ \nonumber \\
 &-&  2
\left[2 q_1q_2 -\eta^+\eta_{22} -\eta\eta_{11}
\right]\mathcal{P}_{12}^+ + \eta^+\eta_{12}\mathcal{P}_{11}^+ +
\eta\eta_{12}\mathcal{P}_{22}^+ \nonumber
\\
&& - 2 \left[2 q_1^+q_2^+ + \eta^+\eta^+_{11} + \eta\eta_{22}^+
\right]\mathcal{P}_{12} -\eta\eta_{12}^+\mathcal{P}_{11} -
\eta^+\eta_{12}^+\mathcal{P}_{22} \nonumber
\\
&&{} + \left[\textstyle\frac{1}{2}\eta_{12}^+\eta_{11}+
\textstyle\frac{1}{2}\eta_{22}^+\eta_{12} - 2q_1q_2^+
\right]\mathcal{P}
  + \left[\textstyle\frac{1}{2}\eta_{11}^+\eta_{12}+
\textstyle\frac{1}{2}\eta_{12}^+\eta_{22} - 2q_2q_1^+
\right]\mathcal{P}^+ \nonumber
\\
&&{} + \left[\textstyle\frac{1}{2}q_1\eta_{12}^+ -
\eta^+q_1^+\right]p_2 + \left[\textstyle\frac{1}{2}q_1^+\eta_{12}
- \eta q_1\right]p_2^+ + \left[\textstyle\frac{1}{2}q_2\eta_{12}^+
- \eta q_2^+\right]p_1\nonumber
\\
&& + \left[\textstyle\frac{1}{2}q_2^+\eta_{12} - \eta^+
q_2\right]p_1^+ + \left[\textstyle\frac{1}{2}\eta_{12}^+\eta_2 -
\eta \eta_2^+\right]\mathcal{P}_1  +
\left[\textstyle\frac{1}{2}\eta_2^+\eta_{12} + \eta^+
\eta_2\right]\mathcal{P}_1^+\nonumber
\\
&&{}
 + \left[\textstyle\frac{1}{2}\eta_{12}^+\eta_1 - \eta^+
\eta_1^+\right]\mathcal{P}_2  +
\left[\textstyle\frac{1}{2}\eta_1^+\eta_{12} - \eta_1
\eta\right]\mathcal{P}_2^+ \,,
\end{eqnarray}
\vspace{-3ex}
\begin{equation}\label{Aical}
\mathcal{A}^i  =  \imath \Bigl\{\eta_{ii}^+\eta_{ii}-2q_iq_i^+ +
\eta\eta^+(\delta^{i1}- \delta^{i2}) +
\textstyle\frac{1}{4}\eta_{12}^+\eta_{12}(\delta^{i1}+
\delta^{i2})\Bigr\}.
\end{equation}
%
%
%
%
%
extended by the ghost Wick-pair variables, has the form
\begin{eqnarray}
\label{sigmai}
  \sigma^i &=& \mathcal{H}^i - h^i +  q_ip_i^+ + q_i^+p_i  - \eta_i \mathcal{P}^+_i +
   \eta_i^+ \mathcal{P}_i - 2\eta_{ii} \mathcal{P}^+_{ii} +
   2\eta_{ii}^+ \mathcal{P}_{ii}\nonumber\\
   &&  (\delta^{i1}+\delta^{i2})[\eta_{12}^+ \mathcal{P}_{12} - \eta_{12}
   \mathcal{P}^+_{12}] + (\delta^{i2}-\delta^{i1})[\eta^+ \mathcal{P} - \eta
   \mathcal{P}^+]\,.
\end{eqnarray}

Second, we choose a representation of the Hilbert space permitting
us to find the BRST cohomology spaces for the first-class
constraint system,
\begin{equation}
\left( p_0, q_i, p_i, \mathcal{P}_0, {\cal{}P}^i_{\mathcal{H}},
\eta_i, {\cal{}P}_i, \eta_{lm}, {\cal{}P}_{lm}, \eta,
\mathcal{P}\right)|0\rangle =0, \label{ghostvac}
\end{equation}
and to extract from $\mathcal{H}_{tot}$ the Hilbert subspace that
does not depend on the $\eta^i_{\mathcal{H}}$ operators (since
$\mathcal{H}^i$ are not first-class constraints as the other
$O_I$),
\begin{eqnarray} \label{chitot}
&  |\chi\rangle = \displaystyle\sum\limits_{k_r}
(q_0)^{k_1}(q_i^+)^{k_{2i}}(p_i^+)^{k_{3i}}(\eta_0)^{k_4}(f^+_i)^{k_{5i}}
  (\eta_i^+)^{k_{6i}}(\mathcal{P}_i^+)^{k_{7i}}(\eta_{lm}^+)^{k^8_{lm}}
  (\mathcal{P}_{lm}^+)^{k^9_{lm}}
  (\eta^+)^{k_{10}}(\mathcal{P}^+)^{k_{11}}\times&\nonumber \\
   & \times (b_{no}^+)^{k^{12}_{no}}(b^+)^{k_{13}}
a^{+\mu_1}_1\ldots\,a^{+\mu_{k_{14}}}_1a^{+\nu_1}_2\ldots\,a^{+\nu_{k_{15}}}_2
\chi^{k_1{}k_{2i}...k_{13}}_{(\mu)_{k_{14}},(\nu)_{k_{15}}}(x)
   |0\rangle. &
\end{eqnarray}
The sum in (\ref{chitot}) is taken  over $k_1, k_{2i}, k_{3i}$,
$k^{12}_{no}$, $k_{13}, k_{14}, k_{15}$, running from 0 to
infinity, for $i,l,m,n,o = 1,2$, $l\leq m, n\leq o$,  and over the
other indices, running from 0 to 1. Next, we derive from the
equations determining the physical vector, $Q'|\chi\rangle$ = $0$,
and from the reducible gauge transformations, $\delta|\chi\rangle$
= $Q'|\Lambda\rangle$, $\delta|\Lambda\rangle =
Q'|\Lambda^{(1)}\rangle$, $\ldots$, $\delta|\Lambda^{(s-1)}\rangle
= Q'|\Lambda^{(s)}\rangle$, a sequence of relations:
\begin{align}
\label{Qchi} & Q|\chi\rangle=0, && (\sigma^i+h^i)|\chi\rangle=0,
&& \left(\varepsilon, {gh}\right)(|\chi\rangle)=(1,0),
\\
& \delta|\chi\rangle=Q|\Lambda\rangle, &&
(\sigma^i+h^i)|\Lambda\rangle=0, && \left(\varepsilon,
{gh}\right)(|\Lambda\rangle)=(0,-1), \label{QLambda}
\\
& \delta|\Lambda\rangle=Q|\Lambda^{(1)}\rangle, &&
(\sigma^i+h^i)|\Lambda^{(1)}\rangle=0, && \left(\varepsilon,
{gh}\right)(|\Lambda^{(1)}\rangle)=(1,-2),\\
& \delta|\Lambda^{(s-1)}\rangle=Q|\Lambda^{(s)}\rangle, &&
(\sigma^i+h^i)|\Lambda^{(s)}\rangle=0, && \left(\varepsilon,
{gh}\right)(|\Lambda^{(s)}\rangle)= (s,-s-1). \label{QLambdai}
\end{align}
The middle set of equations in (\ref{Qchi})--(\ref{QLambdai})
determines the possible values of the parameters $h^i$ and the
eigenvectors of the operators $\sigma^i$. Solving these equations,
we obtain a set of eigenvectors, $|\chi\rangle_{(n_1,n_2)}$,
$|\Lambda\rangle_{(n_1,n_2)}$, $\ldots$,
$|\Lambda^{(s)}\rangle_{(n_1,n_2)}$, $n_1 \geq n_2 \geq 0$, and a
set of eigenvalues,
\begin{eqnarray}
\label{hi} -h^i &=& n^i+\frac{d-4}{2} - \delta^{2i}2\;, \quad
i=1,2\,,\quad n_1 \in \mathbb{Z}, n_2 \in \mathbb{N}_0\,,
\end{eqnarray}
with $(n_1,n_2)$ related to spin,
$\mathbf{s}=(n_1,n_2)+(1/2,1/2)$. The values of $n_1, n_2$ are
related to the spin components $s_1, s_2$  of the field, because
the proper vector $|\chi\rangle_{(n_1,n_2)}$ corresponding to
$(h_1,h_2)$ has the leading term $\chi^{0\cdots
0}_{(\mu)_{n_1},(\nu)_{n_2}}(x)$, independent of the auxiliary and
ghost operators, which corresponds to the field
$\Phi_{(\mu)_{n_1},(\nu)_{n_2}}(x)$ with the initial value of spin
$\mathbf{s} = (s_1,s_2)$ in the decomposition (\ref{chitot}),
\begin{eqnarray}\label{spin s}
    |\chi\rangle_{(n_1,n_2)} & = &
\Bigl[a_1^{+\mu_1}\ldots\,a_1^{+\mu_{n_1}}a_2^{+\nu_1}\ldots\,a_2^{+\nu_{n_2}}
\chi^{0\cdots 0}_{(\mu)_{n_1},(\nu)_{n_2}}(x)  \nonumber \\
& &   +
    b^{+}a^{+\mu_1}\ldots\,a^{+\mu_{n_1+1}}a_2^{+\nu_1}\ldots\,a_2^{+\nu_{n_2-1}}\chi^{0\cdots
    0{}1}_{(\mu)_{n_1+1},(\nu)_{n_2-1}}(x)  \nonumber \\
& & +
b^{+}_{12}a^{+\mu_1}\ldots\,a^{+\mu_{n_1-1}}a_2^{+\nu_1}\ldots\,a_2^{+\nu_{n_2-1}}\chi^{0\cdots
    0{}1{}0{}0}_{(\mu)_{n_1-1},(\nu)_{n_2-1}}(x) + ...
\Bigr]|0\rangle ,
\end{eqnarray}
where the values of $(n_1,n_2)$ can be composed of the set of
coefficients $\{k_r\} \setminus \{k_1,k_4\}$ in (\ref{chitot})
by the formulae
\begin{eqnarray}\label{nidecompos}
n_i &= & k_{2i} + k_{3i} + k_{5i} + k_{6i} + k_{7i} + 2k^8_{ii} +
k^8_{12} +  2k^9_{ii} + k^9_{12} \nonumber \\
&{}& 2k^{12}_{ii} + k^{12}_{12} + (-1)^{i}(k_{10} + k_{11} +
k_{13}) + k_{14}\delta_{i1} + k_{15}\delta_{i2}\,.
\end{eqnarray}
Therefore, relations (\ref{Qchi})--(\ref{QLambdai}) guarantee both
the extraction of vectors with the required value of spin and the
nilpotency of $Q$ in the corresponding Hilbert subspace. If one fixes
the value of spin, then the parameters  $h^i$ are also fixed by
(\ref{hi}). Having fixed the value of $h^i$, we should substitute it
into each of the expressions (\ref{Qchi})--(\ref{QLambdai}).

Third, we should extract the zero-mode ghosts from the operator
$Q$ as follows:
\begin{eqnarray}
\label{strQ} Q &=& q_0\tilde{T}_0+\eta_0{L}_0 +
\imath(\eta_i^+q_i-\eta_iq_i^+)p_0 -
\imath(q_0^2-\eta_i^+\eta_i){\cal{}P}_0 +\Delta{}Q,
\end{eqnarray}
where
\begin{eqnarray}
\label{tildeT0} \tilde{T}_0 &=& T_0 -2q_i^+{\cal{}P}_i
-2q_i{\cal{}P}_i^+\,,
\\
\label{deltaQ} \Delta{}Q & = & q_i^+T^i + T_i^+q^i +\eta_i^+L^i+
L_i^+\eta^i +\eta_{lm}^+L^{lm} +L_{lm}^+\eta^{lm} + \eta^+T +
T^+\eta \nonumber
\\&&
{}  +\eta_{ii}q_i^+p_i^+ +\eta_{ii}^+q_ip_i +
\eta_{ii}^+\eta_i{\cal{}P}_i + \eta_i^+\eta_{ii}{\cal{}P}_i^+
-2q_i^{+2}{\cal{}P}_{ii}
-2q_i^2{\cal{}P}_{ii}^+ \nonumber \\
 &-&  2
\left[2 q_1q_2 -\eta^+\eta_{22} -\eta\eta_{11}
\right]\mathcal{P}_{12}^+ + \eta^+\eta_{12}\mathcal{P}_{11}^+ +
\eta\eta_{12}\mathcal{P}_{22}^+ \nonumber
\\
&& - 2 \left[2 q_1^+q_2^+ + \eta^+\eta^+_{11} + \eta\eta_{22}^+
\right]\mathcal{P}_{12} -\eta\eta_{12}^+\mathcal{P}_{11} -
\eta^+\eta_{12}^+\mathcal{P}_{22} \nonumber
\\
&&{} + \left[\textstyle\frac{1}{2}\eta_{12}^+\eta_{11}+
\textstyle\frac{1}{2}\eta_{22}^+\eta_{12} - 2q_1q_2^+
\right]\mathcal{P}
  + \left[\textstyle\frac{1}{2}\eta_{11}^+\eta_{12}+
\textstyle\frac{1}{2}\eta_{12}^+\eta_{22} - 2q_2q_1^+
\right]\mathcal{P}^+ \nonumber
\\
&&{} + \left[\textstyle\frac{1}{2}q_1\eta_{12}^+ -
\eta^+q_1^+\right]p_2 + \left[\textstyle\frac{1}{2}q_1^+\eta_{12}
- \eta q_1\right]p_2^+ + \left[\textstyle\frac{1}{2}q_2\eta_{12}^+
- \eta q_2^+\right]p_1\nonumber
\\
&& + \left[\textstyle\frac{1}{2}q_2^+\eta_{12} - \eta^+
q_2\right]p_1^+ + \left[\textstyle\frac{1}{2}\eta_{12}^+\eta_2 -
\eta \eta_2^+\right]\mathcal{P}_1  +
\left[\textstyle\frac{1}{2}\eta_2^+\eta_{12} + \eta^+
\eta_2\right]\mathcal{P}_1^+\nonumber
\\
&&{}
 + \left[\textstyle\frac{1}{2}\eta_{12}^+\eta_1 - \eta^+
\eta_1^+\right]\mathcal{P}_2  +
\left[\textstyle\frac{1}{2}\eta_1^+\eta_{12} - \eta_1
\eta\right]\mathcal{P}_2^+ \,.
\end{eqnarray}
Here, $\tilde{T}_0$,  $\Delta{}Q$ are independent of $q_0$, $p_0$,
$\eta_0$, $\mathcal{P}_0$. We also expand the state vector and
gauge parameters in powers of the zero-mode ghosts:
\begin{align}
\label{0chi} |\chi\rangle &=\sum_{k=0}^{\infty}q_0^k(
|\chi_0^k\rangle +\eta_0|\chi_1^k\rangle), &
&gh(|\chi^{k}_{m}\rangle)=-(m+k),
\\
\label{0L} |\Lambda^{(s)}\rangle
&=\sum_{k=0}^{\infty}q_0^k(|\Lambda^{(s)}{}^k_0\rangle
+\eta_0|\Lambda^{(s)}{}^k_1\rangle), &
&gh(|\Lambda^{(s)}{}^k_m\rangle)=-(s+k+m+1) .
\end{align}
Following the procedure described in \cite{Sagnotti,
symferm-flat}, we get rid of all the fields except two,
$|\chi^0_0\rangle$, $|\chi^1_0\rangle$.

Namely, after the extraction of zero-mode ghosts from the BRST
operator  $Q$ (\ref{strQ}),  as well as from
 the state vector  and the gauge parameter (\ref{0chi}), (\ref{0L}),
 the gauge transformation for the fields $|\chi_0^k\rangle$, $k \geq 2$
 has the form
\begin{equation}
    \delta|\chi_0^k\rangle = \Delta{}Q |\Lambda_0^k\rangle + \eta_i\eta_i^+
    |\Lambda_1^k\rangle +
    (k+1)(q_i\eta_i^+-\eta_iq_i^+)|\Lambda_0^{k+1}\rangle +
    \tilde{T}_0|\Lambda_0^{k-1}\rangle+
    |\Lambda_1^{k-2}\rangle\,,
\end{equation}
implying, by induction, that  we can make all the fields
$|\chi_0^k\rangle$, $k \geq 2$  equal to zero by using the gauge
parameters $|\Lambda_1^{k}\rangle$. Then, considering  the
equations of motion for the powers $q_0^k, k\geq 3$ and taking
into account that $|\chi_0^k\rangle = 0$, $k \geq 2$, we can see
that these equations contain the subsystem
\begin{equation}\label{subsystem}
|\chi_1^{k-2}\rangle = \eta_i\eta_i^+ |\chi_1^k\rangle\,, \qquad k
\geq 3\,,
\end{equation}
which permits us to find, by induction, that all the fields
$|\chi_1^k\rangle$, $k \geq 1$ are equal to zero. Finally,
we examine the equations of motion for the power $q_0^2$:
\begin{equation}\label{lasteqs}
|\chi_1^{0}\rangle = - \tilde{T}_0 |\chi_0^1\rangle\,,
\end{equation}
  in order to express the vector $|\chi_1^0\rangle$ in terms of
  $|\chi_0^1\rangle$. Thus, as in the totally symmetric case, there remain
  only  two independent fields: $|\chi^0_0\rangle$, $|\chi^1_0\rangle$.
The first equation in (\ref{Qchi}), (\ref{strQ}), the
decomposition (\ref{0chi}), and the above analysis then imply that
the independent equations of motion for these vectors have the form
\begin{eqnarray}
&& \Delta{}Q|\chi^{0}_{0}\rangle
+\frac{1}{2}\bigl\{\tilde{T}_0,\eta_i^+\eta_i\bigr\}
|\chi^{1}_{0}\rangle =0, \label{EofM1all}
\\&&
\tilde{T}_0|\chi^{0}_{0}\rangle + \Delta{}Q|\chi^{1}_{0}\rangle
=0\,, \label{EofM2all}
\end{eqnarray}
where $\{F,G\}=FG+GF$ for any quantities $F, G$.

Then, due to the fact that the  operators $Q$, $\tilde{T}_0$,
$\eta_i^+\eta_i$ commute with  $\sigma^i$, we obtain from
(\ref{EofM1all}), (\ref{EofM2all}) the equations of
motion for the fields with a fixed value of  spin:
\begin{eqnarray}
&& \Delta{}Q|\chi^{0}_{0}\rangle_{(n_1,n_2)}
+\frac{1}{2}\bigl\{\tilde{T}_0,\eta_i^+\eta_i\bigr\}
|\chi^{1}_{0}\rangle_{(n_1,n_2)} =0, \label{EofM1}
\\&&
\tilde{T}_0|\chi^{0}_{0}\rangle_{(n_1,n_2)} +
\Delta{}Q|\chi^{1}_{0}\rangle_{(n_1,n_2)} =0. \label{EofM2}
\end{eqnarray}
where the fields $|\chi_0^k\rangle_{(n_1,n_2)}$, $k=0,1$ are
assumed to obey the relations
\begin{eqnarray}\label{schi}
\sigma^i|\chi^k_0\rangle_{(n_1,n_2)}
=\bigl(n_i+(d-4)/2-2\delta_{i2}\bigr)|\chi^k_0\rangle_{(n_1,n_2)}\,,
\qquad k=0,1.
\end{eqnarray}

The field equations (\ref{EofM1}), (\ref{EofM2}) are Lagrangian
ones and can be deduced, in view of the invertibility
of the operator $K$, from the following Lagrangian
action:\footnote{As usual, the action is defined up to an overall
factor.}
\begin{eqnarray}
{\cal{}S}_{(n_1, n_2)} &=& {}_{(n_1,
n_2)}\langle\tilde{\chi}^{0}_{0}|K_{(n_1,
n_2)}\tilde{T}_0|\chi^{0}_{0}\rangle_{(n_1, n_2)} +
\frac{1}{2}\,{}_{(n_1, n_2)}\langle\tilde{\chi}^{1}_{0}|K_{(n_1,
n_2)}\bigl\{
   \tilde{T}_0,\eta_i^+\eta_i\bigr\}|\chi^{1}_{0}\rangle_{(n_1,
   n_2)}\,,
\nonumber
\\&&
+ {}_{(n_1, n_2)}\langle\tilde{\chi}^{0}_{0}|K_{(n_1,
n_2)}\Delta{}Q|\chi^{1}_{0}\rangle_{(n_1, n_2)} + {}_{(n_1,
n_2)}\langle\tilde{\chi}^{1}_{0}|K_{(n_1,
n_2)}\Delta{}Q|\chi^{0}_{0}\rangle_{(n_1, n_2)}\,, \label{L1}
\end{eqnarray}
where the standard scalar product for the creation and
annihilation operators is assumed, and $K_{(n_1,
n_2)}$ is the operator $K$ (\ref{tK}) with the following
substitution: $h^i
\to-\bigl(n_i+(d-4)/2-2\delta_{i2}\bigr)$.

The equations of motion (\ref{EofM1}), (\ref{EofM2}) and the
action (\ref{L1}) are invariant with respect to the gauge
transformations
\begin{eqnarray}
\delta|\chi^{0}_{0}\rangle_{(n_1, n_2)} &=&
\Delta{}Q|\Lambda^{0}_{0}\rangle_{(n_1, n_2)}
 +
 \frac{1}{2}\bigl\{\tilde{T}_0,\eta_i^+\eta_i\bigr\}
 |\Lambda^{1}_{0}\rangle_{(n_1, n_2)}\,,
\label{GT1}
\\
\delta|\chi^{1}_{0}\rangle_{(n_1, n_2)} &=&
\tilde{T}_0|\Lambda^{0}_{0}\rangle_{(n_1, n_2)}
 +\Delta{}Q|\Lambda^{1}_{0}\rangle_{(n_1, n_2)}\,
 ,
\label{GT2}
\end{eqnarray}
which are reducible, with the gauge parameters
$|\Lambda^{(s)}{}^{j}_{0}\rangle_{(n_1, n_2)}$, $j=0,1$
 subject to the same conditions
as those for $|\chi^j_0\rangle_{(n_1, n_2)}$ in (\ref{schi}),
\begin{align}
\delta|\Lambda^{(s)}{}^{0}_{0}\rangle_{(n_1, n_2)} &=
\Delta{}Q|\Lambda^{(s+1)}{}^{0}_{0}\rangle_{(n_1, n_2)}
 +
 \frac{1}{2}\bigl\{\tilde{T}_0,\eta_i^+\eta_i\bigr\}
 |\Lambda^{(s+1)}{}^{1}_{0}\rangle_{(n_1, n_2)},
& |\Lambda^{(0)}{}^0_0\rangle =|\chi^0_0\rangle\,, \label{GTi1}
\\
\delta|\Lambda^{(s)}{}^{1}_{0}\rangle_{(n_1, n_2)} &=
\tilde{T}_0|\Lambda^{(s+1)}{}^{0}_{0}\rangle_{(n_1, n_2)}
 +\Delta{}Q|\Lambda^{(s+1)}{}^{1}_{0}\rangle_{(n_1, n_2)},
& |\Lambda^{(0)}{}^1_0\rangle = |\chi^1_0\rangle\,,
\label{GTi2}
\end{align}
and with a finite number of reducibility stages \footnote{In the case of
a spin-tensor field
$\Phi_{(\mu)_{n_1},(\nu)_{n_2},...,(\rho)_{n_k}}(x)$ with the Young
tableaux having $k$ rows, one can show that the stage of reducibility for
the corresponding Lagrangian formulation must be equal to $s_{max} =
\sum_{l=1}^k n_l + k(k-1)/2 - 1$, so that for a totally symmetric
field $\Phi_{(\mu)_{n_1}}(x)$ one has $s_{max} = n_1 -1$, in
accordance with \cite{symferm-flat}.} at $s_{max}=n_1+n_2$
for spin $\mathbf{s} = (n_1+1/2,n_2+1/2)$.

In addition to restrictions (\ref{nidecompos}), the set of coefficients
$\{k_r\} \setminus \{k_1,k_4\}$ in (\ref{chitot}) for fixed values of $n_i$,
satisfies the following equations for $|\chi^j_0\rangle_{(n_1,n_2)}$
$|\Lambda^{(s){}j_0}\rangle_{(n_1,n_2)}$, respectively,
\begin{eqnarray}\label{ghnum}
   \hspace{-0.7em} |\chi^j_0\rangle_{(n_1,n_2)} \hspace{-0.7em} &:& \hspace{-0.7em}
     \sum_{i}\left(k_{2i} - k_{3i} + k_{6i} - k_{7i}\right) +
\sum_{l\leq m}\left(k^8_{lm} - k^9_{lm}\right) + k_{10} - k_{11} =
-j ,
 \\ \label{ghLnum}
\hspace{-0.7em} |\Lambda^{(s){}j}_0\rangle_{(n_1,n_2)}
 \hspace{-0.7em} & : &  \hspace{-0.7em}\sum_{i}\left(k_{2i} - k_{3i} + k_{6i} -
 k_{7i}\right) +
\sum_{l\leq m}\left(k^8_{lm} - k^9_{lm}\right) + k_{10} - k_{11} =
-(s + j + 1),
\end{eqnarray}
due to the ghost number distribution (\ref{0chi}), (\ref{0L}).

Thus, we have constructed, by using the BRST procedure, a
gauge-invariant Lagrangian description of fermionic fields with a
mixed symmetry of any fixed spin $\mathbf{s}$.

A Lagrangian description for a half-integer mixed-symmetry HS field
of mass $m$ in a $d$-dimensional Minkowski space can be deduced
in two ways. First, one should use a modified procedure starting
from the Dirac equation
\begin{equation}\label{Eq-om}
  \big(\imath\gamma^{\mu}\partial_{\mu}-m\big)\Phi_{(\mu)_{n_1},(\nu)_{n_2}}
=0  \Longleftrightarrow \big(\imath \tilde{\gamma}^\mu\partial_\mu- \tilde{\gamma}
m\big)\Phi_{(\mu)_{n_1},(\nu)_{n_2}}  = 0,
\end{equation}
with the relations (\ref{Eq-1}), (\ref{Eq-2}) being unaltered. The equation (\ref{Eq-om})
contains a massive term in both even and odd space-time dimensions (see footnote~3),
so that an equivalent description in terms of the Clifford algebra elements
$\tilde{\gamma}^\mu$, $\tilde{\gamma}$
is possible for both $d=2N$ and $d=2N+1$:
\begin{equation}\label{evengamm}
\tilde{\gamma} = \kappa_{d}\Pi  {\gamma}^{d} \ \ \mathrm{for}\ \
{\gamma}^{d}=\frac{1}{d!}\Big(\prod_{i=1}^{d}{\gamma}_{\mu_i}\Big)\epsilon^{\mu_1\ldots \mu_d}
\ \mathrm{and} \ \kappa_{d} = \left\{\begin{array}{cc}   1 , & d=4M ,\\
\imath , & d=4M+2,
\end{array}
\right.
\end{equation}
with an odd non-degenerate supermatrix $\Pi$, such that $\Pi^2=1$, and
with the Levi-Civita tensor $\epsilon^{\mu_1\ldots \mu_d}$
normalized as $\epsilon^{01\ldots d-1}=1$.
For $d=2N-1$, we enlarge all the spin-tensors $\Phi_{A(\mu)_{n_1},(\nu)_{n_2}}(x)$,...
to those with doubled components
$\widehat{\Phi}_{2A(\mu)_{n_1},(\nu)_{n_2}}=
(\Phi_{A(\mu)_{n_1},(\nu)_{n_2}}, \Phi_{A(\mu)_{n_1},(\nu)_{n_2}})^T$,...,
with the Dirac index $A=1,...,2^{N-1}$, while also enlarging $\gamma^\mu_{AB}$
to $\gamma^\mu_{(2A)(2B)}$=$\rm{antidiag}(\gamma^{\mu}_{AB},\gamma^\mu_{AB})$
(see (\ref{reduction11})), which are now elements of a $d'=d+1$-dimensional
space-time (although with $\gamma^d_{(2A)(2B)}$ being absent), where
${\gamma}^{d+1}={\gamma}^{2N}$. The  Lagrangian formulation is determined
by the same relations as those in the massless case, with some modifications:
first, for the initial operators $t_0, l_0$,
\begin{equation}\label{changeferm}
  \big(t_0,\, l_0\big) \to \big(\check{t}_0,\, \check{l}_0\big) = \big({t}_0 + \tilde{\gamma} m ,\, {l}_0 + m^2\big),
\end{equation}
which, along with the remaining elements of $o_I$,
obey the same HS symmetry superalgebra, except for the additional
commutators
\begin{equation}\label{ll+}
  [t^+_i, l_j] = \delta_{ij}(\check{t}_0 - \tilde{\gamma}m),\qquad  [t_i, l^+_j] = -\delta_{ij}(\check{t}_0 - \tilde{\gamma}m),\qquad   [l_i, l^+_j] = \delta_{ij}(\check{l}_0 -  m^2).
\end{equation}
The additional parts $o'_I$ coincide with those of the massless case,
whereas the converted set of constraints has the form $O^m_I=\check{o}_I+o'_I$,
no longer having a central charge $\tilde{\gamma}m$ for massive HS fields
with $\check{o}_I = {o}_I+{\hat{o}}_I(b_i, b_i^+)$ (additional bosonic
$4$-oscillators $[b_i, b_j^+]=\delta_{ij}$ acting in the Fock space
$\mathcal{H}_m$), determined by adding the terms induced by dimensional reduction
\begin{eqnarray}
&& \left(\hat{t}_0,\, \hat{l}_0\,, \hat{l}_i,\, \hat{l}_i^+, \hat{g}_0^i\right)  = \left(0,\, 0,\,  mb_i, \,   mb_i^+,\,b^+_ib_{i} + \frac{1}{2}\right) , \label{hat0}\\
 && \left(\hat{t}_i,\, \hat{t}_i^+\,, \hat{l}_{ij},\, \hat{l}_{ij}^+,\,\hat{t},\,\hat{t}^+\right) = -\left(\tilde{\gamma}b_i,\, \tilde{\gamma}b_i^+,\,
\frac{1}{2}b_ib_{j}, \,
 \frac{1}{2}b^+_ib^+_{j},\,b^+_{1}b_2,\,b_{1}b^+_2\right) . \label{hat1}
\end{eqnarray}
The set of $O^m_I$ satisfies the same HS symmetry superalgebra as the one
for massless half-integer irreducible representations of the Poincare group).
The generalized spin and BRST operators, $\sigma^i_m = \mathcal{H}^i_m + ...$,
$Q_m= Q\vert_{O\to O^m}$, with an arbitrary vector
$|\chi_{m}\rangle \in \mathcal{H}^m_{tot}=\mathcal{H}_{tot}\otimes \mathcal{H}_m$,
are formally identical with those of the massless case (\ref{sigmai}), (\ref{Q}),
after the replacement $\mathcal{H}^i\to \mathcal{H}^i_m$, whereas
$|\chi_{m}\rangle$ has the vector $|\chi\rangle$ (\ref{chitot}) as the massless
limit for $b^+_i=0$:
\begin{eqnarray}
|\chi_m \rangle &=& \sum_{n'_l\geq (0)_l} \prod_{l=1}^2(b_l^+)^{n'_l}|\chi_{n'_l} (a^+,f^+,q^+,p^+,\eta^+,\mathcal{P}^+ ) \rangle  \ \mathrm{ for} \ |\chi_{n'_l}\rangle \in \mathcal{H}_{tot}. \label{chifm}
\end{eqnarray}
Note that the $b_l^+$-independent vectors $|\chi_{n'_l}\rangle$, for $l=0,1,2$,
have a decomposition in oscillator powers, given by  (\ref{chitot}).
At the same time, one may follow  the  dimensional reduction
\cite{ScherkShcwarz, RINDANISAHDEV-dimred-ferm} of a massless HS field theory
of the same type in a $(d+1)$-dimensional flat space $\mathbb{R}^{1,d}$.
To this end, we make a projection $\mathbb{R}^{1,d}  \to
\mathbb{R}^{1,d-1}$ over the sphere $S^1$ with a simple decomposition,
\vspace{-1ex}\begin{align} \label{reduction}
   &\partial^{M} = (\partial^{\mu}, -\imath m)\,, &&a^{M}_i = (a^{\mu}_i, b_i)\,, &&
   a^{M{}+}_i = (a^{\mu{}+}_i, b_i^+)\,,    \\
   &M=0,1,\ldots ,d\,, && \mu=0,1,\ldots ,d-1\,, && \eta^{MN} =
   diag (1,-1,\ldots,-1,-1)\,,\label{reduction1}
   \end{align}
   \begin{eqnarray}
   && \tilde{\gamma}^M_{\textstyle 2^{\left[\frac{d+1}{2}\right]}} \equiv \tilde{\gamma}^M_{\textstyle 2^{N}} =   \left\{\begin{array}{ll}
                                                                                                                                          \tilde{\gamma}^{\mu}_{\textstyle 2^{\left[\frac{d}{2}\right]}};
   \tilde{\gamma}_{\textstyle 2^{\left[\frac{d}{2}\right]}}\equiv \tilde{\gamma}{}^d_{\textstyle 2^{\left[\frac{d}{2}\right]}} , & \quad d=2N ,\\
                                                                                                                                      \left(                                                                                                                                                  \begin{array}{cc}
                                                                                                                                                    0 & \tilde{\gamma}^{\mu}_{\textstyle 2^{N-1}} \\
                                                                                                                                                       \tilde{\gamma}^{\mu}_{\textstyle2^{N-1} } & 0 \\
                                                                                                                                                  \end{array}\right);                                                                                                                                        \tilde{\gamma}^d_{\textstyle 2^{N}}     , & \quad d=2N-1,
                                                                                                                                        \end{array}
  \right. , \label{reduction11}\\
   && \Phi_{A(M)_{n_1},(N)_{n_2}}(x,x^d )  =   \left\{\begin{array}{ll}
                                                                                                                                          \exp\{i m x^d\}\Phi_{A(\mu)_{n_1-r_1},(\nu)_{n_2-r_2}} (x) , \ r_i=0,...,n_i  , & d=2N \\
                                                                                                                                      \exp\{i m x^d\}\left(\hspace{-0.7ex}\begin{array}{l}\Phi_{L|a(\mu)_{n_1-r_1},(\nu)_{n_2-r_2}} \\
                                                                                                                                         \Phi_{R|b(\mu)_{n_1-r_1},(\nu)_{n_2-r_2}}   \end{array}\hspace{-0.7ex}\right)(x), & \hspace{-0.3em} d=2N-1,
                                                                                                                                        \end{array}
  \right. \label{reduction12}
\end{eqnarray}
for $A=(a,b)$ with $d=2N$ and  $a,b=1,...,2^{N-1}$.
For odd $d=2N-1$, it is only the $2^{N-1}$-dimensional $\gamma$-matrices and spinors
in the left-hand side of (\ref{reduction11}), (\ref{reduction12}) that correspond
to an irreducible representation.
The reduction from $\mathbb{R}^{1,2N}$ to $\mathbb{R}^{1,2N-1}$ leads precisely
to the Dirac equation (\ref{Eq-om}) for $(M,N)=(\mu,\nu)$:
$\imath\tilde{\gamma}^{M}\partial_{M}\Phi_{A(\mu)_{n_1},(\nu)_{n_2}}(x,x^d)=
\exp\{i m x^d\} \big(\imath\tilde{\gamma}^{\mu}\partial_{\mu} +
\imath^2 \tilde{\gamma}m \big)\Phi_{A(\mu)_{n_1},(\nu)_{n_2}}(x)=0$.
To make a projection from $\mathbb{R}^{1,2N-1}$ to $\mathbb{R}^{1,2N-2}$, one may use
the following representation of ${\gamma}^M=\tilde{\gamma}^M\tilde{\gamma}$-matrices,
which can be realized inductively, according to (\ref{reduction11}),
as an outer product of the unity $\mathbf{1}_2$ and Pauli $\sigma_i$-matrices,
$i=1,2,3$, for any even $d$:
\begin{eqnarray}\label{gamma0}
\hspace{-0.8em}   &\hspace{-0.8em}&\hspace{-0.8em} d=2: \  \big({\gamma}^0_{\textstyle 2}, {\gamma}^1_{\textstyle 2}, {\gamma}^2_{\textstyle 2}\big)=  \big(\sigma_1 ,  \imath\sigma_2, \imath{\gamma}^0_{\textstyle 2}{\gamma}^1_{\textstyle 2}\big)\ \mathrm{for}\   \big(\sigma_1,  \sigma_2,  \sigma_3\big)  = \hspace{-0.3em}\left(\hspace{-0.3em}\left(\hspace{-0.3em}\begin{array}{cc}
                                        0 & 1 \\
                                        1 & 0
                       \end{array}\hspace{-0.3em}
   \right)\hspace{-0.3em},\left(\hspace{-0.3em}\begin{array}{cc}
                                        0 & -\imath \\
                                        \imath & 0
                       \end{array}\hspace{-0.3em}
   \right)\hspace{-0.3em},\left(\hspace{-0.3em}\begin{array}{cc}
                                        1 & 0 \\
                                        0 & -1
                       \end{array}\hspace{-0.3em}
   \right) \hspace{-0.3em}\right)\hspace{-0.3em} ,\\
   \label{gammapq}
 \hspace{-0.8em}   &\hspace{-0.8em}&\hspace{-0.8em} d=2N: \  {\gamma}^{\mu}_{\textstyle 2^{N}}   =  \rm{antidiag}\big({\gamma}^{\mu}_{\textstyle 2^{N-1}}, {\gamma}^{\mu}_{\textstyle 2^{N-1}} \big), \mu=0,...,d-2, \ {\gamma}^{d-1}_{\textstyle 2^{N}}=  \left( \hspace{-0.3em}\begin{array}{cc}
                                        0 & 1_{\textstyle 2^{N-1}} \\
                                        -1_{\textstyle 2^{N-1}} & 0                       \end{array}
   \hspace{-0.3em}
   \right)\hspace{-0.2em}, \\
 \label{gammad}
  \hspace{-0.8em}   &\hspace{-0.8em}&\hspace{-0.8em}  {\gamma}^{d}_{\textstyle 2^{N}} \  = \   (-\imath)^{d/2+2}\prod_{i=0}^{d-1}{\gamma}^{i}_{2^N}=\imath \hspace{-0.2em}\left(\hspace{-0.3em}\begin{array}{cc}
                                        - {1}_{2^{N-1}} & 0 \\
                                        0 & {1}_{2^{N-1}}
                       \end{array}\hspace{-0.3em}
   \right)\hspace{-0.2em},
\end{eqnarray}
where the gamma-matrices of a $d=2N$-dimensional space are used for $d=2N+1$,
and the matrix $\gamma^d_{2^N}$ is introduced (\ref{gammad}).
By the Pauli theorem, any set of $\gamma^\mu_{2^{\left[\frac{d}{2}\right]}}$-matrices
for a fixed $d$ can be obtained from the given set by using a similarity transformation.
In this case, a respective massless spinor possesses $2^N$ components and contains
two Weyl spinors in a transposed form $\big(\Phi_{L|a(M)_{n_1},(N)_{n_2}},
\Phi_{R|b(M)_{n_1},(N)_{n_2}}\big)^T(x,x^d )$, being an element of an $ISO(1,2N-1)$
reducible massless representation for a half-integer spin, and having the
property of an eigenvector for ${\gamma}^{d}_{\textstyle 2^{N}}$,
$\Phi_{R,L} = (1/2)(1\mp \imath {\gamma}^{d})\Phi$.
Any element of an irreducible massless representation has to be chiral,
e.g., right-handed, $\big(0,\Phi_{R|b(M)_{n_1},(N)_{n_2}}\big)^T(x,x^d )$,
which transforms by the projection (\ref{reduction12}) into the set
$\Phi_{R|b(\mu)_{n_1-r_1},(\nu)_{n_2-r_2}}(x)$.
According to (\ref{gamma0})--(\ref{gammad}), ${\gamma}^M_{2^N}$, $M=0,...,d$,
become ${\gamma}^\mu_{2^{N-1}}$, $\mu=0,...,d-1$, and
${\gamma}^d_{2^{N-1}}= \mathbf{1}_{2^{N-1}}$.
Once again, for the Dirac equation projected from $\mathbb{R}^{1,2N-1}$
to $\mathbb{R}^{1,2N-2}$ with $(M,N)=(\mu,\nu)$, we have
\begin{equation}\label{Eq-omv}
\imath{\gamma}^{M} \partial_{M}
\big(0,\Phi_{R|b(\mu)_{n_1},(\nu)_{n_2}}\big)^T(x,x^d ) =
\exp\{i m x^d\} \big(\imath{\gamma}^{\mu}\partial_{\mu}
- m \big)\Phi_{R|b(\mu)_{n_1},(\nu)_{n_2}}(x) =0.
\end{equation}
Left-handed massless spin-tensors can be subject to a similar treatment
and thereby represent after projection independent massive fermions.

For the doubled massive components $\widehat{\Phi}_{A(\mu)_{n_1},(\nu)_{n_2}}
= (\Phi_{R|a(\mu)_{n_1},(\nu)_{n_2}},
\Phi_{R|a(\mu)_{n_1},(\nu)_{n_2}})^T$,
the representation (\ref{Eq-om}) with $2^N$-dimensional
$\tilde{\gamma}^{\mu}$-matrices is also valid.

Therefore, the quantities $\tilde{\gamma}T_0\tilde{\gamma}$, identical
with $T_0$ for massless HS fields, transform for massive fields
as $\tilde{\gamma}T_0\tilde{\gamma}$ = $T_0^\ast$,  $T_0^\ast =
-\imath\tilde{\gamma}^\mu \partial_\mu  - \tilde{\gamma} m$.
Hence, for the coefficient functions entering (along with the
matrix $\tilde{\gamma}$) into the decomposition of any vector composed
of $|\chi^j_0\rangle_{(n_1,n_2)}$, $|\Lambda^{(s){}j}_0\rangle_{(n_1,n_2)}$,
the part of a vector $|\Xi\rangle_{(n_1,n_2)}$ homogeneous with respect
to $\tilde{\gamma}$ is subject to
\begin{align} \label{reduction2}
   &\imath{\gamma}^{M} \partial_{M}|\Xi\rangle =
   \Bigl(\imath{\gamma}^{\mu} \partial_{\mu}-(-1)^{\rm{deg}_{\tilde{\gamma}}|\Xi\rangle}m\Bigr)|\Xi\rangle,
    && \rm{deg}_{\tilde{\gamma}}|\Xi\rangle = 0,1 \,,&& |\Xi\rangle \in \{|\chi^j_0\rangle, |\Lambda^{(s){}j}_0\rangle
    \}\,.
\end{align}
Relations (\ref{reduction}) and (\ref{reduction1}) indicate
the presence of four additional second-class constraints, $l_i,
l_i^+$, with the corresponding oscillator operators $b_i, b_i^+$,
$[b_i, b_j^+] = \delta_{ij}$, in comparison with the massless
case.\footnote{Based on the above reasons, one can state
that the procedure of dimensional reduction given by
(\ref{reduction}), (\ref{reduction1}) can be applied to massless
mixed-symmetry bosonic fields \cite{BurdikPashnev} in order to
obtain a Lagrangian description of massive mixed-symmetry bosonic
fields, whereas for fermionic HS fields it is necessary to make
allowance for the matrix structure, according to
(\ref{reduction11}), (\ref{reduction12}), (\ref{Eq-omv}),
(\ref{reduction2}); cf. \cite{symferm-flat}.}

A simultaneous construction of Lagrangian actions describing
the propagation of all massless (massive) fermionic fields
with two rows of the Young tableaux in Minkowski space is similar
to the case of totally symmetric spin-tensors in flat spaces
\cite{symferm-flat}, and we only note that a necessary condition
for solving this problem is to replace in $Q^{\prime}$, $Q$, $K$
the parameters $-h^i$ by the operators $\sigma^i$ in an appropriate
way, and to discard the condition (\ref{schi}) for the fields and
gauge parameters. Amongst other things, this completes both the conversion
procedure for the initial constraint system $\{o_I\}\setminus \{g_0^i\}$
and the construction of a nilpotent BRST operator in the entire Hilbert
space for the set of converted constraints $\{O_I\}\setminus
\{\mathcal{H}^i\}$.

In the section to follow, we outline a proof of the fact that
the action, in fact, reproduces the correct equations of motion
(\ref{Eq-0})--(\ref{Eq-3}).

\section{Reduction to the Initial Irreducible Relations}\label{Proof}
\setcounter{equation}{0}

Let us briefly show the fact that  it is only the solutions of the
equations of motion (\ref{Eq-0})--(\ref{Eq-3}) that determine the
space of BRST cohomologies of the operator $Q$ (\ref{Q}) with a
vanishing ghost number in the Fock space $\mathcal{H}$ for the
basic fermionic field with spin $\mathbf{s} = (n_1+1/2,n_2+1/2)$.
To solve this problem, we can follow two ways: the first one is
realized, for instance, in \cite{Barnich} for massless
totally-symmetric bosonic fields in a flat space-time, and the
second one, for totally-symmetric fermionic fields
\cite{symferm-flat, symferm-ads}. We will use the  technics of
\cite{symferm-flat, symferm-ads}, taking into account the fact
that the spectrum of component fields for an arbitrary vector
$|\chi\rangle_{(n_1,n_2)}$ in (\ref{chitot}) for $k_1 = k_4 = 0$
is essentially larger than the spectrum for a totally-symmetric
fermionic vector $|\chi\rangle_{(n_1+n_2,0)}$, for whose
description one  should not use the operators $q_2^+$, $p_2^+$,
$f_2^+$, $\eta_2^+$, $\mathcal{P}_2^+$, $\eta_{12}^+$,
$\mathcal{P}_{12}^+$, $\eta_{22}^+$, $\mathcal{P}_{22}^+$,
$\eta^+$, $\mathcal{P}^+$, $b^+$, $b^+_{12}$, $b^+_{22}$,
$a^{+\nu}_{2}$ and the corresponding conjugations.\footnote{The
total number of independent ``creation'' operators which are
necessary to compose the vector $|\chi\rangle_{(n_1,n_2)}$ is more
than twice as large as the number required for
$|\chi\rangle_{(n_1+n_2,0)}$: $(2d +22)/(d+8)$.} As a consequence,
the character of proof is more involved even in comparison with
the case of the AdS space \cite{symferm-ads}.

In the standard manner, the proof consists of two steps.
 First, in order to simplify the spectrum of the gauge
parameters $|\Lambda^{(s)}{}^j_{0}\rangle$ and the fields
$|\chi^j_{0}\rangle$, $j=0,1$, we apply to them a gauge-fixing
based  on the structure of gauge transformations
(\ref{GT1})--(\ref{GTi2}) and extract the physical field
$|\Phi\rangle_{(n_1,n_2)}$ alone, by using (\ref{PhysState}) for
$s = (n_1+1/2, n_2+1/2)$:
\begin{equation}\label{relation}
    |\chi^0_0\rangle_{(n_1,n_2)} = |\Phi\rangle_{(n_1,n_2)} + |\Phi_A\rangle_{(n_1,n_2)}, \qquad
\left.\phantom{\Bigl[}|\Phi_A\rangle_{(n_1,n_2)}\right|_{\mathcal{C}=\mathcal{P}=b^+_{ij}=b^+=f^+_i
= 0} = 0.
\end{equation}
Second, we use a part of the Lagrangian equations of motion
(\ref{EofM1}), (\ref{EofM2}) in order to select from them
only the equations of motion for $|\Phi\rangle_{(n_1,n_2)}$,
 and to remove all of the remaining auxiliary fields of lower spins.

Let us now describe the basic sequence of gauge-fixing. Our
strategy consists in a successive elimination of the terms with
$\mathcal{P}^+_{11}$ from the fields $|\chi^j_{0}\rangle$
and gauge parameters $|\Lambda^{(s)}{}^j_{0}\rangle$, starting
from the top of the tower of gauge transformations
(\ref{GT1})--(\ref{GTi2}). For this purpose, it should be noted
that we have a reducible gauge theory of $(n_1+n_2)$-th stage of
reducibility. Because of the restrictions for the spin
(\ref{nidecompos}) and ghost number  (\ref{ghnum}),
(\ref{ghLnum}), the independent parameters of the lowest stage
have the form
\begin{align}
& |\Lambda^{(n_1+n_2)}{}^0_0\rangle_{(n_1,n_2)} \hspace{1ex} =
\hspace{1ex} (p^+_1)^{n_1}\left|A(p_i^+, \mathcal{P}_i^+,
\mathcal{P}^+)\rangle_{(0,n_2)}\right.\,, &&
|\Lambda^{(n_1+n_2)}{}^1_0\rangle_{(n_1,n_2)} \hspace{1ex} \equiv
\hspace{1ex} 0\,,\label{gauge param}
\end{align}
where the vector $\left|A(p_i^+, \mathcal{P}_i^+,
\mathcal{P}^+)\rangle_{(l,n_2-m)}\right.$  $\equiv
\left|A\rangle_{(l,n_2-m)}\right.$, $l < n_1$, $m < n_2$, has the
structure
\begin{eqnarray}
\left|A\rangle_{(l,n_2-m)}\right. &=&  \mathcal{P}^+\Bigl\{
\sum_{k=m+1}^{n_2}
(p_1^+)^k(p_2^+)^{n_2-k}|\omega^{1{}l}_k\rangle_{(-k+l+1,k-1-m)} \nonumber \\
   &&
   +
\mathcal{P}_1^+ \sum_{k=m}^{n_2-1}
(p_1^+)^k(p_2^+)^{n_2-k-1}|\omega^{2{}l}_k\rangle_{(-k+l,k-m)}
 \nonumber
\end{eqnarray}
\begin{eqnarray}
\phantom{\left|A\rangle_{(l,n_2-m)}\right.}
   && +  \mathcal{P}_2^+\sum_{k=m+1}^{n_2-1}
(p_1^+)^k(p_2^+)^{n_2-k-1}|\omega^{3{}l}_k\rangle_{(-k+l+1,k-1-m)}\nonumber \\
   &&
+ \mathcal{P}_1^+ \mathcal{P}_2^+\sum_{k=m}^{n_2-2}
(p_1^+)^k(p_2^+)^{n_2-k-2}|\omega^{4{}l}_k\rangle_{(-k+l,k-m)}\Bigr\}\,.
\label{Aln2m}
\end{eqnarray}

It can be verified directly that one can eliminate the dependence
on the ghost $\mathcal{P}_{11}^+$ from the gauge function
$|\Lambda^{(n_1+n_2-1)}{}^0_0\rangle$ of $(n_1+n_2-1)$-th stage
of reducibility, whereas the vector
$|\Lambda^{(n_1+n_2-1)}{}^1_0\rangle$ has the same structure as
$|\Lambda^{(n_1+n_2)}{}^0_0\rangle$ in (\ref{gauge param}).
Indeed, for $|\Lambda^{(n_1+n_2-1)}{}^0_0\rangle$ we have the
following expansion in the powers of $\mathcal{P}_{11}^+$,
$\mathcal{P}_{12}^+$, $\mathcal{P}_{22}^+$:
\begin{eqnarray}
 |\Lambda^{(n_1+n_2-1)}{}^0_0\rangle_{(n_1,n_2)} & = &
|\Lambda^{(n_1+n_2-1)}{}^0_{00}\rangle_{(n_1,n_2)} +
\mathcal{P}_{11}^+ (p^+_1)^{n_1-2}|\tilde{A}\rangle_{(0,n_2)}\,,
\nonumber\\
&& + \mathcal{P}_{12}^+
(p^+_1)^{n_1-1}|\tilde{A}\rangle_{(0,n_2-1)} +\mathcal{P}_{22}^+
(p^+_1)^{n_1}|\tilde{A}\rangle_{(0,n_2-2)}
\,,\label{gaugeparamn1n21}
\end{eqnarray}
with $|\tilde{A}\rangle_{(0,n_2-k)}$ defined according to (\ref{Aln2m}),
so that the  gauge transformation (\ref{GTi1}) at
$\mathcal{P}_{11}^+$ implies
\begin{equation}\label{firsttransf}
    \delta|\tilde{A}\rangle_{(0,n_2)} =
    -2q_1^2(p_1^+)^2 |{A}\rangle_{(0,n_2)}\,.
\end{equation}
After the vector $|\tilde{A}\rangle_{(0,n_2)}$ has been removed,
the theory is transformed to a theory of $(n_1+n_2-1)$-th stage of reducibility.
Then, it is  possible to verify that one can remove the
dependence of $|\Lambda^{(n_1+n_2-2)}{}^j_0\rangle$ on
$\mathcal{P}_{11}^+$ with the help of the remaining gauge
parameters $|\Lambda^{(n_1+n_2-1)}{}^j_0\rangle$, which do not depend on $\mathcal{P}_{11}^+$.

It then becomes possible to prove by induction that after removing the dependence
on $\mathcal{P}_{11}^+$ from the gauge parameters up to the
$(s+1)$-th stage, $|\Lambda^{(l)}{}^k_0\rangle$, $k=0,1$, $l \geq s+1$
(i.e., we have $\eta_{11}|\Lambda^{(l)}{}^k_0\rangle = 0$), and
applying the restricted vector $|\Lambda^{(s+1)}{}^k_0\rangle$,
one can eliminate the dependence on $\mathcal{P}_{11}^+$ from the
gauge functions $|\Lambda^{(s)}{}^k_0\rangle$. To this end, we introduce the
following notation for the gauge parameters related to their expansion
in the ghosts $\mathcal{P}_{ij}^+$:
\begin{eqnarray}
  |\Lambda^{(l)}{}^k_0\rangle &=& |\Lambda^{(l){}k}_{00}\rangle  +
  \sum_{i\leq j}\mathcal{P}_{ij}^+|\Lambda^{(l){}k}_{0ij}\rangle +\mathcal{P}_{11}^+\Bigl(
  \mathcal{P}_{12}^+|\Lambda^{(l){}k}_{01}\rangle\nonumber \\
   && +
  \mathcal{P}_{22}^+|\Lambda^{(l){}k}_{02}\rangle+
  \mathcal{P}_{12}^+\mathcal{P}_{22}^+|\Lambda^{(l){}k}_{03}\rangle \Bigr) +
  \mathcal{P}_{12}^+\mathcal{P}_{22}^+|\Lambda^{(l){}k}_{04}\rangle\,. \label{decompPij}
\end{eqnarray}
Here and elsewhere, we omit  the vector subscripts
associated with the eigenvalues  of the operators $\sigma^i$
(\ref{schi}). From (\ref{GTi1}), (\ref{GTi2}), we obtain the
gauge transformations for $|\Lambda^{(s){}k}_{011}\rangle$,
$|\Lambda^{(s){}k}_{0p}\rangle$, $p=1,2,3$, being the coefficients
at $\mathcal{P}_{11}^+$, namely,
\begin{eqnarray}
  \delta |\Lambda^{(s){}k}_{011}\rangle &=&  -2q^2_1|\Lambda^{(s+1){}k}_{00}\rangle + \eta^+|\Lambda^{(s+1){}k}_{012}\rangle\,,
  \label{lambdas011}\\
  \delta |\Lambda^{(s){}k}_{01}\rangle &=&  -2q^2_1|\Lambda^{(s+1){}k}_{012}\rangle\,,\label{lambdas01}\\
  \delta |\Lambda^{(s){}k}_{02}\rangle &=&  -2q^2_1|\Lambda^{(s+1){}k}_{022}\rangle+
  \eta^+|\Lambda^{(s+1){}k}_{04}\rangle\,,\label{lambdas02}\\
  \delta |\Lambda^{(s){}k}_{03}\rangle &=&  -2q^2_1|\Lambda^{(s+1){}k}_{04}\rangle \,. \label{lambdas03}
\end{eqnarray}
Then,  a certain choice for $|\Lambda^{(s+1){}k}_{04}\rangle$,
$|\Lambda^{(s+1){}k}_{012}\rangle$ removes
$|\Lambda^{(s){}k}_{03}\rangle$, $|\Lambda^{(s){}k}_{01}\rangle$,
respectively, whereas  a certain choice for
$|\Lambda^{(s+1){}k}_{022}\rangle$,
$|\Lambda^{(s+1){}k}_{00}\rangle$ eliminates
$|\Lambda^{(s){}k}_{02}\rangle$, $|\Lambda^{(s){}k}_{011}\rangle$
by means of the remaining gauge transformations. Thus, we have
shown that the dependence on $\mathcal{P}_{11}^+$ can be
eliminated from $|\Lambda^{(l)}{}^k_0\rangle$. As a consequence of
the above procedure, the theory becomes a gauge theory of $l$-th
stage of reducibility.

This algorithm is valid down to the  vector
$|\Lambda^{(n_2+1)}{}^k_0\rangle$, when there arise terms
linear in  $p_1^+$. When these terms are present, one deals with
gauge parameters that have remained unused after eliminating
the dependence on $\mathcal{P}_{11}^+$.
Therefore, in view of the $\eta^+$-dependent terms in
(\ref{lambdas011}), (\ref{lambdas02}), a gauge
transformation with such parameters may cause some $\mathcal{P}_{11}^+$-dependent
terms to appear in the transformed vector
$|\Lambda^{(n_2)}{}^k_0\rangle$. Consequently, it is necessary
to make a gauge transformation with parameters linear
in $p_1^+$, or independent of it, before removing the
$\mathcal{P}_{11}^+$-dependence. Let us examine a gauge
transformation with the gauge function
$|\Lambda^{(n_2)}{}^k_0\rangle$  more carefully.

Suppose that the dependence on the ghost $\mathcal{P}_{11}^+$ in
$|\Lambda^{(n_2+1)}{}^k_0\rangle$ has been removed by a gauge
transformation, and hence the functions
$|\Lambda^{(n_2+1)}{}^k_0\rangle$, $|\Lambda^{(n_2)}{}^k_0\rangle$
admit the following representation:
\begin{eqnarray}
 |\Lambda^{(n_2+1)}{}^0_0\rangle &=&  p_1^+\Bigl(|\Lambda^{(n_2+1)}{}^0_{00}\rangle + \mathcal{P}_{12}^+
 |\Lambda^{(n_2+1){}0}_{012}\rangle  + p_1^+\mathcal{P}_{22}^+
 |\Lambda^{(n_2+1){}0}_{022}\Bigr)\rangle\,,
  \label{lambdan2110}\\
  |\Lambda^{(n_2+1)}{}^1_{0}\rangle &=& (p_1^+)^2\Bigl(|\Lambda^{(n_2+1)}{}^1_{01}\rangle +
   \mathcal{P}_{12}^+|\Lambda^{(n_2+1)1}_{012}\rangle +p_1^+\mathcal{P}_{22}^+
   |\Lambda^{(n_2+1)1}_{022}\rangle\Bigr) \,,\label{lambdan2111}\\
   |\Lambda^{(n_2)}{}^0_{0}\rangle &=&  |\Lambda^{(n_2)}{}^0_{00}\rangle + \sum_{i\leq j}
   \mathcal{P}_{ij}^+|\Lambda^{(n_2){}0}_{0ij}\rangle
   \,,\label{lambdan210}\\
 |\Lambda^{(n_2)}{}^1_{0}\rangle &=&  p_1^+\Bigl(|\tilde{\Lambda}^{(n_2)}{}^1_{00}\rangle + \mathcal{P}_{12}^+
 |\tilde{\Lambda}^{(n_2){}1}_{012}\rangle  + p_1^+\mathcal{P}_{22}^+
 |\tilde{\Lambda}^{(n_2){}1}_{022}\rangle\Bigr) \,, \label{lambdan211}
\end{eqnarray}
where the vectors $|\Lambda^{(l)}{}^0_{00}\rangle$,
$|\Lambda^{(l){}k}_{0ij}\rangle$, $l = n_2, n_2+1$, $i,j=1,2,
i\leq j$,  possess terms having no dependence on $p_1^+$, except
for $|\Lambda^{(n_2){}0}_{022}\rangle$,  and the vectors in
(\ref{lambdan211}) have a structure analogous to the corresponding
structure  in (\ref{lambdan2110}). Then, one has to make a
transformation with parameters linear in $p^+_1$. We will use
$|\Lambda^{(n_2+1)}{}^0_{00}\rangle$,
$|\Lambda^{(n_2+1){}0}_{012}\rangle$  to make such gauge
transformations. Since
\begin{align}
    & \delta|\Lambda^{(n_2)}{}^0_{00}\rangle \hspace{1ex} =
    \hspace{1ex} T_1^+|\hat{\Lambda}^{(n_2+1)}{}^0_{00}\rangle\,,
    && |\hat{\Lambda}^{(n_2+1)}{}^0_{00}\rangle \hspace{1ex} = \hspace{1ex}
    q_1|{\Lambda}^{(n_2+1)}{}^0_{00}\rangle\,, \label{auxlambdan200}\\
& \delta|\Lambda^{(n_2){}0}_{012}\rangle \hspace{1ex} =
    \hspace{1ex} T_1^+|\hat{\Lambda}^{(n_2+1){}0}_{012}\rangle\,,
    && |\hat{\Lambda}^{(n_2+1){}0}_{012}\rangle\hspace{1ex} = \hspace{1ex}
    q_1|{\Lambda}^{(n_2+1){}0}_{012}\rangle\,, \label{auxlambdan2012}
\end{align}
one can use the vectors $|{\Lambda}^{(n_2+1)}{}^0_{00}\rangle$,
$|{\Lambda}^{(n_2+1){}0}_{012}\rangle$ to eliminate the dependence
on $b_{11}^+$ and $f_1^+$ from $|\Lambda^{(n_2)}{}^0_{00}\rangle$
and $|\Lambda^{(n_2){}0}_{012}\rangle$, respectively, due to the
fact that the $b_{11}^+$- and $f_1^+$-linear components of the
latter vectors are identical to the  corresponding components of
the previous vectors. As a result, we obtain the gauge-fixing
\begin{equation}\label{b11f1}
    b_{11}|\Lambda^{(n_2)}{}^0_{00}\rangle =
    f_1|\Lambda^{(n_2)}{}^0_{00}\rangle =b_{11}|\Lambda^{(n_2){}0}_{012}\rangle
    = f_1|\Lambda^{(n_2){}0}_{012}\rangle = 0\,,
\end{equation}
and then remove the $\mathcal{P}_{11}^+$-dependence from
$|\Lambda^{(n_2)}{}^j_{0}\rangle$, as has been described in the
case of the system (\ref{lambdas011})--(\ref{lambdas03}).

Proceeding by induction, we may use the algorithm which has been applied
to the treatment of the vectors (\ref{lambdan210}), (\ref{lambdan211}) in
order to eliminate the dependence on $\mathcal{P}_{11}^+$  related
to all the vectors down to $|\Lambda^{(0)}{}^j_{0}\rangle$,
whereas the $\mathcal{P}_{22}^+$-independent terms in
$|\Lambda^{(l)}{}^j_{0}\rangle$ for $l\geq n_2$ are restricted by
relations of the form (\ref{b11f1}).

Let us now turn to the gauge-fixing of the fields. We expand the
fields in the powers of the ghosts $\mathcal{P}_{ij}^+$ by analogy
with the gauge parameters:
\begin{eqnarray}
  |\chi^k_0\rangle &=& |\chi^k_{00}\rangle  +
  \sum_{i\leq j}\mathcal{P}_{ij}^+|\chi^{k}_{0ij}\rangle +\mathcal{P}_{11}^+\Bigl(
  \mathcal{P}_{12}^+|\chi^{k}_{01}\rangle\nonumber \\
   && +
  \mathcal{P}_{22}^+|\chi^{k}_{02}\rangle+
  \mathcal{P}_{12}^+\mathcal{P}_{22}^+|\chi^{k}_{03}\rangle \Bigr) +
  \mathcal{P}_{12}^+\mathcal{P}_{22}^+|\chi^{k}_{04}\rangle\,. \label{decompchiij}
\end{eqnarray}
Further, we need to restrict the vectors by the gauge conditions
(\ref{b11f1}),  which follow from the gauge transformations, and then
we eliminate the terms coupled to $\mathcal{P}_{11}^+$.

Having completed the above procedure, we briefly mention that the
remaining gauge ambiguity is sufficient to eliminate the auxiliary
oscillators $b_{ij}^+$, $b^+$, $f^+_i$ from the field
$\hspace{-0.3ex}|\chi^{00000}_{00}\rangle $,
\begin{equation}\label{chik00}
    |\chi^k_{00}\rangle = \sum_{l_i\geq 0} \sum_{m_j = 0}^1
    (p_i^+)^{l_i}(\mathcal{P}_j^+)^{m_j}|\chi^{kl_im_j}_{00}\rangle\,,
\end{equation}
and therefore, in view of $gh(|\chi^{00000}_{00}\rangle) = 0$,
this field has no dependence on the ghost ``coordinates'',  so
that, after the gauge-fixing, we conclude
\begin{eqnarray}
|\chi^{00000}_{00}\rangle&=&|\Phi\rangle\,. \label{PhSt}
\end{eqnarray}

The second step of establishing an equivalence of
equations (\ref{Eq-0})--(\ref{Eq-3}) with the Lagrangian equations
(\ref{EofM1}), (\ref{EofM2}) is more involved and is based on
a detailed expansion of equations (\ref{EofM1}),
(\ref{EofM2}) in the powers of $p_i^+$, $q_i^+$, $\eta_i^+$,
$\eta_{ij}^+$, $\eta^+$ and then in the powers of $b_{ij}^+$, $f_i^+$.
We only state the result that after gauge-fixing
$|\chi^k_{00}\rangle$ and $|\chi^k_{012}\rangle$
$|\chi^k_{022}\rangle$, $|\chi^k_{04}\rangle$, expanded
by analogy with (\ref{chik00}), the only independent equations among
(\ref{EofM1}), (\ref{EofM2}) have the form
\begin{equation}\label{finaleqs}
    t_0|\Phi\rangle = t_i|\Phi\rangle = t|\Phi\rangle = 0\,,
\end{equation}
and all of the auxiliary fields can be made equal to zero.

In what follows, we consider some examples of the Lagrangian formulation
procedure.


\section{Examples}\label{Examples}
\setcounter{equation}{0}

Here, we shall realize the general prescriptions of
our Lagrangian formulation in the case of fermionic fields of lowest spins.

\subsection{Spin-(3/2,1/2) Field}\label{Example3/2}

In the case of a field of spin (3/2,1/2), we have $(n_1,n_2)=(1,0)$,
$(h^1,h^2) = (1-d/2,4-d/2)$. Since $s_{max}=0$, the corresponding
Lagrangian formulation is an irreducible gauge theory and
describes a totally symmetric fermionic field of spin $s = 3/2$.
The nonvanishing fields $|\chi^i_0\rangle_{(1,1)}$ and gauge
parameters $|\Lambda^0_0\rangle_{(1,0)}$,
 (for
$|\Lambda^{1}_0\rangle_{(1,0)} \equiv 0$, due to
${gh}(|\Lambda^{1}_0\rangle_{(1,0)}) = -2$), have the following
Grassmann grading and ghost number distribution:
\begin{align}
& \left(\varepsilon, {gh}\right)(|\chi_0^i\rangle_{(1,0)})=(1,
-i)\,, && \left(\varepsilon,
{gh}\right)(|\Lambda_0^0\rangle_{(1,0)})=(0,-1)\, .
\end{align}
These conditions determine the dependence of the fields and gauge
parameters on the oscillator variables in a unique form, with the
help of the operators corresponding only to the first row of the
Young tableaux,
\begin{align}\label{chi320}
& |\chi_0^0\rangle_{(1,0)} = \left[-i a_1^{+\mu}\psi_{\mu}(x) +
f_1^+ \tilde{\gamma}\psi(x)\right]|0\rangle, &&
|\chi_0^1\rangle_{(1,0)} =
\left[\mathcal{P}^{+}_1\tilde{\gamma}\chi(x) +p_1^+\chi_1(x)
\right]|0\rangle,\\
& \label{brachi320} {}_{(1,0)}\langle\tilde{\chi}_0^0| =
\langle0|\left[i \psi_{\mu}^+(x)a_1^{\mu} +
\psi^+(x)\tilde{\gamma} f_1 \right]\tilde{\gamma}^0,  &&
{}_{(1,0)}\langle\tilde{\chi}_0^1| = \langle 0| \left[
\chi^+(x)\tilde{\gamma}\mathcal{P}_1 +
    \chi^+_1p_1
\right]\tilde{\gamma}^0, \\
& |\Lambda_0^0\rangle_{(1,0)} = \left[ \mathcal{P}^{+}_1\xi_1(x)+
p_1^+\tilde{\gamma}\xi_2(x) \right]|0\rangle ,  \label{l001}
\end{align}
Substituting (\ref{chi320}), (\ref{brachi320}) into (\ref{L1}), we
find the action   (up to an overall factor) for a free massless field
of spin $(3/2,1/2)$ on a flat background:
\begin{eqnarray}
{\cal{}S}_{(1,0)} &=& \int d^dx  \left[\bar{\psi}^\mu\Bigl\{
i\gamma^\nu\partial_\nu\psi_\mu -\partial_\mu\chi - i \gamma_\mu \chi_1 \Bigr\} +
(d-2)\bar{\psi} \Bigl\{ i\gamma^\mu\partial_\mu\psi
 +\chi_1 \Bigr\} \right.\nonumber
\\
&&{} \left.+  \bar{\chi}\Bigl\{ i\gamma^\mu\partial_\mu \chi
- \chi_1 +\partial^\mu\psi_\mu  \bigr\}+\bar{\chi}_1\Bigl\{
i\gamma^\mu\psi_\mu +(d-2)\psi -\chi\Bigr\}\right]
 \label{L3/21/2} .
\end{eqnarray}
In deriving the action
(\ref{L3/21/2}), we have used the expressions (\ref{K}), (\ref{tK})
for the operators
$K_{(1,0)}$\footnote{For $n_2=0$, we have the case of
totally symmetric spin-tensors in a $d$-dimensional flat space
\cite{symferm-flat}, so that the total Hilbert space
$\mathcal{H}_{tot}$ and all of the operators acting on it can be
factorized from $q^+_2,\eta^+_2, \eta^+_{12},
\eta^+_{22}, \eta^+$, $q_2,\eta_2, \eta_{12}, \eta_{22}, \eta$,
$f^+_2, b^+_{12}, b^+_{12}, b^+$ and their canonically conjugate
operators. In the expressions for the action (\ref{L1}) and
the sequence of gauge transformations (\ref{GT1})--(\ref{GTi2}), we
must set $n_2 =0$ and  use the above restrictions for
$\mathcal{H}_{tot}$. In particular, the operator
$K'_{(n_1,0)}$ has an exact form \cite{symferm-flat}, $K'_{(n_1,0)}=
\sum_{n_{11}=0}\frac{1}{n_{11}!}
  \Bigl(\,
     |n_{11}\rangle{}\langle{}n_{11}|\,C(n_{11},h_{n_1})
     -
     2f_1^+|n_{11}\rangle\langle{}n_{11}|f_1\,C(n_{11}+1,h_{n_1})\,
  \,\Bigr)$, for
$C(n,h)=h(h+1)\cdots(h+n-1),  C(0,h)=1$.}.
A substitution of (\ref{chi320})--(\ref{l001}) into (\ref{GT1}),
(\ref{GT2}) permits one to find the gauge transformations (\ref{GT1}),
(\ref{GT2}) in the form
\begin{eqnarray}
&& \delta\psi_\mu =\partial_\mu\xi_1+i\gamma_\mu\xi_2,\quad
\delta\psi=\xi_2,\quad  \delta\chi=  i\gamma^\mu\partial_\mu \xi_1
-2\xi_2, \quad  \delta\chi_1 = - i\gamma^\mu\partial_\mu \xi_2 .
\label{gtr}
\end{eqnarray}

Let us present the action in terms of the physical field
$\psi_\mu$ alone. To this end, we get rid of the field $\psi$ by using
its gauge transformation and the gauge parameter $\xi_2$. Having
expressed the field $\chi$ by using the equation of motion
$\chi=i\gamma^\mu\psi_\mu$, we can see that the terms with the
Lagrangian multiplier $\chi_1$ turn to zero. As a result, we
obtain the action
\begin{eqnarray} \mathcal{S}_{(1,0)} &=& \int d^dx \Bigl\{
i\bar{\psi}^\mu\gamma^\nu\partial_\nu\psi_\mu
-i\bar{\psi}^\mu(\gamma_\nu\partial_\mu+\gamma_\mu\partial_\nu)\psi^\nu
+i\bar{\psi}^\nu\gamma_\nu\gamma^\sigma\partial_\sigma\gamma^\mu\psi_\mu\Bigr\}
, \label{S3/21/2}
\end{eqnarray}
which is invariant with respect to the residual gauge
transformation $\delta\psi_\mu =\partial_\mu\xi_1$.

To obtain a Lagrangian description for a massive fermionic field, we may take two ways,
either following the dimensional reduction procedure (\ref{reduction})--(\ref{Eq-omv}),
starting directly from the action (\ref{S3/21/2}) presented for a $(d+1)$-dimensional
Minkowski space, or following the prescription (\ref{changeferm})--(\ref{chifm}),
starting from the expansion (\ref{chi320}), (\ref{brachi320}), where
it is only $|\chi_0^0\rangle$ that changes to $|\chi_{0m}^0\rangle$,
\begin{align}\label{chi3200m}
& |\chi_{0m}^0\rangle_{(1,0)} = |\chi_{0}^0\rangle_{(1,0)} + b_1^+\varphi(x) |0\rangle, &&{}_{(1,0)}\langle\tilde{\chi}_{0m}^0| ={}_{(1,0)}\langle\tilde{\chi}_{0}^0| +
\langle0|\varphi^+(x)b_1\tilde{\gamma}^0.
\end{align}
In the latter case, the Lagrangian action and the gauge transformations for $ h^1_m = (1-d)/2$
read as follows:
\begin{eqnarray}
{\cal{}S}_{(1,0)}^m &=&\int d^dx \left[\bar{\psi}^\mu\Bigl\{
\bigl[i\gamma^\nu\partial_\nu-m\bigr]\psi_\mu -\partial_\mu\chi - i \gamma_\mu \chi_1 \Bigr\} +
(d-1)\bar{\psi}\Bigl\{ \bigl[i\gamma^\nu\partial_\nu+m\bigr]\psi
 +\chi_1 \Bigr\} \right.\nonumber
\\
&&{} \left.+  \bar{\chi}\Bigl\{\bigl[i\gamma^\nu\partial_\nu+m\bigr]\chi
- \chi_1 +\partial^\mu\psi_\mu + m\varphi \Bigr\}+\bar{\chi}_1\Bigl\{
i\gamma^\mu\psi_\mu +(d-1)\psi -\chi-\varphi\Bigr\}
  \right.\nonumber\\
&&{} \left. - \bar{\varphi}\Bigl\{
\bigl[i\gamma^\nu\partial_\nu-m\bigr]\varphi  -m\chi  +  \chi_1 \Bigr\}\right] , \label{L3/21/2m}\\
\delta\psi_\mu &
=& \partial_\mu\xi_1+i\gamma_\mu\xi_2,
\hspace*{4em}
\delta\psi=\xi_2,
\hspace*{4em}
\delta\varphi=m\xi_1+\xi_2, \nonumber
\\
\delta\chi  &= &
\bigl[
i\gamma^\mu\partial_\mu
-m
\bigr]\xi_1
-2\xi_2,
\hspace*{4em}
\delta\chi_1
= -
\bigl[
i\gamma^\mu\partial_\mu
+m
\bigr]p
\xi_2.
\label{gtrmass}
\end{eqnarray}
Then one gets rid of the fields $\psi$, $\varphi$ by using the respective
gauge transformations and gauge parameters $\xi_2$, $\xi_1$.
Once again, using the corresponding equation of motion, we express the field
$\chi$ as $\chi=i\gamma^\mu\psi_\mu$ and obtain the Rarita--Schwinger Lagrangian
in a $d$-dimensional flat space \cite{symferm-flat},
\begin{eqnarray} \mathcal{L}_{RS}&=&
\bar{\psi}{}^\mu(i\gamma^\nu\partial_\nu-m)\psi_\mu
-i\bar{\psi}{}^\mu
 (\gamma^\nu\partial_\mu+\gamma_\mu\partial^\nu)
 \psi_\nu
+ \bar{\psi}{}_\mu\gamma^\mu
 (i\gamma^\sigma\partial_\sigma+m)\gamma^\nu\psi_\nu.
\end{eqnarray}
The same result follows from dimensional reduction applied to
a massless theory in $R^{1,d}$ (e.g., with an odd $d+1$),
for a spin-$3/2$ massive fermionic field ${\psi}^M = ({\psi}^\mu, \varphi)$,
which contains the Stueckelberg field $\varphi$, with
$i\gamma^M\partial_M{\psi}^N =
(i\gamma^\mu\partial_\mu-m){\psi}^N$,
$i\gamma^M\partial_M\gamma_N{\psi}^N =
(i\gamma^\mu\partial_\mu+m)\gamma_N{\psi}^N$, in view of
(\ref{reduction2}) and due to the relation $\chi=i\gamma_N\psi^N$ before
Eq. (\ref{S3/21/2}), with allowance for the structure of $|\chi_0^1\rangle_{(1,0)}$
in (\ref{chi320}) and the resolution of the respective gauge transformations
$\delta(\psi_\mu, \varphi) =(\partial_\mu, m)\xi$.

\subsection{Rank-2 antisymmetric spin-tensor  field}\label{Example3/23/2}

In the case of a spin-(3/2,3/2) field, we have $n_i=1$, $(h^1,h^2) =
(1-d/2,3-d/2)$. Since $s_{max}= 2$, the corresponding Lagrangian
description is a reducible gauge theory of second-stage
reducibility. The nonvanishing fields $|\chi^0_0\rangle_{(1,1)}$,
$|\chi^1_0\rangle_{(1,1)}$,  gauge parameters
$|\Lambda^k_0\rangle_{(1,1)}$, first-stage gauge parameters
$|\Lambda^{(1){}k}_0\rangle_{(1,1)}$, and second-stage gauge
parameters (for $|\Lambda^{(2){}1}_0\rangle_{(1,1)} \equiv 0$, due
to ${gh}(|\Lambda^{(2){}1}_0\rangle_{(1,1)}) = -4$), have the
following Grassmann grading and ghost number distribution:
\begin{align}
& \left(\varepsilon, {gh}\right)(|\chi_0^i\rangle_{(1,1)})=(1,
-i)\,, && \left(\varepsilon,
{gh}\right)(|\Lambda_0^k\rangle_{(1,1)})=(0,-1-k)\,, \\
& \left(\varepsilon,
{gh}\right)(|\Lambda_0^{(1){}k}\rangle_{(1,1)})=(1,-2-k) \,, &&
\left(\varepsilon,
{gh}\right)(|\Lambda_0^{(2){}0}\rangle_{(1,1)})=(0,-3)\,.
\end{align}
These conditions allow one, first, to extract the  dependence  on
the ghost variables from the fields and gauge parameters:
\begin{eqnarray}
  |\chi_0^0\rangle_{(1,1)} &=&  |\Psi\rangle_{(1,1)} +  \eta_1^+\mathcal{P}^+_2|\Psi_1\rangle_{(0,0)}
+ \mathcal{P}^+_1\eta_2^+|\Psi_2\rangle_{(0,0)} +
q_1^+p_2^+|\Psi_3\rangle_{(0,0)}  \label{chi00} \\
   && + p_1^+q_2^+|\Psi_4\rangle_{(0,0)} +   \eta_1^+p^+_2\tilde{\gamma}|\Psi_5\rangle_{(0,0)}+
q_1^+\mathcal{P}^+_2\tilde{\gamma}|\Psi_6\rangle_{(0,0)}+
\mathcal{P}^+_1q_2^+\tilde{\gamma}|\Psi_7\rangle_{(0,0)}
   \nonumber \\
  \phantom{|\chi_0^0\rangle_{(1,1)} } &&  +  p^+_1\eta_2^+\tilde{\gamma}|\Psi_8\rangle_{(0,0)} + \mathcal{P}^+_{11}
   \eta^+|\Psi_9\rangle_{(0,0)} +  \eta^+_{11}
   \mathcal{P}^+|\Psi_{10}\rangle_{(0,0)}
   \nonumber  \\
   && + q_1^+\mathcal{P}^+\tilde{\gamma}|\varphi_1\rangle_{(1,0)} +
   \eta_1^+\mathcal{P}^+|\varphi_2\rangle_{(1,0)} +
   \mathcal{P}^+_1\eta^+|\varphi_3\rangle_{(1,0)} +
    {p}^+_1\eta^+\tilde{\gamma}|\varphi_4\rangle_{(1,0)}
     \nonumber\\
   &&  + q_1^+p_1^+|\rho_1\rangle_{(-1,1)}  +
   q_1^+\mathcal{P}_1^+\tilde{\gamma}|\rho_2\rangle_{(-1,1)}
    + \eta_1^+p_1^+\tilde{\gamma}|\rho_3\rangle_{(-1,1)} +
    \eta_1^+\mathcal{P}_1^+|\rho_4\rangle_{(-1,1)}\,, \nonumber
  \\
  |\chi_0^1\rangle_{(1,1)} &=&
p^+_1
  |\chi_1\rangle_{(0,1)} + p^+_2
   |\chi_2\rangle_{(1,0)} + \mathcal{P}^+_1\tilde{\gamma}
   |\chi_3\rangle_{(0,1)} + \mathcal{P}^+_2\tilde{\gamma}
   |\chi_4\rangle_{(1,0)}    \label{chi01}\\
   && + \mathcal{P}^+_{12} \tilde{\gamma}|\chi_5\rangle_{(0,0)} + q_1^+p_1^+
   \mathcal{P}^+\tilde{\gamma}|\chi_6 \rangle_{(0,0)} + \eta_1^+\mathcal{P}_1^+
   \mathcal{P}^+\tilde{\gamma}|\chi_7 \rangle_{(0,0)} \nonumber \\
   &&
+ q_1^+\mathcal{P}_1^+
   \mathcal{P}^+ |\chi_8 \rangle_{(0,0)} +  p_1^+\eta_1^+
   \mathcal{P}^+ |\chi_9 \rangle_{(0,0)} + p_1^+
   \mathcal{P}^+_1 \eta^+|\chi_{10} \rangle_{(0,0)}
\nonumber     \\
&&  + (p_1^+)^2 \eta^+ \tilde{\gamma}|\chi_{11} \rangle_{(0,0)} +
\mathcal{P}^+\tilde{\gamma}|\chi_{12} \rangle_{(2,0)} +
\mathcal{P}^+_{11}\tilde{\gamma}|\chi_{13} \rangle_{(-1,1)}\,,
\nonumber \\
  |\Lambda_0^0\rangle_{(1,1)}  &=& p^+_1 \tilde{\gamma}
  |\xi_1\rangle_{(0,1)} + p^+_2\tilde{\gamma}
   |\xi_2\rangle_{(1,0)} + \mathcal{P}^+_1
   |\xi_3\rangle_{(0,1)} + \mathcal{P}^+_2
   |\xi_4\rangle_{(1,0)}    \label{L00}\\
   && + \mathcal{P}^+_{12} |\xi_5\rangle_{(0,0)} + q_1^+p_1^+
   \mathcal{P}^+|\xi_6 \rangle_{(0,0)} + \eta_1^+\mathcal{P}_1^+
   \mathcal{P}^+|\xi_7 \rangle_{(0,0)} \nonumber \\
   &&
+ q_1^+\mathcal{P}_1^+
   \mathcal{P}^+ \tilde{\gamma}|\xi_8 \rangle_{(0,0)} +  p_1^+\eta_1^+
   \mathcal{P}^+ \tilde{\gamma}|\xi_9 \rangle_{(0,0)} + p_1^+
   \mathcal{P}^+_1 \eta^+\tilde{\gamma}|\xi_{10} \rangle_{(0,0)}
    \nonumber  \\
&&  + (p_1^+)^2 \eta^+ |\xi_{11} \rangle_{(0,0)} +
\mathcal{P}^+|\xi_{12} \rangle_{(2,0)} +
\mathcal{P}^+_{11}|\xi_{13} \rangle_{(-1,1)}\,,\nonumber
\\
  |\Lambda_0^1\rangle_{(1,1)} &=&  p^+_1p^+_2 \tilde{\gamma}
  |\lambda_1\rangle_{(0,0)} + p^+_1 \mathcal{P}^+_2
   |\lambda_2\rangle_{(0,0)} + \mathcal{P}^+_1 p^+_2
   |\lambda_3\rangle_{(0,0)} \\
   &&  +  \mathcal{P}^+_1 \mathcal{P}^+_2\tilde{\gamma}
   |\lambda_4\rangle_{(0,0)}
   +   \mathcal{P}^+_{11} \mathcal{P}^+ \tilde{\gamma}
   |\lambda_5\rangle_{(0,0)} +  \mathcal{P}^+_{1} \mathcal{P}^+ \tilde{\gamma}
   |\lambda_6\rangle_{(1,0)} \nonumber \\
   && +   p^+_{1} \mathcal{P}^+  |\lambda_7\rangle_{(1,0)} +
 ({p}^+_{1})^2   \tilde{\gamma}|\lambda_8\rangle_{(-1,1)} + {p}^+_{1}\mathcal{P}^+_{1}
   |\lambda_9\rangle_{(-1,1)}\,,                    \label{L01}
   \nonumber     \\
   |\Lambda_0^{(1){}0}\rangle_{(1,1)} &=&
   p^+_1p^+_2 |\xi^{(1)}_1\rangle_{(0,0)} + p^+_1 \mathcal{P}^+_2 \tilde{\gamma}
   |\xi^{(1)}_2\rangle_{(0,0)} + \mathcal{P}^+_1 p^+_2 \tilde{\gamma}
   |\xi^{(1)}_3\rangle_{(0,0)}  \\
   && +  \mathcal{P}^+_1 \mathcal{P}^+_2
   |\xi^{(1)}_4\rangle_{(0,0)} +   \mathcal{P}^+_{11} \mathcal{P}^+
   |\xi^{(1)}_5\rangle_{(0,0)} +  \mathcal{P}^+_{1} \mathcal{P}^+
   |\xi^{(1)}_6\rangle_{(1,0)}
   \nonumber \\
   && +   p^+_{1} \mathcal{P}^+ \tilde{\gamma} |\xi^{(1)}_7\rangle_{(1,0)} +
 ({p}^+_{1})^2   |\xi^{(1)}_8\rangle_{(-1,1)} + {p}^+_{1}\mathcal{P}^+_{1}\tilde{\gamma}
   |\xi^{(1)}_9\rangle_{(-1,1)}\,,                    \label{L10}
   \nonumber     \\
   |\Lambda_0^{(1){}1}\rangle_{(1,1)} &=&
   (p^+_1)^2\mathcal{P}^+ \tilde{\gamma}|\lambda^{(1)}_1\rangle_{(0,0)} +
   p^+_1 \mathcal{P}^+_1 \mathcal{P}^+
   |\lambda^{(1)}_2\rangle_{(0,0)}\,,
   \label{L11}  \\
   |\Lambda_0^{(2){}0}\rangle_{(1,1)}  &=&
   (p^+_1)^2\mathcal{P}^+ |\xi^{(2)}_1\rangle_{(0,0)} + p^+_1 \mathcal{P}^+_1
   \mathcal{P}^+ \tilde{\gamma}
   |\xi^{(2)}_2\rangle_{(0,0)} \label{L20}\,,
\end{eqnarray}
where the coefficient fermionic fields  and gauge parameters in
the right-hand side of equations (\ref{chi00})--(\ref{L20})
 are independent of ghost operators.
The bra-vectors ${}_{(1,1)}\langle\tilde{\chi}_0^k|$ corresponding
to expansion (\ref{chi00}), (\ref{chi01})
 have the form
\begin{eqnarray}
  {}_{(1,1)}\langle\tilde{\chi}_0^0| &=&  {}_{(1,1)}\langle\tilde{\Psi}| +
  {}_{(0,0)}\langle\tilde{\Psi}_1|\mathcal{P}_2\eta_1
+ {}_{(0,0)}\langle\tilde{\Psi}_2|\eta_2\mathcal{P}_1 +
{}_{(0,0)}\langle\tilde{\Psi}_3|q_1p_2   \label{brachi00}
\\
   && + {}_{(0,0)}\langle\tilde{\Psi}_4|q_2p_1 +   {}_{(0,0)}\langle\tilde{\Psi}_5|\tilde{\gamma}p_2\eta_1 +
{}_{(0,0)}\langle\tilde{\Psi}_6|\tilde{\gamma}\mathcal{P}_2 q_1 +
{}_{(0,0)}\langle\tilde{\Psi}_7|\tilde{\gamma}q_2\mathcal{P}_1
   \nonumber \\
   &&  + {}_{(0,0)}\langle\tilde{\Psi}_8|\tilde{\gamma}\eta_2p_1 +
  {}_{(0,0)}\langle\tilde{\Psi}_9| \eta\mathcal{P}_{11} +
 {}_{(0,0)}\langle\tilde{\Psi}_{10}|\mathcal{P} \eta_{11}
   \nonumber  \\
   && + {}_{(1,0)}\langle\tilde{\varphi}_1| \tilde{\gamma}\mathcal{P}q_1 +
{}_{(1,0)}\langle\tilde{\varphi}_2|  \mathcal{P} \eta_1 +
 {}_{(1,0)}\langle\tilde{\varphi}_3| \eta\mathcal{P}_1 +
   {}_{(1,0)}\langle\tilde{\varphi}_4|\tilde{\gamma}\eta {p}_1
     \nonumber\\
   &&  + {}_{(-1,1)}\langle\tilde{\rho}_1| p_1q_1  +
 {}_{(-1,1)}\langle\tilde{\rho}_2|\tilde{\gamma}\mathcal{P}_1 q_1
    + {}_{(-1,1)}\langle\tilde{\rho}_3|\tilde{\gamma} p_1\eta_1 +
   {}_{(-1,1)}\langle\tilde{\rho}_4| \mathcal{P}_1\eta_1\,,
   \nonumber
  \\
  {}_{(1,1)}\langle\tilde{\chi}_0^1| &=&
{}_{(0,1)}\langle\tilde{\chi}_1| p_1
   +
 {}_{(1,0)}\langle\tilde{\chi}_2|p_2 + {}_{(0,1)}\langle\tilde{\chi}_3|\tilde{\gamma}\mathcal{P}_1
    + {}_{(1,0)}\langle\tilde{\chi}_4|\tilde{\gamma}\mathcal{P}_2
       \label{brachi01}\\
   && + {}_{(0,0)}\langle\tilde{\chi}_5| \tilde{\gamma}\mathcal{P}_{12} +
  {}_{(0,0)}\langle\tilde{\chi}_6 |\tilde{\gamma}\mathcal{P}p_1q_1
    +  {}_{(0,0)}\langle\tilde{\chi}_7 |\tilde{\gamma}
   \mathcal{P} \mathcal{P}_1\eta_1\nonumber \\
   &&
+ {}_{(0,0)}\langle\tilde{\chi}_8 |\mathcal{P}\mathcal{P}_1q_1
     +  {}_{(0,0)}\langle\tilde{\chi}_9 | \mathcal{P}\eta_1p_1
    + {}_{(0,0)}\langle\tilde{\chi}_{10}| \eta \mathcal{P}_1p_1
\nonumber     \\
&&  + {}_{(0,0)}\langle\tilde{\chi}_{11}| \tilde{\gamma} \eta
p_1^2
 + {}_{(2,0)}\langle\tilde{\chi}_{12}| \tilde{\gamma}\mathcal{P} +
{}_{(-1,1)}\langle\tilde{\chi}_{13} |
\tilde{\gamma}\mathcal{P}_{11}\,. \nonumber
\end{eqnarray}
Substituting (\ref{chi00}), (\ref{chi01}), (\ref{brachi00}),
(\ref{brachi01}) into (\ref{L1}), we find the action   (up to an
overall factor) for a spin-$(3/2,3/2)$ free massless field on a
flat background in the form of a scalar product for vectors
defined only in $\mathcal{H}\otimes \mathcal{H}'$,
\begin{eqnarray}
{\cal{}S}_{(1,1)} &=& \Bigl[\langle
\tilde{\Psi}|K_{(1,1)}\left\{\textstyle\frac{1}{2}T_0 |\Psi\rangle
+ T^+_1|\chi_1\rangle + T^+_2|\chi_2\rangle +
\tilde{\gamma}L^+_1|\chi_3\rangle\right.
 \label{L3/23/2} \nonumber\\
&& +\left.\tilde{\gamma}L^+_2|\chi_4\rangle
+\tilde{\gamma}L^+_{12}|\chi_5\rangle +
\tilde{\gamma}T^+|\chi_{12}\rangle
+\tilde{\gamma}L^+_{11}|\chi_{13}\rangle\right\}
\nonumber \\
&& + \langle \tilde{\Psi}_1|K_{(1,1)}\left\{T_0 |\Psi_2\rangle -
2\tilde{\gamma}|\Psi_8\rangle - \tilde{\gamma} L_2|\chi_3\rangle +
\textstyle\frac{1}{2}\tilde{\gamma}|\chi_5\rangle  +\tilde{\gamma}
|\chi_7\rangle \right\}
 \nonumber \\
&& + \langle \tilde{\Psi}_2|K_{(1,1)}\left\{
2\tilde{\gamma}|\Psi_5\rangle + \tilde{\gamma} L_1|\chi_4\rangle -
\textstyle\frac{1}{2}\tilde{\gamma}|\chi_5\rangle  -\tilde{\gamma}
|\chi_7\rangle \right\}
\nonumber \\
&& +\langle \tilde{\Psi}_3|K_{(1,1)}\left\{T_0 |\Psi_4\rangle-
2\tilde{\gamma}|\Psi_8\rangle +  T_2|\chi_1\rangle +
\textstyle\frac{1}{2}\tilde{\gamma}|\chi_5\rangle  -\tilde{\gamma}
|\chi_6\rangle -4\tilde{\gamma} |\chi_{11}\rangle \right\}
\nonumber \\
&& + \langle \tilde{\Psi}_4|K_{(1,1)}\left\{-
2\tilde{\gamma}|\Psi_5\rangle +  T_1|\chi_2\rangle +
\textstyle\frac{1}{2}\tilde{\gamma}|\chi_5\rangle  -\tilde{\gamma}
|\chi_6\rangle  \right\}
 \nonumber \\
&& + \langle \tilde{\Psi}_5|K_{(1,1)}\left\{- T_0|\Psi_7\rangle -
T_2|\chi_3\rangle + \tilde{\gamma}|\chi_8\rangle  +2\tilde{\gamma}
|\chi_{10}\rangle\right\}
 \nonumber\\
&&
 + \langle \tilde{\Psi}_6|K_{(1,1)}\left\{- T_0|\Psi_8\rangle +
\tilde{\gamma}L_2|\chi_1\rangle +
\tilde{\gamma}|\chi_9\rangle\right\}
\nonumber \\
&& + \langle \tilde{\Psi}_7|K_{(1,1)}\left\{
\tilde{\gamma}L_1|\chi_2\rangle +
\tilde{\gamma}|\chi_9\rangle\right\}+ \langle
\tilde{\Psi}_8|K_{(1,1)}\left\{- T_1|\chi_4\rangle +
\tilde{\gamma}|\chi_8\rangle\right\}
\nonumber\\
&&
+\langle \tilde{\Psi}_9|K_{(1,1)}\left\{T_0|\Psi_{10}\rangle
-\textstyle\frac{1}{2} \tilde{\gamma}|\chi_5\rangle +
\tilde{\gamma}|\chi_6\rangle + \tilde{\gamma}|\chi_7\rangle
+\tilde{\gamma}L_{11}|\chi_{12}\rangle\right\}
\nonumber \\
&& + \langle
\tilde{\Psi}_{10}|K_{(1,1)}\left\{\tilde{\gamma}|\chi_5\rangle -
4\tilde{\gamma}|\chi_{11}\rangle -
\tilde{\gamma}T|\chi_{13}\rangle\right\}
 \nonumber \\
&+&  \langle
\tilde{\varphi}_{1}|K_{(1,1)}\left\{-T_0|\varphi_4\rangle +
\tilde{\gamma}T|\chi_1\rangle - \tilde{\gamma}|\chi_2\rangle +
\tilde{\gamma}L^+_1 |\chi_{10}\rangle -
2T_1^+|\chi_{11}\rangle\right\}
\nonumber \\
&& + \langle
\tilde{\varphi}_{2}|K_{(1,1)}\left\{T_0|\varphi_3\rangle
-2\tilde{\gamma} |\varphi_4\rangle - \tilde{\gamma}T|\chi_3\rangle
+ \tilde{\gamma}|\chi_4\rangle +  T_1^+|\chi_{10}\rangle\right\}
 \nonumber \\
&&
 + \langle
\tilde{\varphi}_{3}|K_{(1,1)}\left\{ -
\tilde{\gamma}L_1^+|\chi_7\rangle + T_1^+|\chi_9\rangle +
\tilde{\gamma}L_1|\chi_{12}\rangle\right\}
 \nonumber \\
&& + \langle \tilde{\varphi}_{4}|K_{(1,1)}\left\{ -
2\tilde{\gamma}|\chi_2\rangle - T_1^+|\chi_6\rangle  +
\tilde{\gamma}L_1^+|\chi_8\rangle - T_1|\chi_{12}\rangle\right\}
 \nonumber
 \\
\phantom{{\cal{}S}_{(1,1)} } &+&  \langle
\tilde{\rho}_{1}|K_{(1,1)}\left\{
\textstyle\frac{1}{2}T_0|\rho_1\rangle - 2
\tilde{\gamma}|\rho_3\rangle + T_1|\chi_1\rangle +
\tilde{\gamma}T^+|\chi_{6}\rangle +
\tilde{\gamma}|\chi_{13}\rangle\right\}
\nonumber \\
 && + \langle
\tilde{\rho}_{2}|K_{(1,1)}\left\{ - T_0|\rho_3\rangle +
\tilde{\gamma}L_1|\chi_1\rangle -\tilde{\gamma}T^+|\chi_{9}\rangle
\right\} \nonumber\\
&&
+ \langle \tilde{\rho}_{3}|K_{(1,1)}\left\{ -
2\tilde{\gamma}|\rho_4\rangle - T_1|\chi_3\rangle
-\tilde{\gamma}T^+|\chi_{8}\rangle \right\}
\nonumber\\
&& + \langle \tilde{\rho}_{4}|K_{(1,1)}\left\{
-\textstyle\frac{1}{2} T_0|\rho_4\rangle -
\tilde{\gamma}L_1|\chi_3\rangle -\tilde{\gamma}T^+|\chi_{7}\rangle
+ \tilde{\gamma}|\chi_{13}\rangle\right\}
\nonumber\\
&+&  \langle \tilde{\chi}_{1}|K_{(1,1)}\left\{
-\tilde{\gamma}|\chi_3\rangle\right\} + \langle
\tilde{\chi}_{2}|K_{(1,1)}\left\{
-\tilde{\gamma}|\chi_4\rangle\right\} +\langle
\tilde{\chi}_{3}|K_{(1,1)}\left\{
-\textstyle\frac{1}{2}T_0|\chi_3\rangle\right\} \nonumber \\
&&
 +
 \langle
\tilde{\chi}_{4}|K_{(1,1)}\left\{
-\textstyle\frac{1}{2}T_0|\chi_4\rangle\right\} + \langle
\tilde{\chi}_{6}|K_{(1,1)}\left\{
-\tilde{\gamma}|\chi_{10}\rangle\right\} + \langle
\tilde{\chi}_{7}|K_{(1,1)}\left\{
-\tilde{\gamma}|\chi_{10}\rangle\right\}
\nonumber\\
&& +\langle \tilde{\chi}_{8}|K_{(1,1)}\left\{T_0|\chi_{10}\rangle
-2\tilde{\gamma}|\chi_{11}\rangle\right\} \Bigr] + c.c.\,,
\end{eqnarray}
where we have omitted the lower spin subscripts of the component
fields. In deriving the action (\ref{L3/23/2}), we have used the
expressions for the operators $K_{(1,1)}$ (\ref{K}), (\ref{tK}),
and then, substituting (\ref{chi320})--(\ref{l001}) into
(\ref{GT1}), (\ref{GT2}), we find the gauge transformations for the
vectors $|\Psi\rangle, |\Psi_k\rangle$,
\begin{eqnarray}
  \delta|\Psi\rangle
  &=&
  - \tilde{\gamma}(T_1^+|\xi_1\rangle +T_2^+|\xi_2\rangle)+
  L_1^+|\xi_3\rangle +L_2^+|\xi_4\rangle + L_{12}^+|\xi_5\rangle +  L_{11}^+|\xi_{13}\rangle
   + T^+|\xi_{12}\rangle,   \label{dpsi}\\
  \delta|\Psi_1\rangle
   &=&
   L_1|\xi_4\rangle -\textstyle\frac{1}{2}|\xi_5\rangle -|\xi_7\rangle
  -|\lambda_2\rangle + 2|\lambda_3\rangle - \tilde{\gamma}T_0 |\lambda_4\rangle,
\label{dpsi1}
  \\
  \delta|\Psi_2\rangle
   &=&
  -L_2|\xi_3\rangle +\textstyle\frac{1}{2}|\xi_5\rangle + |\xi_7\rangle
  -2|\lambda_2\rangle + |\lambda_3\rangle - \tilde{\gamma}T_0 |\lambda_4\rangle,
  \label{dpsi2}\\
  \delta|\Psi_3\rangle
  &=&
   -\tilde{\gamma}T_1|\xi_2\rangle +\textstyle\frac{1}{2}|\xi_5\rangle - |\xi_6\rangle
  - |\lambda_3\rangle ,
  \label{dpsi3} \\
\delta|\Psi_4\rangle
   &=&
   -\tilde{\gamma}T_2|\xi_1\rangle +\textstyle\frac{1}{2}|\xi_5\rangle -
   |\xi_6\rangle -4   |\xi_{11}\rangle
  - |\lambda_2\rangle ,     \label{dpsi4} \\
\delta|\Psi_5\rangle
   &=&
L_1|\xi_2\rangle +|\xi_9\rangle - |\lambda_1\rangle +
\tilde{\gamma}T_0|\lambda_3\rangle, \label{dpsi5}
   \\
\delta|\Psi_6\rangle
   &=&
\tilde{\gamma}T_1|\xi_4\rangle +|\xi_8\rangle - |\lambda_4\rangle
\label{dpsi6}
,    \\
\delta|\Psi_7\rangle
   &=&
\tilde{\gamma}T_2|\xi_3\rangle +|\xi_8\rangle + 2|\xi_{10}\rangle
+ |\lambda_4\rangle , \label{dpsi7}
   \\
\delta|\Psi_8\rangle
   &=&
L_2|\xi_1\rangle +|\xi_9\rangle +
\tilde{\gamma}T_0|\lambda_2\rangle - |\lambda_{1}\rangle
,\label{dpsi8}
   \\
\delta|\Psi_9\rangle
   &=&
|\xi_5\rangle - 4|\xi_{11}\rangle - T |\xi_{13}\rangle ,
\label{dpsi9}
   \\
  \delta|\Psi_{10}\rangle
  &=&
-  \textstyle\frac{1}{2}|\xi_5\rangle+|\xi_6\rangle +|\xi_7\rangle
+ L_{11} |\xi_{12}\rangle ,\label{dpsi10}
\end{eqnarray}
for the vectors $|\varphi_l\rangle, |\rho_l\rangle$,
\begin{eqnarray}
  && \delta|\varphi_1\rangle \hspace{1ex}
=\hspace{1ex}
   -2|\xi_2\rangle + \tilde{\gamma}T_1^+ |\xi_{6}\rangle +
L_1^+
   |\xi_8\rangle + \tilde{\gamma}T_1|\xi_{12}\rangle -
   |\lambda_{6}\rangle\,, \phantom{+\tilde{\gamma}T_1|\xi_{12}\rangle - |\lambda_{6}\rangle}
  \label{dvphi1}
  \\
   && \delta|\varphi_2\rangle\hspace{1ex} =\hspace{1ex}
-L_1^+|\xi_7\rangle - \tilde{\gamma}T_1^+ |\xi_{9}\rangle + L_1
   |\xi_{12}\rangle  - \tilde{\gamma}T_0|\lambda_{6}\rangle - |\lambda_{7}\rangle  ,
   \label{dvphi2}
    \\
   &&
   \delta|\varphi_3\rangle
   \hspace{1ex} =\hspace{1ex} -T|\xi_3\rangle + |\xi_4\rangle - \tilde{\gamma}T_1^+
   |\xi_{10}\rangle\,, \label{dvphi3}
   \\
   && \delta|\varphi_4\rangle
   \hspace{1ex} =\hspace{1ex}
T|\xi_1\rangle - |\xi_2\rangle +L_1^+|\xi_{10}\rangle + 2
\tilde{\gamma}T_1^+  |\xi_{11}\rangle\,,\label{dvphi4} \\
     &&
\delta|\rho_1\rangle\hspace{1ex} =\hspace{1ex}
-\tilde{\gamma}T_1|\xi_1\rangle + T^+ |\xi_6\rangle +
|\xi_{13}\rangle -
|\lambda_{9}\rangle\,,\phantom{\delta|\rho_1\rangle\hspace{1ex}
=\hspace{1ex}\delta|\rho_1\rangle\hspace{5,5ex} =\hspace{5,5ex}}
\label{drho1}
   \\
   && \delta|\rho_2\rangle \hspace{1ex} =\hspace{1ex}
\tilde{\gamma}T_1|\xi_3\rangle - T^+ |\xi_8\rangle\,,\label{drho2}
   \\
   && \delta|\rho_3\rangle \hspace{1ex} =\hspace{1ex}
   L_1|\xi_{1}\rangle
-T^+|\xi_9\rangle - 2|\lambda_8\rangle +  \tilde{\gamma}T_0
|\lambda_{9}\rangle\,, \label{drho3}
   \\
   && \delta|\rho_4\rangle\hspace{1ex} =\hspace{1ex}
L_1|\xi_{3}\rangle + T^+|\xi_7\rangle - |\xi_{13}\rangle
+|\lambda_{9}\rangle\,
   , \label{drho4}
   \end{eqnarray}
and for the vectors $|\chi_m\rangle$,
\begin{eqnarray}
   && \delta|\chi_1\rangle \hspace{1ex} =\hspace{1ex}
- \tilde{\gamma}T_0|\xi_1\rangle -
\tilde{\gamma}T_2^+|\lambda_{1}\rangle +
 L_2^+|\lambda_2\rangle +
T^+ |\lambda_{7}\rangle -2 \tilde{\gamma}T_1^+|\lambda_{8}\rangle
+ L_1^+ |\lambda_{9}\rangle \,, \label{dchi1}
   \\
   && \delta|\chi_2\rangle
   \hspace{1ex} =\hspace{1ex}
- \tilde{\gamma}T_0|\xi_2\rangle -
\tilde{\gamma}T_1^+|\lambda_{1}\rangle +
 L_1^+|\lambda_3\rangle - |\lambda_{7}\rangle \,,
 \label{dchi2}
   \\
   && \delta|\chi_3\rangle
   \hspace{1ex} =\hspace{1ex}
   - 2|\xi_1\rangle +
 \tilde{\gamma}T_0|\xi_3\rangle +
\tilde{\gamma}T_2^+|\lambda_{3}\rangle -
 L_2^+|\lambda_4\rangle -
T^+ |\lambda_{6}\rangle + \tilde{\gamma}T_1^+|\lambda_{9}\rangle
\,, \label{dchi3}
   \\
   && \delta|\chi_4\rangle
   \hspace{1ex} =\hspace{1ex}
- 2|\xi_2\rangle +
 \tilde{\gamma}T_0|\xi_4\rangle +
\tilde{\gamma}T_1^+|\lambda_{2}\rangle +
 L_1^+|\lambda_4\rangle +|\lambda_{6}\rangle
\,, \label{dchi4}
   \\
   && \delta|\chi_5\rangle
   \hspace{1ex} =\hspace{1ex}
\tilde{\gamma}T_0|\xi_5\rangle -4|\lambda_{1}\rangle +
2|\lambda_{5}\rangle \,, \label{dchi5}
   \\
   && \delta|\chi_6\rangle
   \hspace{1ex} =\hspace{1ex}
 \tilde{\gamma}T_0|\xi_6\rangle - 2|\xi_9\rangle -2|\lambda_{1}\rangle
+|\lambda_{5}\rangle +
 \tilde{\gamma}T_1|\lambda_7\rangle
\,, \phantom{-2 \tilde{\gamma}T_1^+|\lambda_{8}\rangle
T_1^+|\lambda_{8}\rangle }\label{dchi6}
   \\
   && \delta|\chi_7\rangle
   \hspace{1ex} =\hspace{1ex}
 \tilde{\gamma}T_0|\xi_7\rangle + 2|\xi_9\rangle -|\lambda_{5}\rangle
+  L_1|\lambda_6\rangle \,, \label{dchi7}
   \\
   && \delta|\chi_8\rangle
   \hspace{1ex} =\hspace{1ex}
- \tilde{\gamma}T_0|\xi_8\rangle - 2|\xi_6\rangle - 2|\xi_7\rangle
+ 2|\lambda_{3}\rangle -
 \tilde{\gamma}T_1|\lambda_6\rangle
\,,\label{dchi8}
   \\
   && \delta|\chi_9\rangle
   \hspace{1ex} =\hspace{1ex}
- \tilde{\gamma}T_0|\xi_9\rangle + L_1|\lambda_{7}\rangle\,,
\label{dchi9}
   \\
   && \delta|\chi_{10}\rangle
   \hspace{1ex} =\hspace{1ex}
- \tilde{\gamma}T_0|\xi_{10}\rangle - 4|\xi_{11}\rangle +
|\lambda_{2}\rangle + |\lambda_{3}\rangle - T|\lambda_9\rangle \,,
\label{dchi10}
   \\
   && \delta|\chi_{11}\rangle
   \hspace{1ex} =\hspace{1ex}
    \tilde{\gamma}T_0|\xi_{11}\rangle -
|\lambda_{1}\rangle +  T|\lambda_8\rangle \,, \label{dchi11}
\\
     && \delta|\chi_{12}\rangle
     \hspace{1ex} =\hspace{1ex}
 \tilde{\gamma}T_0|\xi_{12}\rangle  +
L^+_{11}|\lambda_{5}\rangle + L^+_1|\lambda_{6}\rangle +
 \tilde{\gamma}T^+_1|\lambda_7\rangle \,, \label{dchi12}
     \\
     &&\delta|\chi_{13}\rangle
     \hspace{1ex} =\hspace{1ex}
     \tilde{\gamma}T_0|\xi_{13}\rangle  -T^+|\lambda_{5}\rangle -4
|\lambda_{8}\rangle  \,. \label{dchi13}
\end{eqnarray}
Then, substituting expressions (\ref{L00})--(\ref{L11}) into
(\ref{GTi1}), (\ref{GTi2}), we obtain the gauge transformations for
the zero-stage gauge vectors $|\xi_m\rangle, |\lambda_n\rangle$,
\begin{eqnarray}
   && \delta|\xi_1\rangle \hspace{1ex} =\hspace{1ex}
- \tilde{\gamma}T_2^+|\xi^{(1)}_1\rangle + L_2^+
|\xi^{(1)}_{2}\rangle +
 T^+|\xi^{(1)}_7\rangle - 2
\tilde{\gamma}T_1^+|\xi^{(1)}_{8}\rangle + L_1^+
|\xi^{(1)}_{9}\rangle \,, \label{dxi1}
   \\
   && \delta|\xi_2\rangle
   \hspace{1ex} =\hspace{1ex}
- \tilde{\gamma}T_1^+|\xi^{(1)}_1\rangle + L_1^+
|\xi^{(1)}_{3}\rangle - |\xi^{(1)}_7\rangle \,,
 \label{dxi2}
   \\
   && \delta|\xi_3\rangle
   \hspace{1ex} =\hspace{1ex}
 \tilde{\gamma}T_2^+|\xi^{(1)}_3\rangle - L_2^+
|\xi^{(1)}_{4}\rangle -
 T^+|\xi^{(1)}_6\rangle + \tilde{\gamma}T_1^+|\xi^{(1)}_{9}\rangle\,, \label{dxi3}
   \\
   && \delta|\xi_4\rangle
   \hspace{1ex} =\hspace{1ex}
\tilde{\gamma}T_1^+|\xi^{(1)}_2\rangle + L_1^+
|\xi^{(1)}_{4}\rangle + |\xi^{(1)}_6\rangle \,, \label{dxi4}
   \\
   && \delta|\xi_5\rangle
   \hspace{1ex} =\hspace{1ex}
-4|\xi^{(1)}_1\rangle + 2 |\xi^{(1)}_{5}\rangle  \,, \label{dxi5}
   \\
   && \delta|\xi_6\rangle
   \hspace{1ex} =\hspace{1ex}
-2|\xi^{(1)}_2\rangle +  |\xi^{(1)}_{5}\rangle +
 \tilde{\gamma}T_1|\xi^{(1)}_7\rangle - |\lambda^{(1)}_{2}\rangle \,, \label{dxi6}
   \\
   && \delta|\xi_7\rangle
   \hspace{1ex} =\hspace{1ex}
- |\xi^{(1)}_{5}\rangle +
 L_1|\xi^{(1)}_6\rangle + |\lambda^{(1)}_{2}\rangle \,, \label{dxi7}
   \\
   && \delta|\xi_8\rangle
   \hspace{1ex} =\hspace{1ex}
2|\xi^{(1)}_3\rangle -  \tilde{\gamma}T_1|\xi^{(1)}_6\rangle\,,
\label{dxi8}
   \\
   && \delta|\xi_9\rangle
   \hspace{1ex} =\hspace{1ex}
   L_1|\xi^{(1)}_7\rangle -2|\lambda^{(1)}_{1}\rangle-  \tilde{\gamma}T_0|\lambda^{(1)}_2\rangle
\,, \label{dxi9}
   \\
   && \delta|\xi_{10}\rangle
   \hspace{1ex} =\hspace{1ex}
|\xi^{(1)}_2\rangle +  |\xi^{(1)}_{3}\rangle -
T|\xi^{(1)}_9\rangle\,, \label{dxi10}
  \\
   && \delta|\xi_{11}\rangle
   \hspace{1ex} =\hspace{1ex}
    -|\xi^{(1)}_1\rangle + T|\xi^{(1)}_8\rangle \,,\label{dxi11}
\\
     && \delta|\xi_{12}\rangle
     \hspace{1ex} =\hspace{1ex}
L_{11}^+|\xi^{(1)}_{5}\rangle + L_1^+|\xi^{(1)}_6\rangle +
 \tilde{\gamma}T_1^+|\xi^{(1)}_7\rangle \,, \label{dxi12}
    \\
     &&\delta|\xi_{13}\rangle
     \hspace{1ex} =\hspace{1ex}
-T^+|\xi^{(1)}_5\rangle -4 |\xi^{(1)}_8\rangle  \,, \label{dxi13}
\end{eqnarray}
\vspace{-4ex}
\begin{eqnarray}
 && \delta|\lambda_1\rangle \hspace{1ex} =\hspace{1ex} -
\tilde{\gamma}T_0|\xi^{(1)}_1\rangle - 2
|\lambda^{(1)}_{1}\rangle\,,  \phantom{++
 T^+|\xi^{(1)}_7\rangle - 2
\tilde{\gamma}T_1^+|\xi^{(1)}_{8}\rangle + L_1^+
|\xi^{(1)}_{9}\rangle} \label{dlambd1}
   \\
   && \delta|\lambda_2\rangle
   \hspace{1ex} =\hspace{1ex}
- 2|\xi^{(1)}_1\rangle + \tilde{\gamma}T_0|\xi^{(1)}_2\rangle
+|\lambda^{(1)}_{2}\rangle \,,
 \label{dlambd2}
   \\
   && \delta|\lambda_3\rangle
   \hspace{1ex} =\hspace{1ex}
- 2|\xi^{(1)}_1\rangle + \tilde{\gamma}T_0|\xi^{(1)}_3\rangle
+|\lambda^{(1)}_{2}\rangle \,, \label{dlambd3}
   \\
   && \delta|\lambda_4\rangle
   \hspace{1ex} =\hspace{1ex}
- 2|\xi^{(1)}_2\rangle - \tilde{\gamma}T_0|\xi^{(1)}_4\rangle
+2|\xi^{(1)}_{3}\rangle \,, \label{dlambd4}
\\
   && \delta|\lambda_5\rangle
   \hspace{1ex} =\hspace{1ex}
- \tilde{\gamma}T_0|\xi^{(1)}_5\rangle -4|\lambda^{(1)}_{1}\rangle
\,,\phantom{\delta|\lambda_5\rangle
   \hspace{6em} =\hspace{6,5em}} \label{dlambd5}
   \\
   && \delta|\lambda_6\rangle
   \hspace{1ex} =\hspace{1ex}
 -\tilde{\gamma}T_0|\xi^{(1)}_6\rangle - 2|\xi^{(1)}_7\rangle  -\tilde{\gamma}T_1^+|\lambda^{(1)}_2\rangle
 \,, \label{dlambd6}
   \\
   && \delta|\lambda_7\rangle
   \hspace{1ex} =\hspace{1ex}
 \tilde{\gamma}T_0|\xi^{(1)}_7\rangle + 2\tilde{\gamma}T_1^+|\lambda^{(1)}_1\rangle +
 L_1^+|\lambda^{(1)}_2\rangle\,, \label{dlambd7}
   \\
   && \delta|\lambda_8\rangle
   \hspace{1ex} =\hspace{1ex}
- \tilde{\gamma}T_0|\xi^{(1)}_8\rangle +
T^+|\lambda^{(1)}_1\rangle \,,\label{dlambd8}
   \\
   && \delta|\lambda_9\rangle
   \hspace{1ex} =\hspace{1ex}
 \tilde{\gamma}T_0|\xi^{(1)}_9\rangle -4|\xi^{(1)}_8\rangle -
 T^+|\lambda^{(1)}_{2}\rangle\,.
\label{dlambd9}
\end{eqnarray}
Finally, using the equations (\ref{GTi1}), (\ref{GTi2}) for the
vectors  (\ref{L10})--(\ref{L20}), we find the gauge
transformations for the first-stage gauge vectors
$|\xi^{(1)}_n\rangle, |\lambda^{(1)}_o\rangle$,
\begin{align}
   & \delta|\xi^{(1)}_1\rangle \hspace{1ex} =\hspace{1ex}
- 2|\xi^{(2)}_1\rangle \,, &&
   \delta|\xi^{(1)}_2\rangle
   \hspace{1ex} =\hspace{1ex}
|\xi^{(2)}_2\rangle \,,
 \label{d1xi2}
   \\
& \delta|\xi^{(1)}_3\rangle
   \hspace{1ex} =\hspace{1ex}
 |\xi^{(2)}_2\rangle \,,
   && \delta|\xi^{(1)}_4\rangle
   \hspace{1ex} =\hspace{1ex}
0 \,, \label{d1xi4}
   \\
   & \delta|\xi^{(1)}_5\rangle
   \hspace{1ex} =\hspace{1ex}
-4|\xi^{(2)}_1\rangle   \,, &&
    \delta|\xi^{(1)}_6\rangle
   \hspace{1ex} =\hspace{1ex}
- \tilde{\gamma}T_1^+|\xi^{(2)}_2\rangle  \,, \label{d1xi6}
   \\
   & \delta|\xi^{(1)}_7\rangle
   \hspace{1ex} =\hspace{1ex}
2 \tilde{\gamma}T_1^+|\xi^{(2)}_{1}\rangle +
 L_1^+|\xi^{(2)}_2\rangle  \,,
   && \delta|\xi^{(1)}_8\rangle
   \hspace{1ex} =\hspace{1ex}
T^+|\xi^{(2)}_1\rangle\,, \label{d1xi8}\\
   & \delta|\xi^{(1)}_9\rangle
   \hspace{1ex} =\hspace{1ex}
   -T^+|\xi^{(2)}_1\rangle
\,, \label{d1xi9}
   \\
  & \delta|\lambda^{(1)}_1\rangle \hspace{1ex} =\hspace{1ex}
 \tilde{\gamma}T_0|\xi^{(2)}_1\rangle  \,,
   && \delta|\lambda^{(1)}_2\rangle
   \hspace{1ex} =\hspace{1ex}
- \tilde{\gamma}T_0|\xi^{(2)}_2\rangle -4|\xi^{(2)}_{1}\rangle .
 \label{d1lambd2}
\end{align}

In order to derive the action  $\mathcal{S}_{(1,1)}$
(\ref{L3/23/2}) only in terms of the component fields, we, first
of all, present the vectors $|\Psi\rangle$, $|\Psi_k\rangle$,
$|\varphi_l\rangle, |\rho_l\rangle$, $|\chi_m\rangle$ and
the corresponding bra-vectors as expansions with respect to the
initial and auxiliary creation operators:
\begin{eqnarray}
  |\Psi\rangle_{(1,1)} &=& \Bigl(\textstyle a^{+\mu}_1a^{+\nu}_2\psi_{\mu,\nu}(x)
  +f^+_2a^{+\mu}_1\tilde{\gamma}\psi^1_{\mu}(x)+
  f^+_1a^{+\nu}_2\tilde{\gamma}\psi^2_{\nu}(x) + b_{12}^+ \psi(x) \label{expPsi} \\
&&+f_1^+f_2^+ \psi^1_1(x) +\textstyle
  \frac{1}{2}a^{+\mu}_1a^{+\nu}_1b^+\psi_{\mu\nu}(x) +
  a^{+\mu}_1f^+_1b^+\tilde{\gamma}\psi^3_{\mu}(x)+b^+_{11}b^+\psi^2_2(x)
   \Bigr)|0\rangle\,, \label{Psibas}\nonumber\\
   {}_{(1,1)}\langle\tilde{\Psi}| &=& \langle0|\Bigl(\textstyle \psi^+_{\mu,\nu}(x)a^{\nu}_2a^{\mu}_1
  + \psi^{1+}_{\mu}(x)\tilde{\gamma}a^{\mu}_1f_2 +
 \psi^{2+}_{\nu}(x) \tilde{\gamma}a^{\nu}_2f_1 +  \psi^+(x)b_{12} \label{expbraPsi} \\
&&+ \psi^{1{}+}_1(x)f_2f_1 +\textstyle
  \frac{1}{2}\psi^+_{\mu\nu}(x)ba^{\nu}_1a^{\mu}_1 +
 \psi^{3+}_{\mu}(x)\tilde{\gamma} bf_1a^{\mu}_1 + \psi^{2{}+}_2(x)b b_{11}
   \Bigr)\tilde{\gamma}_0 \,, \nonumber\\
   |\chi_m\rangle_{(0,1)}   &=&
\Bigl(\hspace{-0.1em}a^{+\mu}_1b^+\chi^{1{}m}_{\mu}(x)+
a^{+\mu}_2\chi^{2{}m}_{\mu}(x)
  + f_1^+b^+ \tilde{\gamma}\chi^1_m(x) +f_2^+ \tilde{\gamma}\chi^2_m(x)\hspace{-0.1em}
  \Bigr)|0\rangle , m=1,3,\label{chi1 3}\\
  {}_{(0,1)}\langle\tilde{\chi}_m|   &=& \langle0|
\Bigl(\hspace{-0.1em}\chi^{1{}m+}_{\mu}(x) ba^{\mu}_1 +
\chi^{2{}m+}_{\mu}(x)a^{\mu}_2
  + \chi^{1+}_m(x)\tilde{\gamma}b f_1  + \chi^{2+}_m(x)\tilde{\gamma}f_2 \hspace{-0.1em}
  \Bigr)\tilde{\gamma}_0 , m=1,3,\\
|\chi_{12}\rangle_{(2,0)} &=& \Bigl(\textstyle
\frac{1}{2}a^{+\mu}_1a^{+\nu}_1\chi^{12}_{\mu\nu}(x)
  + a^{+\mu}_1f_1^+ \tilde{\gamma}\chi_{\mu}^{12}(x) +
  b_{11}^+\chi_{12}(x)\Bigr)|0\rangle\,,\label{chi12}\\
{}_{(2,0)}\langle\tilde{\chi}_{12}| &=& \langle 0|\Bigl(\textstyle
\frac{1}{2}\chi^{12+}_{\mu\nu}(x)a^{\nu}_1a^{\mu}_1
  + \chi_{\mu}^{12+}(x) \tilde{\gamma} f_1a^{\mu}_1 +
 \chi^+_{12}(x) b_{11}\Bigr)\tilde{\gamma}_0\,,\\
  |\chi_m\rangle_{(1,0)}
   &=&
   \Bigl(a^{+\mu}_1\chi^m_{\mu}(x)
  + f_1^+ \tilde{\gamma}\chi_m(x)\Bigr)|0\rangle  ,\ m=2,4,  \\
 {}_{(1,0)} \langle\tilde{\chi}_m|
   &=& \langle 0|
   \Bigl(\chi^{m+}_{\mu}(x)a^{\mu}_1
  + \chi^+_m(x) \tilde{\gamma}f_1\Bigr)\tilde{\gamma}_0  ,
  \label{chim}
  \end{eqnarray}
  \vspace{-4ex}
  \begin{align}
   & |\Psi_k\rangle_{(0,0)} \hspace{1ex} =\hspace{1ex} \psi_k(x)|0\rangle,
   k=1,...,10, &&
|\varphi_l\rangle_{(1,0)} \hspace{1ex} =   \hspace{1ex}
\Bigl(a^{+\mu}_1\varphi^l_{\mu}(x)
  + f_1^+ \tilde{\gamma}\varphi_l(x)\Bigr)|0\rangle , \label{Psik}
 \\
&  {}_{(0,0)}\langle\tilde{\Psi}_k| \hspace{1ex} = \hspace{1ex}
\langle 0|\psi^+_k(x)\tilde{\gamma}_0, &&
{}_{(1,0)}\langle\tilde{\varphi}_l| \hspace{1ex} =   \hspace{1ex}
  \langle 0| \Bigl(\varphi^{l+}_{\mu}(x) a^{\mu}_1
  + \varphi^+_l(x) \tilde{\gamma}f_1\Bigr)\tilde{\gamma}_0 ,
 \\
&  |{\rho}_l\rangle_{(-1,1)} \hspace{1ex}=\hspace{1ex}
b^+\rho_l(x)|0\rangle,\ l=1,..,4, &&
  |{\chi}_m\rangle_{(0,0)} \hspace{1ex} =
  \hspace{1ex} \chi_m(x)|0\rangle,\ m=5,...,11\,, \\
&  {}_{(-1,1)}\langle\tilde{\rho}_l| \hspace{1ex}=\hspace{1ex}
\langle 0|\rho^+_l(x)b\tilde{\gamma}_0, &&
  {}_{(0,0)}\langle\tilde{\chi}_m| \hspace{1ex} =
  \hspace{1ex} \langle 0|\chi^+_m(x)\tilde{\gamma}_0\,, \\
&  |\chi_{13}\rangle_{(-1,1)} \hspace{1ex} = \hspace{1ex}
   b^+\chi_{13}(x)|0\rangle\,, && {}_{(-1,1)}\langle\tilde{\chi}_{13}|
\hspace{1ex}  =\hspace{1ex}
   \langle 0|\chi^+_{13}(x)b\tilde{\gamma}_0\,. \label{chi13}
\end{align}
Second, let us fix preliminarily a part of the gauge ambiguity
starting from the first-stage gauge parameters. To this end, we
can use the choice of the second-stage independent vectors
  $|\xi^{(2)}_1\rangle, |\xi^{(2)}_2\rangle$, entering
relations (\ref{d1xi2})--(\ref{d1lambd2}) as shift parameters, in
order to get rid, for instance, of the vectors
$|\xi^{(1)}_k\rangle, k=1,3$, so that the description of the model
is transformed to a first-stage reducible theory with independent
first-stage gauge parameters, $|\lambda^{(1)}_l\rangle$, $l=1,2$
$|\xi^{(1)}_n\rangle$, $n=2,4,..., 9$, and without restrictions
(\ref{d1xi2})--(\ref{d1lambd2}).

By the same argument, we can make the zero-stage gauge vectors
$|\xi_m\rangle$, $m=2,4, 5, 6, 10,  13$, $|\lambda_l\rangle$,
$l=1,3$ to vanish by using a choice for the parameters
$|\xi^{(1)}_m\rangle$, $m=7,6, 5, 2, 9, 8$,
$|\lambda^{(1)}_l\rangle$, $l=1,2$, respectively, in the gauge
transformations (\ref{dxi1})--(\ref{dlambd9}). As a result, the
remaining gauge transformations with the independent first-stage
gauge parameters $|\xi^{(1)}_{4}\rangle$ for the remaining
zero-stage vectors have the form
\begin{align}
   & \delta|\xi_k\rangle \hspace{1ex} =\hspace{1ex}
0,\ \ k=1, 9, 11 \,, && \delta|\xi_3\rangle \hspace{1ex}
=\hspace{1ex}
         (-L_2^+ + T^+L_1^+)|\xi^{(1)}_{4}\rangle \,,\label{dxi1red} \\
         &
           \delta|\xi_7\rangle \hspace{1ex} =\hspace{1ex}
         -L_0|\xi^{(1)}_{4}\rangle\,,
         && \delta|\xi_8\rangle \hspace{1ex} =\hspace{1ex}
                  -\tilde{\gamma}T_0|\xi^{(1)}_{4}\rangle\,,
                  \label{dxi7red} \\
                  & \delta|\xi_{12}\rangle \hspace{1ex} =\hspace{1ex}
         -(L_1^+)^2|\xi^{(1)}_{4}\rangle \,,
         &&  \delta|\lambda_4\rangle \hspace{1ex} =\hspace{1ex}
         -\tilde{\gamma}T_0|\xi^{(1)}_{4}\rangle\,,\label{dxi12red} \\
         & \delta|\lambda_6\rangle
   \hspace{1ex} =\hspace{1ex} \tilde{\gamma}T_0
  L_1^+|\xi^{(1)}_4\rangle  \,,&&
 \delta|\lambda_l\rangle
   \hspace{1ex} =\hspace{1ex} 0, \ \ l=  2,  5, 7, 8, 9\,.  \label{dlambd69}
  \end{align}

Finally, in the same manner, we can get rid of the fields
$|\Psi_k\rangle$, $k=4,..., 7, 9, 10$, $|\varphi_1\rangle$,
$|\rho_1\rangle$, $|\chi_l\rangle$, $l=2,5,11$, with the help of a
corresponding choice for the independent (except $|\xi_7\rangle$,
$|\xi_8\rangle $, $|\lambda_4\rangle$, $|\lambda_6\rangle$ which
may be used in pairs in order to take account of its reducibility)
gauge parameters $|\lambda_2\rangle$, $|\xi_9\rangle$,
$|\xi_8\rangle$, $|\lambda_4\rangle$, $|\xi_{11}\rangle$,
$|\xi_7\rangle$, $|\lambda_l\rangle$, $l=6, 9, 7, 5,  8$
respectively, in the gauge transformations
(\ref{dpsi})--(\ref{dchi13}).  The resultant gauge transformations
for the remaining  fields $|\Psi\rangle$, $|\Psi_k\rangle$,
$k=1,2,3, 8$, $|\varphi_l\rangle$, $l=2,3,4$, $|\rho_m\rangle$, $m
= 2,3,4$, $|\chi_k\rangle$, $k = 1,3,4,6, 7, 8, 9, 10,12,13$, with
the zero-stage gauge vectors $|\xi_{k}\rangle$, $k = 1, 3, 12$,
 that have not been used previously, are reduced to
\begin{align}
&  \delta|\Psi\rangle
  \hspace{1ex} = \hspace{1ex} - \tilde{\gamma}T_1^+|\xi_1\rangle +
      L_1^+|\xi_3\rangle   + T^+|\xi_{12}\rangle\,,&& \label{dpsired}\\
  &\delta|\Psi_1\rangle
   \hspace{1ex} = \hspace{1ex}
 L_{11}|\xi_{12}\rangle + \tilde{\gamma}T_2|\xi_1\rangle +
 \frac{1}{2}T_0T_2
 |\xi_3\rangle \,, &&\delta|\Psi_3\rangle
 \hspace{1ex}  =  \hspace{1ex}
0 \,,
 \label{dpsi12red}\\
& \delta|\Psi_2\rangle
   \hspace{1ex}  =  \hspace{1ex}2 \tilde{\gamma}T_2|\xi_1\rangle
  - L_{11}|\xi_{12}\rangle -  \frac{1}{2}T_2T_0
 |\xi_3\rangle\,,  && \delta|\Psi_{8}\rangle
   \hspace{1ex} =  \hspace{1ex}
 \bigl(L_{2} - T_0T_2\bigl) |\xi_{1}\rangle\,,
   \label{dpsi8red}
      \\
  &\delta|\varphi_2\rangle
   \hspace{1ex} =\hspace{1ex}\bigl(L_{11}L_1^++T_1T_0\bigr)
   |\xi_{12}\rangle  + \frac{1}{2}
   T_0L_1^+T_2|\xi_3\rangle \,,
   && \delta(|\varphi_3\rangle\,,|\varphi_4\rangle)
   \hspace{1ex} =\hspace{1ex} T(-|\xi_3\rangle\,,|\xi_1\rangle) \,,
   \label{dvphi12red}\\
  & \delta|\rho_2\rangle \hspace{1ex} =\hspace{1ex}
\tilde{\gamma}\bigl(T_1 + \frac{1}{2}T^+T_2\bigr)|\xi_3\rangle \,,
   && \delta|\rho_3\rangle\hspace{1ex} =\hspace{1ex}
\bigl(L_1-T_0T_{1}\bigr)|\xi_{1}\rangle  \,
   ,  \label{dvphi34red}
   \\
   & \delta|\rho_4\rangle\hspace{1ex} =\hspace{1ex}
L_1|\xi_{3}\rangle-T^+L_{11}|\xi_{12}\rangle
-\tilde{\gamma}T_1|\xi_1\rangle\,
   ,  &&\delta|\chi_k\rangle \hspace{1ex} =\hspace{1ex}
0 ,\ k=6, 9, 13\,,  \label{drho23red}
    \\
   & \delta|\chi_1\rangle
   \hspace{1ex} =\hspace{1ex}-
 \tilde{\gamma}\bigl(T_0 +L_2^+T_2+ L_1^+T_1\bigr)|\xi_1\rangle\,, \label{dchi4red}
 \end{align}
 \begin{equation} \delta|\chi_{3}\rangle
     \hspace{1ex} =\hspace{1ex}
 -\bigl(2 + T_1^+T_1\bigr)|\xi_{1}\rangle + \tilde{\gamma}\bigl(T_0  +
 \frac{1}{2} L_1^+T^+T_2\bigr)
 \xi_3\rangle  -\tilde{\gamma}T^+T_1|\xi_{12}\rangle\,,
\end{equation}
\begin{align}
  & \delta|\chi_{4}\rangle
     \hspace{1ex} =\hspace{1ex}
 - T_1^+T_2|\xi_{1}\rangle
 -\tilde{\gamma}L^+_1T_2 |\xi_3\rangle
 + \tilde{\gamma}T_1|\xi_{12}\rangle\,,&&  \delta|\chi_{8}\rangle
     \hspace{1ex} =\hspace{1ex}
 \frac{1}{2} L_1^+T_1T_2|\xi_{3}\rangle\,, \label{dchi34red}\\
 & \delta|\chi_{7}\rangle
     \hspace{1ex} =\hspace{1ex}
 \tilde{\gamma}\bigl(L_1T_1-T_0L_{11}\bigr)|\xi_{12}\rangle
 -\frac{\tilde{\gamma}}{2} L_1L_1^+T_2|\xi_3\rangle\,, &&
  \delta|\chi_{10}\rangle
     \hspace{1ex} =\hspace{1ex}
  -\tilde{\gamma}T_1T|\xi_{1}\rangle\,,
 \label{dchi78red}\\
&  \delta|\chi_{12}\rangle
     \hspace{1ex} =\hspace{1ex}
  \tilde{\gamma}\bigl(T_0 + L_1^+T_1\bigr)|\xi_{12}\rangle
  -\frac{\tilde{\gamma}}{2}(L_1^+)^2T_2|\xi_3\rangle\,. &&  {} \label{dchi12sred}
\end{align}
One can easily prove that the  gauge transformations
(\ref{dpsired})--(\ref{dchi12sred}) are invariant with respect to
 their gauge transformations (\ref{dxi1red})--(\ref{dlambd69}) for the gauge
 parameters $|\xi_m\rangle$, $m= 1, 3, 12$.
 We then take into account the internal structure of the
above gauge parameters $|\xi_m\rangle$ (having the same respective
form as that for the fields $|\chi_m\rangle$, $m= 1, 3, 12$  in
(\ref{chi1 3}), (\ref{chi12})) in order to gauge away the fields
$|\varphi_l\rangle$, $l=3,4$, and to simplify the structure of the
basic field $|\Psi\rangle$. As a result, the gauge transformations
have the form
\begin{align}
&  \delta|\Psi\rangle
  \hspace{1ex} = \hspace{1ex} - \tilde{\gamma}T_1^+|\xi^r_1\rangle +
      L_1^+|\xi^r_3\rangle   + T^+|\xi_{12}\rangle\,,\label{ndpsired}\\
   &\delta|\Psi_1\rangle
   \hspace{1ex} = \hspace{1ex}
 L_{11}|\xi_{12}\rangle + \tilde{\gamma}T_2|\xi^r_1\rangle +
 \frac{1}{2}T_0T_2
 |\xi^r_3\rangle \,, &&\delta|\Psi_3\rangle
 \hspace{1ex}  =  \hspace{1ex}
0 \,,  \label{ndpsi12red}\\
&\delta|\Psi_2\rangle
   \hspace{1ex}  =  \hspace{1ex}2 \tilde{\gamma}T_2|\xi^r_1\rangle
  - L_{11}|\xi_{12}\rangle -  \frac{1}{2}T_2T_0
 |\xi^r_3\rangle\,,&&
  \delta|\Psi_{8}\rangle
   \hspace{1ex} =  \hspace{1ex}
 \bigl(L_{2} - T_0T_2\bigl) |\xi^r_{1}\rangle\,,  \label{ndpsi8red}
      \\
  &
   \delta|\varphi_2\rangle
   \hspace{1ex} =\hspace{1ex}\bigl(L_{11}L_1^++T_1T_0\bigr)
   |\xi_{12}\rangle  + \frac{1}{2}
   T_0L_1^+T_2|\xi^r_3\rangle \,,&&
   \delta|\rho_2\rangle \hspace{1ex} =\hspace{1ex}
\tilde{\gamma}\bigl(T_1 +
\frac{1}{2}T^+T_2\bigr)|\xi^r_3\rangle\equiv 0 \,,
     \label{ndvphi12red}\\
   &\delta|\rho_4\rangle\hspace{1ex} =\hspace{1ex}
L_1|\xi^r_{3}\rangle-T^+L_{11}|\xi_{12}\rangle
-\tilde{\gamma}T_1|\xi^r_1\rangle \,
   , &&  \delta|\rho_3\rangle\hspace{1ex} =\hspace{1ex}
\bigl(L_1-T_0T_{1}\bigr)|\xi^r_{1}\rangle  \,
   , \label{ndrho4red} \\
   & \delta|\chi_k\rangle \hspace{1ex} =\hspace{1ex}
0 ,\ k=6, 9, 10, 13\,,
   && \delta|\chi_1\rangle
   \hspace{1ex} =\hspace{1ex}-
 \tilde{\gamma}\bigl(T_0 +L_2^+T_2+ L_1^+T_1\bigr)|\xi^r_1\rangle\,, \label{ndchi4red}
 \end{align}
 \begin{equation} \delta|\chi_{3}\rangle
     \hspace{1ex} =\hspace{1ex}
 -\bigl(2 + T_1^+T_1\bigr)|\xi^r_{1}\rangle + \tilde{\gamma}\bigl(T_0  +
 \frac{1}{2} L_1^+T^+T_2\bigr)|
 \xi^r_3\rangle  -\tilde{\gamma}T^+T_1|\xi_{12}\rangle\,,
\end{equation}
\begin{align}
  & \delta|\chi_{4}\rangle
     \hspace{1ex} =\hspace{1ex}
 - T_1^+T_2|\xi^r_{1}\rangle
 -\tilde{\gamma}L^+_1T_2 |\xi^r_3\rangle
 + \tilde{\gamma}T_1|\xi_{12}\rangle\,,\qquad \delta|\chi_{8}\rangle
     \hspace{1ex} =\hspace{1ex}
 \frac{1}{2} L_1^+T_1T_2|\xi^r_{3}\rangle=0\,, \label{ndchi34red}\\
& \delta(|\chi_{7}\rangle\,,|\chi_{12}\rangle)
     \hspace{1ex} =\hspace{1ex}\tilde{\gamma}\Bigl(
 \bigl(L_1T_1-T_0L_{11}\bigr)\,, \bigl(T_0 + L_1^+T_1\bigr)\Bigr)
  |\xi_{12}\rangle
 -\frac{\tilde{\gamma}}{2} \Bigl(L_0T_2\,,(L_1^+)^2T_2\Bigr)
 |\xi^r_3\rangle\,,&&\label{ndchi12sred}
\end{align}
where the vectors $|\xi_m^r\rangle$, $m=1,3$, are  solutions of
the equations $T|\xi_m\rangle = 0$,
\begin{eqnarray}
  |\xi_m^r\rangle_{(0,1)}   &=&
\bigl(\hspace{-0.1em}a^{+\mu}_1b^+-2
a^{+\mu}_2\bigr)\xi^{1{}m}_{\mu}(x)
  - \tilde{\gamma}\bigl(f_1^+b^+  -2
  f_2^+ \bigr)\xi^1_m(x)\hspace{-0.1em}
  \Bigr)|0\rangle , m=1,3,\label{xi1 3red}
\end{eqnarray}
and $|\xi_{12}\rangle_{(2,0)}$ has the form
\begin{equation} |\xi_{12}\rangle_{(2,0)} = \Bigl(
\frac{1}{2}a^{+\mu}_1a^{+\nu}_1\xi^{12}_{\mu\nu}(x)
  + a^{+\mu}_1f_1^+ \tilde{\gamma}\xi_{\mu}^{12}(x) +
  b_{11}^+\xi_{12}(x)\Bigr)|0\rangle\,.
\end{equation}
 As
a consequence, the action (\ref{L3/23/2}) is simplified as
follows:
\begin{eqnarray}
{\cal{}S}_{(1,1)} &=& \Bigl[\langle
\tilde{\Psi}|K_{(1,1)}\left\{\textstyle\frac{1}{2}T_0 |\Psi\rangle
+ T^+_1|\chi_1\rangle + \tilde{\gamma}L^+_1|\chi_3\rangle +
\tilde{\gamma}L^+_2|\chi_4\rangle
 + \tilde{\gamma}T^+|\chi_{12}\rangle + \tilde{\gamma}L^+_{11}|\chi_{13}\rangle
\right\}
 \nonumber\\
&&  + \langle \tilde{\Psi}_1|K_{(1,1)}\left\{T_0 |\Psi_2\rangle -
2\tilde{\gamma}|\Psi_8\rangle - \tilde{\gamma} L_2|\chi_3\rangle
+\tilde{\gamma} |\chi_7\rangle \right\}
 \nonumber \\
&& + \langle \tilde{\Psi}_2|K_{(1,1)}\left\{
 \tilde{\gamma} L_1|\chi_4\rangle
-\tilde{\gamma} |\chi_7\rangle \right\}  + \langle
\tilde{\Psi}_8|K_{(1,1)}\left\{ -T_1|\chi_4\rangle +
\tilde{\gamma}|\chi_8\rangle\right\}
\nonumber \\
&& +\langle \tilde{\Psi}_3|K_{(1,1)}\left\{-
2\tilde{\gamma}|\Psi_8\rangle +  T_2|\chi_1\rangle
 -\tilde{\gamma}
|\chi_6\rangle  \right\}
\nonumber \\
 &+&
  \langle
\tilde{\varphi}_{2}|K_{(1,1)}\left\{-
\tilde{\gamma}T|\chi_3\rangle + \tilde{\gamma}|\chi_4\rangle +
T_1^+|\chi_{10}\rangle\right\}
 \nonumber
  \end{eqnarray}
  \vspace{-4ex}
  \begin{eqnarray}
\phantom{{\cal{}S}_{(1,1)}}  &+&
 \langle \tilde{\rho}_{2}|K_{(1,1)}\left\{ - T_0|\rho_3\rangle +
\tilde{\gamma}L_1|\chi_1\rangle -\tilde{\gamma}T^+|\chi_{9}\rangle
\right\} \nonumber\\
&& + \langle \tilde{\rho}_{3}|K_{(1,1)}\left\{ -
2\tilde{\gamma}|\rho_4\rangle - T_1|\chi_3\rangle
-\tilde{\gamma}T^+|\chi_{8}\rangle \right\}
\nonumber\\
&& + \langle \tilde{\rho}_{4}|K_{(1,1)}\left\{
-\textstyle\frac{1}{2} T_0|\rho_4\rangle -
\tilde{\gamma}L_1|\chi_3\rangle -\tilde{\gamma}T^+|\chi_{7}\rangle
+ \tilde{\gamma}|\chi_{13}\rangle\right\}
\nonumber\\
&+&  \langle \tilde{\chi}_{1}|K_{(1,1)}\left\{
-\tilde{\gamma}|\chi_3\rangle\right\}
-\textstyle\frac{1}{2}\langle \tilde{\chi}_{3}|K_{(1,1)}
T_0|\chi_3\rangle
 -\textstyle\frac{1}{2}
 \langle
\tilde{\chi}_{4}|K_{(1,1)} T_0|\chi_4\rangle \nonumber \\
&& + \langle \tilde{\chi}_{6}|K_{(1,1)}\left\{
-\tilde{\gamma}|\chi_{10}\rangle\right\} + \langle
\tilde{\chi}_{7}|K_{(1,1)}\left\{
-\tilde{\gamma}|\chi_{10}\rangle\right\} +\langle
\tilde{\chi}_{8}|K_{(1,1)}T_0|\chi_{10}\rangle \Bigr] + c.c.\,  \label{L3/23/2red}
\end{eqnarray}

Then, we gauge away the symmetric part $\psi_{(\mu,\nu)} =
\frac{1}{2}(\psi_{\mu,\nu}+\psi_{\nu,\mu})$ from the  basic
spin-tensor $\psi_{\mu\nu}$ as well as
 the fields $ \psi^k_{\mu}, k=2,3, \psi_l^l, l=1,2, $ from
 the basic field $|\Psi\rangle_{(1,1)}$ (\ref{expPsi}) using
 all the components of the
 gauge parameters $|\xi_{12}\rangle_{(2,0)}$ and
 $|\xi^r_{1}\rangle_{(0,1)}$  in the  gauge transformations
 (\ref{ndpsired})--(\ref{ndchi12sred}),
so that only the  following nontrivial relations for the used
gauge parameters,
\begin{equation}
\xi^{12}_{\mu\nu}(x) = 2\imath \partial_{(\mu}\xi_{\nu)}^{13}(x) +
2\imath
 \gamma_{(\mu}\partial_{\nu)}
 \xi_{3}^{1}(x),\quad \xi^{12}_\mu(x) = 2\imath\partial_{\mu}\xi_{3}^{1}(x)\,,
 \quad \xi^{11}_\mu (x)=
  -\imath\partial_{\mu}\xi_{3}^{1}(x)\,,
\end{equation}
hold true. The resulting spectrum of the fields,
 $\psi_{[\mu\nu]}, \psi_{\mu\nu}$, $\psi^1_{\mu}, \psi,
 \psi_k, k =1,2,3,8$, $\varphi^\mu_2, \varphi_2$, $\rho_l,
 l=2,3,4$,
 $\chi^{1{}m}_\mu, \chi^{2{}m}_\mu,  \chi^{1}_m, \chi^{2}_m$, $m=1, 3$, $ \chi^{4}_\mu$, $\chi_k$, $k=4,6,...,10, 12, 13$,
 $\chi_{\mu\nu}^{12}, \chi_{\mu}^{12}, $ and their
 nontrivial gauge transformations have the
 component form, for $\gamma_{\rho\sigma} =
\frac{1}{2}(\gamma_{\rho}\gamma_\sigma-\gamma_\sigma\gamma_{\rho})$,
\begin{align}
    & \delta\psi_{[\mu\nu]}(x)\hspace{1ex} = 2\imath \partial_{[\mu}\xi^{13}_{\nu]}(x)+2\imath \gamma_{[\mu}\partial_{\nu]}\xi^{
    1}_{3}(x)\,,\label{psimu,nu}\\
    & \delta\psi_1(x)\hspace{1ex} = {\imath}\gamma^{\mu\nu}\partial_{\nu}
    \xi^{13}_{\mu}(x)
+\imath(d-1)\gamma^{\nu}\partial_{\nu} \xi^{1}_3(x)
    \,,
    \\
    & \delta\psi_2(x)\hspace{1ex} = {\imath}\gamma^{\mu\nu}\partial_{\nu}
    \xi^{13}_{\mu}(x)
+\imath(d-1)\gamma^{\nu}\partial_{\nu} \xi^{1}_3(x)\hspace{1ex}=
\hspace{1ex}\delta\psi_1(x)
    \,,
    \\
    & \delta\chi^4_\mu(x)\hspace{1ex} =
    \hspace{1ex}\imath \Bigl(
    2\gamma^{\nu}\partial_{[\nu} \xi_{\mu]}^{13} (x)+ [\gamma_{\mu\nu}\partial^{\nu}+(d-1)\partial_{\mu}]\xi_3^1(x)\Bigr) \,,
    \label{rho10mu}\\
    & \delta\chi^{23}_\mu(x)\hspace{1ex} =  \hspace{1ex}- \imath \Bigl(
    2\gamma^{\nu}\partial_{[\nu} \xi_{\mu]}^{13} (x)+
    [\gamma_{\mu\nu}\partial^{\nu}+(d-1)\partial_{\mu}]\xi_3^1(x)\Bigr)
    \hspace{1ex}=\hspace{1ex}-\delta\chi^4_\mu(x)
    \,.\label{chi2310mu}
    \end{align}
whereas the nontrivial gauge transformation for the gauge
parameter is written as follows:
\begin{align}
\delta\xi^{13}_\mu(x)\hspace{1ex} =  \hspace{1ex} -\imath
  \partial_\mu\xi^{(1)}_4 (x)\,. \label{dlambd4red}
  \end{align}
Then let us remove the  additional  fields from the spectrum of
the above fields by means of their equations of motion. Thus, for
the extremals $\frac{\delta_l {\cal{}S}_{(1,1)}}{\delta
\langle\chi_{k}|}= \frac{\delta_l {\cal{}S}_{(1,1)}}{\delta
\langle\rho_{3}|}=0 $, with the vectors $|\chi_{k}\rangle$, $k= 1,
6, 7, 9, 12, 13$, $|\rho_{3}\rangle$ considered as Lagrangian
multipliers,
 we have the solutions
\begin{align}\label{algsol}
& K^{-1}_{(1,1)}\frac{\delta_l {\cal{}S}_{(1,1)}}{\delta \langle\chi_{12}|}=
\tilde{\gamma}T |\Psi\rangle =0  && \Longrightarrow && \psi_{\mu\nu} = \psi^1_{\mu} = \psi= 0,\\
& K^{-1}_{(1,1)}\frac{\delta_l {\cal{}S}_{(1,1)}}{\delta \langle\chi_{13}|}=
\tilde{\gamma}L_{11} |\Psi\rangle + \tilde{\gamma}|\rho_4\rangle =
0  && \Longrightarrow &&
\rho_4 = 0\,, \label{auxEQ1}\\
& K^{-1}_{(1,1)}\frac{\delta_l {\cal{}S}_{(1,1)}}{\delta \langle\chi_{9}|}=
-\tilde{\gamma}T|\rho_2\rangle = 0  && \Longrightarrow &&
\rho_2 = 0\,,\\
& K^{-1}_{(1,1)}\frac{\delta_l {\cal{}S}_{(1,1)}}{\delta \langle\chi_{6}|}=
-\tilde{\gamma} (|\Psi_3\rangle + |\chi_{10}\rangle)  = 0  &&
\Longrightarrow &&
\chi_{10} = - \psi_3\,,\\
& K^{-1}_{(1,1)}\frac{\delta_l {\cal{}S}_{(1,1)}}{\delta \langle\chi_{7}|}=
\tilde{\gamma} (|\Psi_1\rangle -|\Psi_2\rangle -
|\chi_{10}\rangle)  = 0 && \Longrightarrow &&
\chi_{10} = \psi_1-\psi_2\,,\\
& K^{-1}_{(1,1)}\frac{\delta_l {\cal{}S}_{(1,1)}}{\delta \langle\chi_{1}|}=
T_1|\Psi\rangle +T^+_2|\Psi_3\rangle -
\tilde{\gamma}|\chi_{3}\rangle  = 0 && \Longrightarrow &&
\gamma^\mu \psi_{[\mu\nu]} = \gamma_\nu \chi^2_3
- \chi^{23}_\nu\,, \ \psi_3 = \chi^2_3\,, \label{auxEQf-1}\\
& K^{-1}_{(1,1)}\frac{\delta_l {\cal{}S}_{(1,1)}}{\delta \langle\rho_{3}|}=-
T_1|\chi_3\rangle -\tilde{\gamma}T^+|\chi_{8}\rangle =0 &&
\Longrightarrow && \chi_8 = -2\chi^2_3\,,\ \chi_{\mu}^{13} =
\chi^1_3=0\,,\label{auxEQf}
\end{align}
(with $K^{-1}_{(1,1)}$ being the inverse of $K_{(1,1)}$) and thus the basic vector $|\Psi\rangle$ contains only the
antisymmetric spin-tensor  $\psi_{[\mu\nu]}(x)$.  After a
substitution of expressions (\ref{expPsi})--(\ref{chi13}) into
(\ref{L3/23/2red}), we find the action for a field of spin
$(3/2,3/2)$ in a manifest form, with the remaining  auxiliary
fields:
\begin{eqnarray}
{\cal{}S}_{(1,1)} &=& \int d^d x \Biggl[\bar{\psi}_{[\mu\nu]}
\Bigl\{-\frac{\imath}{2}\gamma^{\rho}\partial_{\rho}
{\psi}^{[\mu\nu]}  -
\imath\partial^{[\mu}\Bigl[\gamma^{\nu]}({\psi}_2- {\psi}_1)
+\gamma_\rho{\psi}^{[\nu]\rho]}\Bigr] -
\imath\partial^{[\nu}\chi^{4{}\mu]} \Bigr\}
 \label{expL3/23/2} \nonumber\\
&& + \bar{\psi}_1\Bigl\{-{\imath}\gamma^{\rho}\partial_{\rho}
{\psi}_2-2{\psi}_8 -\imath \partial^\mu
\Bigl[\gamma_{\mu}({\psi}_2- {\psi}_1)
+\gamma^\rho{\psi}_{[\mu\rho]}\Bigr]\Bigr\}
\nonumber \\
&& + {\imath}\bar{\psi}_2\partial^{\mu} {\chi}_\mu^4 -2
(\bar{\psi}_2- \bar{\psi}_1)\psi_8
+ \bar{\psi}_8\Bigl\{\gamma^\mu\chi^{4}_\mu +(2-d)\chi_4 - 2(\psi_2- \psi_1)\Bigr\}\nonumber \\
&& - \bar{\varphi}^2_\mu \Bigl(\gamma_\rho{\psi}^{[\mu\rho]}+
\chi^{4{}\mu} \Bigr) - (2-d)\bar{\varphi}^2\chi^{4}
\nonumber \\
&& -\frac{\imath}{2}\Bigl[(\bar{\psi}_2- \bar{\psi}_1)\gamma^{\mu}
+\bar{\psi}^{[\mu\tau]}\gamma_\tau\Bigr]\gamma^{\sigma}
\partial_{\sigma} \Bigl[\gamma_{\mu}({\psi}_2-
{\psi}_1) +\gamma^\rho{\psi}_{[\mu\rho]}\Bigr] \nonumber
 \\
\phantom{{\cal{}S}_{(1,1)}} && + \frac{\imath}{2}(6-d)
(\bar{\psi}_2- \bar{\psi}_1)\gamma^{\rho}\partial_{\rho}({\psi}_2-
{\psi}_1)-\frac{\imath}{2}\bar{\chi}^4_\mu
\gamma^{\nu}\partial_{\nu}{\chi}^{4{}\mu}\nonumber
 \\
\phantom{{\cal{}S}_{(1,1)}} &&
+\imath(1-\frac{d}{2})
\bar{\chi}^4 \gamma^{\nu}\partial_{\nu}{\chi}^4   - 2\imath
(\bar{\psi}_2- \bar{\psi}_1)\gamma^{\nu}\partial_{\nu}({\psi}_2-
{\psi}_1)
 \Biggr] + c.c.
\end{eqnarray}
 Then, from the extremals for the fields $\bar{\varphi}^2, \bar{\varphi}^2_\mu,
\bar{\psi}_8$, we have their respective solutions
\begin{align}\label{algsol2} & \chi_4 =0 \,,&&
\chi^{4{}\mu} = \gamma_\rho{\psi}^{[\rho\mu]}  && {\psi}_1 =
2{\psi}_2 +
 \frac{1}{2}\gamma^{\rho\mu}{\psi}_{[\rho\mu]}\,,
\end{align}
so that the action (\ref{expL3/23/2}) is transformed  as follows:
\begin{eqnarray}
{\cal{}S}_{(1,1)} &=& \int d^d x \Biggl[\bar{\psi}_{[\mu\nu]}
\Bigl\{-\frac{\imath}{2}\gamma^{\rho}\partial_{\rho}
{\psi}^{[\mu\nu]} -
\imath\partial^{[\mu}\Bigl[\gamma^{\nu]}(-{\psi}_2-
 \frac{1}{2}\gamma^{\rho\sigma}{\psi}_{[\rho\sigma]})
+\gamma_\rho{\psi}^{[\nu]\rho]}\Bigr]
 \label{expL3/23/2pr} \nonumber\\
&&  - \imath\partial^{[\nu}\gamma_\rho{\psi}^{[\rho\mu]]} \Bigr\}
+ \Bigl\{2\bar{\psi}_2 +
 \frac{1}{2}\bar{\psi}_{[\rho\nu]}\gamma^{\nu\rho}\Bigr\}\Bigl\{-\imath
 \partial^\mu \Bigl[\gamma_{\mu}
 \frac{1}{2}\gamma^{\sigma\rho}{\psi}_{[\rho\sigma]}
+\gamma^\rho{\psi}_{[\mu\rho]}\Bigr]\Bigr\}
\nonumber \\
&& -\frac{\imath}{2}\Bigl[\Bigl\{-\bar{\psi}_2 -
 \frac{1}{2}\bar{\psi}_{[\rho\nu]}\gamma^{\nu\rho}\Bigr\}\gamma_{\mu}
+\bar{\psi}_{[\mu\rho]}\gamma^\rho\Bigr]\gamma^{\sigma}
\partial_{\sigma}\Bigl[\gamma^{\mu}\Bigl\{-{\psi}_2 -
 \frac{1}{2}\gamma^{\sigma\tau}{\psi}_{[\sigma\tau]}\Bigr\}
+\gamma_\tau{\psi}^{[\mu\tau]}\Bigr] \nonumber\\
&&+ \frac{\imath}{2}(2-d) \Bigl\{\bar{\psi}_2 +
 \frac{1}{2}\bar{\psi}_{[\rho\nu]}\gamma^{\nu\rho}\Bigr\}\gamma^{\sigma}
 \partial_{\sigma}
 \Bigl\{{\psi}_2
 +
 \frac{1}{2}\gamma^{\tau\mu}{\psi}_{[\tau\mu]}\Bigr\} - {\imath}\bar{\psi}_2\partial^{\mu}
\gamma^\rho{\psi}_{[\mu\rho]} \nonumber
 \\
\phantom{{\cal{}S}_{(1,1)}} &&
-\frac{\imath}{2}\bar{\psi}_{[\rho\mu]}\gamma^\rho
\gamma^{\nu}\partial_{\nu}\gamma_\sigma{\psi}^{[\sigma\mu]}
 \Biggr] + c.c.\,
\end{eqnarray}
One can show that the terms in (\ref{expL3/23/2pr}) with the
auxiliary spinor $\psi_2$ vanish identically, so that we have the
final form of the action and reducible gauge transformations for
the spin-tensor ${\psi}_{[\mu\nu]}$,
\begin{eqnarray}
{\cal{}S}_{(1,1)} &=& \int d^d x \bar{\psi}_{[\mu\nu]}
\Bigl\{-{\imath}\gamma^{\rho}\partial_{\rho} {\psi}^{[\mu\nu]} +
{\imath}\partial^{[\mu}\gamma^{\nu]}
\gamma^{\rho\sigma}{\psi}_{[\rho\sigma]} +
2\imath\partial^{[\nu}\gamma_\rho{\psi}^{[\mu]\rho]}+
2\imath\gamma^{\nu}\partial_\rho{\psi}^{[\mu\rho]}
 \label{expL3/23/2fint} \nonumber\\
&& -\frac{\imath}{2}\gamma^{\nu\mu}
 \gamma^{\sigma}
\partial_{\sigma}\gamma_{\rho\tau}{\psi}^{[\rho\tau]}
+{\imath}\gamma^{\nu\mu}
\partial_{[\rho}\gamma_{\tau]}{\psi}^{[\rho\tau]}
-2{\imath}{}\gamma^\mu
\gamma^{\rho}\partial_{\rho}\gamma_\sigma{\psi}^{[\sigma\nu]}
 \Bigr\}\,,\\
\label{gaugephys} \delta\psi_{[\mu\nu]}  &=& 2\imath
\partial_{[\mu}\xi_{\nu]}+2\imath
\gamma_{[\mu}\partial_{\nu]}\eta\,,\quad \delta\xi_{\mu} = \imath
\partial_{\mu}\xi^{(1)},\quad (\xi_{\mu}, \eta, \xi^{(1)}) \equiv
(\xi^{13}_{\mu},\xi^{1}_3 ,\xi^{(1)}_4).
\end{eqnarray}

To obtain a Lagrangian description of the massive
rank-2 antisymmetric spin-tensor  $\psi_{[\mu\nu]}$, having
the Young tableaux (\ref{Young k2}) and subject to the conditions
(\ref{Eq-1}), (\ref{Eq-2}) and the requirement
$(\imath\gamma^{\mu}\partial_{\mu}-m)\Psi_{\mu,\nu}(x)
=0$, instead of (\ref{Eq-0}), we may follow the example of Section
\ref{Example3/2} with $(h^1_m,h^2_m) =
([1-d]/2,[5-d]/2)$ and apply the prescription (\ref{changeferm})--(\ref{chifm}),
i.e., starting from the expansion (\ref{chi00})--(\ref{brachi01}), (\ref{expPsi})--(\ref{chi13}),
where it is only the vectors in
$|\chi_0^i\rangle$,  $|\Lambda_0^i\rangle$, $|\Lambda_0^{(1)0}\rangle$
that change to $|\chi_{0m}^i\rangle$, $|\Lambda_{0m}^i\rangle$, $|\Lambda_{0m}^{(1)0}\rangle$,
respectively,
\begin{eqnarray}\label{chi320m}
  |\Psi_{m}\rangle_{(1,1)}& = &|\Psi\rangle_{(1,1)} +
 \Bigl(\textstyle
  a^{+\mu}_1\big(b^+_2\psi^1_{m|\mu}(x) + b^+_1b^+\psi^2_{m|\mu}(x)\big)+b^+_1a^{+\mu}_2\psi^3_{m|\mu}(x)   \\
&& +b_1^+\big(b^+_2\psi^1_{m}(x)+ b^+_1b^+\psi^2_{m}(x)\big)+
  f^+_1\tilde{\gamma}\big(b^+_2\psi^3_{m}(x) + b^+_1b^+\psi^4_{m}(x)\big)+ f^+_2\tilde{\gamma} b^+_1\psi^5_{m}(x)
   \Bigr)|0\rangle, \nonumber \\
      {}_{(1,1)}\langle\tilde{\Psi}_m| &=&{}_{(1,1)}\langle\tilde{\Psi}| + \langle0|\Bigl(\textstyle \big(\psi^{1+}_{m|\mu}(x)b_2 + \psi^{2+}_{m|\mu}(x)b_1b\big)a^{\mu}_1 +  +\psi^{3+}_{m|\mu}(x)b_1a^{\mu}_2 \label{expbraPsim} \\
&& + \big(\psi^{1+}_{m}(x)b_2+ \psi^{2+}_{m}(x)bb_1\big)b_1  +  \big(\psi^{3+}_{m}(x)b_2 + \psi^{4+}_{m}(x)b_1b\big)\tilde{\gamma}f_1
   +  \psi^{5+}_{m}(x)b_1\tilde{\gamma}f_2
   \Bigr)\tilde{\gamma}_0 \,, \nonumber\\
    |\chi^k_m\rangle_{(0,1)}   &=&|\chi^k\rangle_{(0,1)}+
\bigl(\hspace{-0.1em }b^+_1b^+\chi^{1k}_m(x)+
b^+_2\chi^{2k}_{m}(x)
  \bigr)|0\rangle , \label{chi13m}\\
  {}_{(0,1)}\langle\tilde{\chi}^k_m|   &=& {}_{(0,1)}\langle\tilde{\chi}^k| +\langle0|
\bigl(\hspace{-0.1em} \chi^{1k+}_{m}(x) bb_1 +
\chi^{2k+}_{m}(x)b_2
 \bigr)\tilde{\gamma}_0 ,\ \  k=1,3,\\
|\chi_{12|m}\rangle_{(2,0)} &=&|\chi_{12}\rangle_{(2,0)}+  \bigl(\textstyle
a^{+\mu}_1b^{+}_1\chi^{12}_{m|\mu}(x)
  + b^{+}_1f_1^+ \tilde{\gamma}\chi_{m}^{12}(x) +
  (b_{1}^+)^2\chi^{12}_{1|m}(x)\bigr)|0\rangle\,,\label{chi12m}\\
{}_{(2,0)}\langle\tilde{\chi}_{12|m}| &=& {}_{(2,0)}\langle\tilde{\chi}_{12}| + \langle 0|\bigl(\textstyle
\chi^{12^+}_{m|\mu}(x)a^{\mu}_1b_1
  + \chi_{m}^{12+}(x) \tilde{\gamma} f_1b_1 +
 \chi^{12+}_{1|m}(x) b^2_{1}\bigr)\tilde{\gamma}_0\,,
  \end{eqnarray}
  \vspace{-4ex}
  \begin{align}
 \hspace{-0.3em} &\hspace{-0.3em} |\chi_{m|k}\rangle_{(1,0)} \hspace{1ex} =   \hspace{1ex}
  |\chi_{k}\rangle_{(1,0)}+
   b ^{+}_1\chi^k_{m}(x)
  |0\rangle,  \hspace{-0.1em} &&\hspace{-0.em} {}_{(1,0)} \langle\tilde{\chi}_{m|k}|
   \hspace{1ex} =   \hspace{1ex} {}_{(1,0)} \langle\tilde{\chi}_{k}|+ \langle 0|
   \chi^{k+}_{m}(x)b_1\tilde{\gamma}_0  ,\ k=2,4,  \label{chim3232}
  \\  \hspace{-0.3em} &\hspace{-0.3em}
|\varphi_{l|m}\rangle_{(1,0)} \hspace{1ex} =   \hspace{1ex}|\varphi_{l}\rangle_{(1,0)}+
b_1^+ \varphi_{l|m}(x)|0\rangle , \hspace{-0.2em} &&\hspace{-0.2em} {}_{(1,0)}\langle\tilde{\varphi}_{l|m}| \hspace{0.9ex} =   \hspace{0.9ex}{}_{(1,0)}\langle\tilde{\varphi}_{l}|+
  \langle 0| \varphi^{l+}_{m}(x) b_1
  \tilde{\gamma}_0,\
   l=1,...,4,  \label{Psikm}.
\end{align}
Then all the terms in (\ref{L3/23/2}), except for the action ${\cal{}S}^m_{(1,1)}$,
have the same form, with a change of the massless $O_I$ to the massive $O_I^m$,
except for the vectors $|\Psi_k\rangle$, $k=5,...,8$, $|\varphi_1\rangle$,
$|\varphi_4\rangle$, $|\rho_2\rangle $, $|\rho_3\rangle$, $|\chi_l\rangle$,
$l=3, 4, 6, 11, 12$, with ``$\tilde{\gamma}$-conjugated'' massive Grassmann-odd
operators $T_0^{*m}, T_i^{*(+)m}$ for $\big(T_0^m, T_i^{(+)m}\big) \equiv \big(\check{t}_0,
T_i^{(+)}-\tilde{\gamma}b_i^{(+)}\big)$ for $i=1,2$, due to the new property
 \begin{align}\label{new_property}
  & \tilde{\gamma}T_0^m\tilde{\gamma} = T_0^{*m}=t_0- \tilde{\gamma}m \ne T_0^m && \mathrm{with} \ \tilde{\gamma}T_0^{*m}\tilde{\gamma}=T_0^m, \\
 \label{new_property1}
  & \tilde{\gamma}T_i^{(+)m}\tilde{\gamma} = T_i^{*(+)m}=T_i^{(+)}+\tilde{\gamma}b_i^{(+)} \ne T_i^{(+)m} && \mathrm{with} \ \tilde{\gamma}T_i^{*(+)m}\tilde{\gamma}=T_i^{(+)m}.
  \end{align}
In the lowest gauge transformations (\ref{d1xi2})--(\ref{d1lambd2}),
it is only the relations (\ref{d1xi6}), (\ref{d1lambd2}) for
$\delta|\xi^{(1)}_6\rangle$, $\delta|\lambda^{(1)}_2\rangle$
that are modified, respectively,
$- \tilde{\gamma}T_1^+|\xi^{(2)}_2\rangle \to T_1^{+ m}\tilde{\gamma}\xi^{(2)}_2\rangle $
and $-\tilde{\gamma}T_0|\xi^{(2)}_1\rangle+... \to T_0^m\tilde{\gamma}|\xi^{(2)}_1\rangle+... $.
The same change should be made in the gauge transformations for
$|\xi^{}_k\rangle $, $k= 3, 4, 6, 12$, with the parameters $|\xi^{(1)}_3\rangle$,
$|\xi^{(1)}_9\rangle$, $|\xi^{(1)}_2\rangle$, $|\xi^{(1)}_4\rangle$, $|\xi^{(1)}_7\rangle$
in (\ref{dxi3}), (\ref{dxi4}), (\ref{dxi6}), (\ref{dxi12}); for $|\lambda^{}_e\rangle $,
$e= 2, 3, 7, 9$, with the parameters $|\xi^{(1)}_e\rangle$, $|\lambda^{(1)}_1\rangle$
in (\ref{dlambd2}), (\ref{dlambd3}), (\ref{dlambd7}), (\ref{dlambd9}); and also in the gauge
transformations for the fields $|\Psi\rangle$, $|\Psi_k\rangle$, $k=1, ...,4$,
$|\varphi_2\rangle$, $|\varphi_3\rangle$,  $|\rho_1\rangle$, $|\chi_e\rangle$,
$e=1, 2, 8, 9, 10$, with the respective parameters $|\lambda^{}_o\rangle$, $o=1,4,6,8$,
$|\xi_e\rangle$ in (\ref{dpsi})--(\ref{dpsi4}), (\ref{dvphi2}), (\ref{dvphi3}),
(\ref{drho1}), (\ref{dchi1}), (\ref{dchi2}), (\ref{dchi8})--(\ref{dchi10}).
The gauge-fixing (\ref{dxi1red})--(\ref{dchi12sred}) remains valid so far as one works
only with the ghost-independent structure of the modified vectors $|\chi_{m}^i\rangle$,...
with a change of the massless $O_I$ to the massive $O_I^m$, featuring the mentioned properties.

From Eqs. (\ref{dxi1red})--(\ref{dlambd69}), we gauge away, using
$|\xi^{(1)}_{4|m}\rangle = |\xi^{(1)}_{4}\rangle$, the $(b_1^+)^2$-dependent
vector of the gauge parameter $|\xi_{12|m}\rangle_{(2,0)}$ in (\ref{dxi12red}),
which has the same form as $|\chi_{12|m}\rangle_{(2,0)}$ in (\ref{chi12m}).
The theory becomes an irreducible one, with independent $|\xi^r_{12|m}\rangle_{(2,0)}$,
$|\xi_{k|m}\rangle_{(0,1)}$, $k=1,3$, so that the gauge transformations for the field
vectors (\ref{dpsired})--(\ref{dchi12sred}) hold true, albeit in the massive case.

Once again,  the fields $|\varphi_{l|m}\rangle_{(1,0)}$, $l=3,4$,  are removed by partially
using $|\xi_{k|m}\rangle_{(0,1)}$, $k= 1,3$, from the massive counterpart
of (\ref{dvphi12red}), so that the gauge transformations (\ref{ndpsired})--(\ref{ndchi12sred})
with the changes (here, the index ``m'' in the operators is omitted)
\begin{align}
&  \delta|\Psi_m\rangle
  \hspace{1ex} = \hspace{1ex} - \tilde{\gamma}T_1^{*+}|\xi^r_{1|m}\rangle +
      L_1^{+}|\xi^r_{3|m}\rangle   + T^{+}|\xi^r_{12|m}\rangle\,, &&
  \delta|\Psi_{8|m}\rangle
   \hspace{1ex} =  \hspace{1ex}
 \bigl(L_{2} - T_0^{*}T_2^{*}\bigl) |\xi^r_{1|m}\rangle\,,\label{ndpsiredm}\\
        &\delta|\rho_{4|m}\rangle\hspace{1ex} =\hspace{1ex}
L_1|\xi^r_{3|m}\rangle-T^{+}L_{11}|\xi^r_{12|m}\rangle
-\tilde{\gamma}T^*_1|\xi^r_{1|m}\rangle \,
   , &&  \delta|\rho_{3|m}\rangle\hspace{1ex} =\hspace{1ex}
\bigl(L_1-T^{*}_0T^{*}_1\bigr)|\xi^r_{1|m}\rangle  \,
   , \label{ndrho4redm} \\
   & \delta|\chi_{1|m}\rangle
   \hspace{1ex} =\hspace{1ex}-
 \tilde{\gamma}\bigl(T_0^{*} +L_2^{+}T_2^{*}+ L_1^{+}T_1^{*}\bigr)|\xi^r_{1|m}\rangle\,, \label{ndchi4redm}\end{align}
 \vspace{-1.5ex}
 \begin{equation}
 \delta|\chi_{3|m}\rangle
     \hspace{1ex} =\hspace{1ex}
 -\bigl(2 + T_1^{*+}T^{*}_1\bigr)|\xi^r_{1|m}\rangle + \tilde{\gamma}\bigl(T_0^{*}  +
 \frac{1}{2} L_1^{+}T^{+}T_2\bigr)|
 \xi^r_{3|m}\rangle  -\tilde{\gamma}T^{+}T_1|\xi^r_{12|m}\rangle\,,
 \end{equation}
 \vspace{-1.5ex} \begin{align} & \delta|\chi_{4|m}\rangle
     \hspace{1ex} =\hspace{1ex}
 - T_1^{*+}T_2^{*}|\xi^r_{1|m}\rangle
 -\tilde{\gamma}L^{+}_1T_2 |\xi^r_{3|m}\rangle
 + \tilde{\gamma}T_1|\xi^r_{12|m}\rangle \label{ndchi34redm}
\end{align}
are valid for the vectors $|\xi_{k|m}^r\rangle$, $k=1,3$, being
solutions of the equations $T^m|\xi_{k|m}\rangle = 0$,
\begin{eqnarray}
  |\xi_{k|m}^r\rangle_{(0,1)}   &=&|\xi_{k}^r\rangle_{(0,1)}+
\bigl(b^{+}_1b^+-2
b^+_2\bigr)\xi^{1}_{k|m}(x)
  \bigr)|0\rangle .\label{xi1 3redm}
\end{eqnarray}
The simplified representation (\ref{L3/23/2red}) for ${\cal{}S}^m_{(1,1)}$
in terms of the massive objects holds true, with the following changes:
\begin{eqnarray}
 \label{L3/23/2redm}\hspace{-0.5em} &\hspace{-0.5em}&   \hspace{-1em} \langle
\tilde{\Psi}_8|K_{(1,1)}T_1|\chi_4\rangle  +
 \langle \tilde{\rho}_{2}|K_{(1,1)} T_0|\rho_3\rangle +  \langle \tilde{\rho}_{3}|K_{(1,1)}
T_1|\chi_3\rangle +\textstyle\frac{1}{2}\sum_{i=3}^4\langle \tilde{\chi}_{i}|K_{(1,1)}
T_0|\chi_i\rangle +c.c.
\\
\hspace{-0.5em}&\hspace{-0.5em}& \hspace{-1em} \to \  \langle\tilde{\Psi}_{8|m}|K_{(1,1)}T^*_1|\chi_{4|m}\rangle  +
 \langle \tilde{\rho}_{2|m}|K_{(1,1)} T^*_0|\rho_{3|m}|\rangle +  \langle \tilde{\rho}_{3|m}|K_{(1,1)}
T^*_1|\chi_{3|m}\rangle \nonumber \\
&\hspace{-0.5em}&  \hspace{-1em} \quad +\textstyle\frac{1}{2}\sum_{i=3}^4\langle \tilde{\chi}_{i|m}|K_{(1,1)}
T_0^*|\chi_{i|m}\rangle +c.c. \nonumber
\end{eqnarray}
In addition to removing the symmetric part $\psi_{(\mu,\nu)}$ from the basic
spin-tensor $\psi_{\mu\nu}$, and the fields $ \psi^k_{\mu}, k=2,3, \psi_l^l, l=1,2,$
from $|\Psi_m\rangle_{(1,1)}$ (\ref{chi320m}), we also gauge away the new fields
$\psi^k_{m}(x)$, $k=2,4,5$, and $\psi^l_{m|\mu}(x)$, $l=2,3$, except for
$\psi^1_{m}(x)$, $\psi^3_{m}(x), \psi^1_{m|\mu}(x)$, by using the respective
gauge parameters $\xi^{1}_{3|m}(x)$, $\xi^{1}_{1|m}(x)$, $\xi^{12}_{m}(x)$,
$\xi^{12}_{m|\mu}(x) $ and the ``massless'' parameters $\xi^{13}_{\mu}(x), \xi^{1}_{3}(x) $,
from the gauge parameters $|\xi^r_{12|m}\rangle_{(2,0)}$ and
$|\xi^r_{k|m}\rangle_{(0,1)}$, k=1,3, in the gauge transformations
(\ref{ndpsiredm})--(\ref{ndchi34redm}). The following restrictions hold true:
   \begin{align}
 \hspace{-0.3em} &\hspace{-0.3em} \big(\xi^{1}_{3|m},\,\xi^{1}_{1|m},\, \xi^{12}_{m} \big) \hspace{1ex}  =   \hspace{1ex}
 \big( -1,\, m,\,-2m\big)\xi^{1}_{3},  && \delta\psi^1_{m|\mu}(x)=0 , \ \delta\psi^k_{m}(x)=0, \ k=1,3, \label{chim3232res}\\
  \hspace{-0.3em} &\hspace{-0.3em} \big(\xi^{12}_{m|\mu},\,\xi^{13}_{\mu} \big)\ = \ \big(2\big[\imath\partial_\mu-m \gamma_\mu \big],\, \gamma_{\mu} \big)\xi^{1}_{3} \quad \Rightarrow  && \delta\psi_{[\mu\nu]}\hspace{1ex} = 2\imath \partial_{[\mu}\xi^{13}_{\nu]}+2\imath \gamma_{[\mu}\partial_{\nu]}\xi^{
    1}_{3}\vert_{(\xi^{13}_{\mu}=\gamma_{\mu} \xi^{1}_{3})} =0.
  \label{chim3232res1}
  \end{align}
From the gauge transformations for the spinors $\Psi_{1|m}$, $\Psi_{2|m}$ (\ref{ndpsi12red}),
(\ref{ndpsi8red}),
 \begin{equation}\label{fingpsi12}
   \delta\big(\Psi_{1|m},\,|\Psi_{2|m}\big)  \hspace{1ex}  =   \hspace{1ex}2\big(\imath \gamma^{\mu}\partial_{\mu}-2m,\, \imath \gamma^{\mu}\partial_{\mu}-4m\big)\xi^{
    1}_{3},
 \end{equation}
it follows that the difference $\big(\Psi_{1|m}-|\Psi_{2|m}\big)$ obeys a Stueckelberg-type
gauge symmetry, so that in the new basis of the fields
$\widehat{\Psi}_{1|m},\,\widehat{\Psi}_{2|m}$ we have
  \begin{equation}\label{fingpsi12n}
     \big( \widehat{\Psi}_{1|m},\,\widehat{\Psi}_{2|m}\big)  \hspace{0.5ex}  =   \hspace{0.5ex} \frac{1}{2} \big(\Psi_{1|m}\pm \Psi_{2|m}\big):\ \ \delta\big( \widehat{\Psi}_{1|m},\,\widehat{\Psi}_{2|m}\big)  \hspace{0.5ex}  =   \hspace{0.5ex} 2\big(\imath\gamma^{\mu}\partial_{\mu}-3m,\, m\big)\xi^{    1}_{3}.
 \end{equation}
The field $\widehat{\Psi}_{2|m}$ may be gauged away by using the last gauge parameter
$\xi^{1}_{3}$, resulting in a non-gauge theory. Once again, the resolution of the ``massive''
analog of equations of motion (\ref{auxEQ1})--(\ref{auxEQf}) leads to the same solutions,
augmented by $\psi^1_{m|\mu}(x)= \psi^k_{m}(x)=0$, for $k=1,3$, in the first equation,
so that it is only the initial massive field ${\psi}^{[\mu\nu]}$ that survives
in $|\Psi_{m}\rangle$, and also augmented by the solutions $\chi^{13}_{m}(x)=0$,
$\chi^{23}_{m}(x)=-\chi^{2}_3(x)=-\psi_3(x)$ in the last two (adapted) equations
(\ref{auxEQf-1}), (\ref{auxEQf}), with allowance for (\ref{L3/23/2redm}).
In addition, $\widehat{\Psi}_{2|m} =0$ implies that $\chi_8=\chi_{10}$ =
$\psi_3 = \chi^2_3 =\chi^{23}_{m}=0$. Substituting the expressions (\ref{chi320m})--(\ref{Psikm})
into (\ref{L3/23/2red}), albeit for ${\cal{}S}^m_{(1,1)}$, we find an explicit action
for a massive field of spin $(3/2,3/2)$, with the remaining auxiliary fields:
\begin{eqnarray}
{\cal{}S}^m_{(1,1)} &=& \int d^d x \Biggl[\bar{\psi}_{[\mu\nu]}
\Bigl\{-\frac{1}{2}\big({\imath}\gamma^{\rho}\partial_{\rho}-m\big)
{\psi}^{[\mu\nu]}  -
\imath\partial^{[\mu}\Bigl[з
\gamma_\rho{\psi}^{[\nu]\rho]} -
\chi^{4{}\nu]}\Bigr] \Bigr\}
 \label{expL3/23/2m} \\
&& + \bar{\widehat{\psi}}_1\Bigl\{-\big({\imath}\gamma^{\rho}\partial_{\rho}-m\big)
{\widehat{\psi}}_1-2{\psi}_8 -\imath \partial^\mu\Bigl[
\gamma^\rho{\psi}_{[\mu\rho]} -{\chi}_\mu^4\Bigr]  +m{\chi}_m^4 \Bigr\}
\nonumber \\
&&
+ \bar{\psi}_8\Bigl\{\gamma^\mu\chi^{4}_\mu +(1-d)\chi_4 - m \chi^4_m \Bigr\} - \bar{\varphi}^2_\mu \Bigl(\gamma_\rho{\psi}^{[\mu\rho]}+
\chi^{4{}\mu} \Bigr) \nonumber \\
&&  - (1-d)\bar{\varphi}^2\chi^{4}+\bar{\varphi}^2_m\chi^{4}_m
 -\frac{1}{2}\bar{\psi}^{[\mu\tau]}\gamma_\tau\big(\imath\gamma^{\sigma}
\partial_{\sigma} + m\big)\gamma^\rho{\psi}_{[\mu\rho]} \nonumber
 \\
\phantom{{\cal{}S}_{(1,1)}} &&-\frac{1}{2}\bar{\chi}^4_\mu
\big(\imath\gamma^{\nu}\partial_{\nu}+ m\big){\chi}^{4{}\mu}
 +\frac{1}{2}(1-d)
\bar{\chi}^4 \big(\imath\gamma^{\nu}\partial_{\nu}- m\big){\chi}^4  +\frac{1}{2}
\bar{\chi}^4_m \big(\imath\gamma^{\nu}\partial_{\nu}+m\big){\chi}^4_m \nonumber
 \Biggr] + c.c.
\end{eqnarray}
From the extremals for the fields $ \varphi^2_\mu,
\psi_8$, $\varphi^2, \varphi^2_m$, we find the respective solutions
$\chi^{4{}\mu} = \gamma_\rho{\psi}^{[\rho\mu]}$,  ${\widehat{\psi}}_1 =
\frac{1}{2}\gamma^{\mu\rho}{\psi}_{[\rho\mu]}$,
${\chi}^4={\chi}^4_m =0$, which allow one to present the
Lagrangian for a massive antisymmetric spin-tensor $(3/2,3/2)$
field in a flat $d$-dimensional space-time:
\begin{eqnarray} {\cal{}L}^m_{\psi_{[\mu\nu]}} &=&
\bar{\psi}_{[\mu\nu]}
\Bigl\{-({\imath}\gamma^{\rho}\partial_{\rho}-m) {\psi}^{[\mu\nu]}
+ {\imath}\partial^{[\mu}\gamma^{\nu]}
\gamma^{\rho\sigma}{\psi}_{[\rho\sigma]} +
2\imath\partial^{[\nu}\gamma_\rho{\psi}^{[\mu]\rho]}+
2\imath\gamma^{\nu}\partial_\rho{\psi}^{[\mu\rho]}
 \label{mexpL3/23/2fint} \nonumber\\
&& \hspace{-1em}-\frac{1}{2}\gamma^{\nu\mu}
 (\imath\gamma^{\sigma}
\partial_{\sigma}-m)\gamma_{\rho\tau}{\psi}^{[\rho\tau]}
+{\imath}\gamma^{\nu\mu}
\partial_{[\rho}\gamma_{\tau]}{\psi}^{[\rho\tau]}
-2{}\gamma^\mu
({\imath}\gamma^{\rho}\partial_{\rho}+m)\gamma_\sigma{\psi}^{[\sigma\nu]}
 \Bigr\}\,.
\end{eqnarray}
Note that one can obtain ${\cal{}L}^m_{\psi_{[\mu\nu]}}$ by using the dimensional
reduction procedure (\ref{reduction})--(\ref{Eq-omv}), starting directly from
the action (\ref{expL3/23/2fint}) presented for a $(d+1)$-dimensional Minkowski space
(for simplicity, we consider an odd $d+1$, whereas for an even $d+1$ one should
work with doubled spin-tensors in order to have $\tilde{\gamma}$, as mentioned after
(\ref{Eq-omv})). First, we have the following representation for the fields:
${\psi}^{[MN]} = ({\psi}^{[\mu\nu]},{\psi}^{[\mu d]})$, with the Stueckelberg field
${\psi}^{[\mu d]} = -{\psi}^{\mu}{}_d \equiv {\varphi}^{\mu}$ and with the gauge
parameters $(\xi_N; \eta) = (\xi_\nu, \xi; \eta)$. Second, the corresponding action
can be obtained from (\ref{expL3/23/2fint}) by using the dimensional projection
$\mathbb{R}^{1,d} \to \mathbb{R}^{1,d-1}$, and so it must be invariant with respect
to the gauge transformations
\begin{align}
    & \delta\psi_{[\mu\nu]}\hspace{1ex} = \hspace{1ex}2\imath
\partial_{[\mu}\xi_{\nu]}+2\imath
\gamma_{[\mu}\partial_{\nu]}\eta\,,
&&\delta{\varphi}_{\mu}\hspace{1ex} = \hspace{1ex}-\imath
\partial_\mu\xi - m\xi_\mu +m \gamma_\mu\eta +\imath
\partial_\mu \eta
    \,,
    \label{psi[munu]red1}
\end{align}
which, in turn, are reducible:
\begin{align}
    & \delta\xi_\mu(x)\hspace{1ex} = \hspace{1ex}\imath \partial_{\mu}\xi^{(1)}(x)
    \,,&&\delta\xi(x)\hspace{1ex} =
\hspace{1ex} - m\xi^{(1)}(x)
    \,.
    \label{ximuxi1red1}
\end{align}
Third, due to (\ref{reduction2}), we use the identity $\gamma^{RS}\psi_{[RS]}
= \gamma^{\rho\sigma}\psi_{[\rho\sigma]}$, with the following identification
implied by (\ref{new_property}), (\ref{new_property1}):
\begin{equation}\label{reduc3/23/2}
i\gamma^M\partial_M{\psi}^{[NK]} =
(i\gamma^\mu\partial_\mu-m){\psi}^{[NK]},\qquad
i\gamma^M\partial_M\gamma_N{\psi}^{[NK]} =
(i\gamma^\mu\partial_\mu+m)\gamma_N{\psi}^{[NK]},
\end{equation}
which is valid if one replaces the quantities ${\psi}^{[NK]}$ [$\gamma_N{\psi}^{[NK]}$]
by $(\gamma_L)^{2k}{\psi}^{[NK]}$ [$(\gamma_L)^{2k+1}{\psi}^{[NK]}$],
for $k \in \mathbb{N}_0$.
After removing the gauge parameter $\xi(x)$ by using the shift transformation,
and then (in the same manner) removing the field ${\varphi}_{\mu}$, by using
the already independent gauge transformation with the parameter $\xi_\mu(x)$,
we finally obtain the Lagrangian (\ref{mexpL3/23/2fint}).

\section{Conclusion}\label{summary}

In the present work, we have constructed a gauge-invariant
Lagrangian description of free half-integer  HS fields belonging
to an irreducible representation of the Poincare group $ISO(1,d-1)$
with the corresponding Young tableaux having two rows in the
``metric-like" formulation. The results of this study are
the most general ones and apply to both massive and massless
fermionic HS fields with a mixed symmetry in a Minkowski  space
of any dimension.

In the standard manner, starting from an embedding of fermionic
HS fields into vectors of an auxiliary Fock space, we treat the
fields as coordinates of Fock-space vectors and reformulate
the theory in such terms. We realize the conditions that determine
an irreducible Poincare-group representation with a given mass
and generalized spin in terms of differential operator constraints
imposed on the Fock space vectors. These  constraints generate
a closed Lie superalgebra of HS symmetry, which contains, with
the exception of two basis generators of its Cartan subalgebra,
a system of first- and second-class constraints.

We demonstrate  that the construction  of a correct Lagrangian
description requires a deformation of the initial symmetry
superalgebra, in order to obtain from the system of mixed-class
constraints a converted system with the same number of first-class
constraints alone, whose structure provides the appearance of the
necessary number of auxiliary spin-tensor fields with lower
generalized spins. We have shown that this purpose can be achieved
with the help of an additional Fock space, by constructing an additive
extension of a symmetry subsuperalgebra which consists of the subsystem
of second-class constraints alone and of the generators of the Cartan
subalgebra, which form an invertible even operator supermatrix
composed of supercommutators of the second-class constraints.

We have realized the Verma module construction \cite{Dixmier}
in order to obtain an auxiliary representation in Fock space
for the above superalgebra with second-class constraints. As a
consequence, the converted Lie superalgebra of HS symmetry has the
same algebraic relations as the initial superalgebra; however,
these relations are realized in an enlarged Fock space. The generators of the
converted Cartan subalgebra contain linearly two auxiliary
independent number parameters, whose choice provides the vanishing
of these generators in the corresponding subspaces of the total
Hilbert space extended by the ghost operators in accordance with
the minimal BFV--BRST construction for the converted HS symmetry
superalgebra. Therefore, the above generators, enlarged by the
ghost contributions up to the ``particle number'' operators  in
the total Hilbert space, covariantly determine Hilbert subspaces
in each of which the converted  symmetry superalgebra consists of
 the first-class constraints alone, labeled by the values
of the above parameters, and constructed from the initial irreducible
Poincare-group relations.

It is shown that the Lagrangian
description corresponding to the BRST operator, which encodes the
converted HS symmetry superalgebra, yields a consistent Lagrangian
dynamics for fermionic fields of any generalized spin.  The
resulting Lagrangian description, realized concisely in
terms of the total Fock space, presents a set of generating
relations for the action and the sequence of gauge transformations
for given fermionic HS fields with a sufficient set of auxiliary
fields, and  proves to be a reducible gauge theory with a finite
number of reducibility stages, increasing  with the value of
generalized spin. We elaborate a dimensional reduction procedure
used to obtain a gauge-invariant Lagrangian description for a massive
fermionic HS field in a $d$-dimensional Minkowski space $R^{1,d}$
starting from a Lagrangian description for a massless fermionic
HS field of the same generalized spin, albeit in $R^{1,d-1}$.

We have outlined a proof of the fact that the solutions of
the Lagrangian equations of motion (\ref{EofM1}), (\ref{EofM2}), after
a partial gauge-fixing,
 correspond to the BRST cohomology space with a vanishing ghost number,
 which is determined only by the relations that extract the fields of
 an irreducible Poincare-group representation with a given value of
 generalized spin.

As examples  demonstrating the applicability of the general
scheme, we have derived  gauge-invariant Lagrangian formulations
for the field of spin $(3/2,1/2)$ and for the rank-2 antisymmetric
spin-tensor   in a manifest form in both massless and massive
cases\footnote{Lagrangian formulations for the case of
antisymmetric spin-tensors of arbitrary rank $n$ in
$\mathbb{R}^{1,d-1}$ and AdS${}_d$-spaces, for $n \leq
\left[\frac{d}{2}\right]$, was considered recently within BRST and
algebraic approach respectively, in Refs. \cite{0902BKRT,
0903Zinoviev}, so that the Lagrangian formulations both for the
massive and massless fields of spin $(3/2,3/2)$ coincide  with
ones in~\cite{0902BKRT, 0903Zinoviev}.}. In principle, the
suggested algorithm permits one to derive manifest actions for any
other half-integer spin fields characterized by two rows of the
corresponding Young tableaux.

The basic results of the present work are given by relations
(\ref{L1}), where the action for a field with an arbitrary
generalized half-integer spin is constructed, as well as by
relations (\ref{GT1})--(\ref{GTi2}), where the gauge transformations
for the fields are presented, along with the sequence of reducible
gauge transformations and gauge parameters.

Concluding, we would like to discuss a number of additional points.
First, the gauge-invariant description of massless and massive HS
field theories with a mixed symmetry is an interesting
starting point for a systematic construction of a
Lagrangian formulation for HS interacting vertices with
mixed-symmetry fermionic HS fields, including the case of the AdS space,
in order to
provide a description of the high-energy limit for open superstrings;
see the arguments in favor of this suggestion in
\cite{Tsulaia}. Second, the role of fermionic HS
fields in the above limit of superstring theory in connection
with the AdS/CFT correspondence signals the importance of extending
the obtained results to the case of fermionic HS fields with
a mixed symmetry in the AdS space. Thus, the present Lagrangian
description takes a first step towards an interacting theory
with mixed-symmetry fermionic HS fields, including the case of  curved
backgrounds, and then towards a covariant construction (following,
e.g., the BV formalism) of the
generating functionals of Green's functions, including the quantum
effective action; examples of such calculations can be found, e.g.,
in \cite{DuffBuchbinder}. Third, we estimate an extension of
the obtained results to the case of arbitrary fermionic HS fields  with
any number of rows in the corresponding Young tableaux as
a challenging technical problem. One of the possible approaches
to this problem may rely on creating a computer
algorithm which would permit one to obtain the HS symmetry
superalgebra and calculate the action with the sequence of gauge
transformations in an analytic component form for fermionic fields
of a given generalized spin. Finally, the example of a field
of spin $(3/2,3/2)$ in Section \ref{Example3/23/2}
has demonstrated a possibility of extracting a large number of
auxiliary fields until the point when the component form of
the action and gauge transformations can be derived in a manifest
form. In our forthcoming
work \cite{ReshMosh}, we plan to realize this possibility,
which should permit one to significantly reduce the spectrum of
fields and gauge parameters in order to simplify the component
structure of the basic results of the present work,
however, with a possible appearance of additional off-shell
constraints for the fields and gauge parameters.

\section*{Acknowledgments}
The authors are grateful to D.M. Gitman, P.M. Lavrov and Yu.M.
Zinoviev for useful discussions, as well as to K.V. Stepanyantz
for advice and discussions on dimensional reduction as applied
to fermionic fields. A.A.R. thanks M.A. Vasiliev and R.R. Metsaev
for discussions on the possibility of extracting algebraic constraints
from the entire set of constraints in order to reduce the spectrum
of auxiliary fields. P.Yu.M. is grateful to CNPq for support.


\begin{thebibliography}{9}

\bibitem{flatin}M. Fierz, W. Pauli, On relativistic wave equations for particles
of arbitrary spin in an electromagnetic field, Proc. R. Soc. London,
Ser. A, 173 (1939) 211--232.

\bibitem{Singh}L.P.S. Singh, C.R. Hagen, Lagrangian
formulation for arbitrary spin. 1. The bosonic case, Phys. Rev. D9
(1974) 898--909; Lagrangian formulation for arbitrary spin. 2. The
fermionic case, Phys. Rev. D9 (1974) 910--920.

\bibitem{Fronsdal}C. Fronsdal, Massless fields with integer spin,
 Phys. Rev. D18 (1978) 3624--3629; J. Fang, C.
Fronsdal, Massless fields with half integral spin, Phys. Rev. D18
(1978) 3630--3633;

\bibitem{Curtright}T. Curtright, Massless field supermultiplets with arbitrary
spin, Phys. Lett. B85 (1979) 219--224; Generalized gauge fields
 Phys. Lett. B165 (1985) 304--308.

\bibitem{Ouvry}C.S. Aulakh, I.G. Koh, S. Ouvry, Higher spin fields with
mixed symmetry, Phys. Lett. B173 (1986) 284--288; S. Ouvry, J.
Stern, Gauge fields of any spin and symmetry, Phys. Lett. B177
(1986) 335--340; A.K.H. Bengtsson, A unified action for higher
spin gauge bosons from covariant string theory, Phys. Lett. B182
(1986) 321--325.

\bibitem{Siegel}W. Siegel, B. Zwiebach, Interacting BRST from the light cone,
Nucl. Phys. B282 (1987) 125--230; W. Siegel, Introduction to
string field theory, [arXiv:hep-th/0107094].

\bibitem{Labastida}J.M.F. Labastida, T.R. Morris, Massless mixed symmetry
bosonic free fields, Phys. Lett. B180 (1986) 101--106; J.M.F.
Labastida, Massless fermionic free fields, Phys. Lett. B186 (1987)
365--369; Massless bosonic free fields,
 Phys. Rev. Lett. 58 (1987) 531--534; Massless
particles in arbitrary representations of the Lorentz group, Nucl.
Phys. B322 (1989) 185--209.

\bibitem{Pashnev1}A. Pashnev,  M.M. Tsulaia, Description of the higher massless
irreducible integer spins in the BRST approach, Mod. Phys. Lett.
A13 (1998) 1853--1864, [arXiv:hep-th/9803207].

\bibitem{BurdikPashnev}C. Burdik, A.
Pashnev, M. Tsulaia, On the mixed symmetry irreducible
representations of the Poincare group in the BRST approach, Mod.
Phys. Lett. A16 (2001) 731--746, [arXiv:hep-th/0101201]; The
Lagrangian description of representations of the Poincare group,
Nucl. Phys. Proc. Suppl. 102 (2001) 285--292, [arXiv:hep-th/0103143].

\bibitem{Francia}D. Francia, A. Sagnotti, Free geometric equations for higher
spins,  Phys. Lett. B543 (2002) 303--310, [arXiv:hep-th/0207002];
D. Francia, J. Mourad, A. Sagnotti, Current exchanges and unconstrained higher
spins,  [arXiv:hep-th/0701163].

\bibitem{Bekaert}X. Bekaert, N. Boulanger, Tensor gauge fields in arbitrary
representations of GL(D,R): duality and Poincare lemma, Commun.
Math. Phys. 245 (2004) 27--67, [arXiv:hep-th/0208058]; On
geometric equations and duality for free higher spins, Phys. Lett.
B561 (2003) 183--190, [arXiv:hep-th/0301243]; Mixed symmetry gauge
fields in a flat background, [arXiv:hep-th/0310209]; Tensor gauge
fields in arbitrary representations of GL(D,R). II. Quadratic
actions, Commun. Math. Phys. 271 (2007) 723--773,
[arXiv:hep-th/0606198]; X. Bekaert, N. Boulanger, S. Cnockaert, No
self-interaction for two-column massless fields, J. Math. Phys. 46
(2005) 012303, [arXiv:hep-th/0407102].

\bibitem{Medeiros}P. de Medeiros,  C. Hull, Geometric second order field
equations for general tensor gauge fields, JHEP 0305 (2003) 019,
[arXiv:hep-th/0303036].

\bibitem{Zinoviev}Yu.M. Zinoviev, Massive N=1 supermultiplets with arbitrary
superspins, [arXiv:0704.1535]; Massive supermultiplets with spin
3/2, [arXiv:hep-th/0703118].

\bibitem{BoulangerKirsch}N. Boulanger, I. Kirsch,
A Higgs mechanism for gravity. Part II: higher spin connections
Phys. Rev. D73 (2006) 124023, [arXiv:hep-th/0602225].

\bibitem{Bonelli2}G. Barnich, G. Bonelli, M. Grigoriev,
From BRST to light-cone description of higher spin gauge fields,
Proceedings of the Workshop ``Quantum Field Theory and Hamiltonian Systems'',
Caciulata, Romania, 16--21 Oct, 2004, [arXiv:hep-th/0502232].

\bibitem{AdSin}C. Fronsdal,
Singletons and massless, integer-spin fileds on de Sitter space,
Phys. Rev. D20 (1979) 848-856; J. Fang, C. Fronsdal, Massless,
half-integer-spin fields in de Sitter space, Phys. Rev. D22 (1980)
1361--1367.

\bibitem{Vasiliev ads}M.A. Vasiliev, Free massless fermionic fields of arbitrary spin
in D-dimensional anti-de~Sitter space, Nucl. Phys. B301 (1988)
26--51; V.E. Lopatin, M.A. Vasiliev, Free massless bosonic fields
of arbitrary spin in D-dimensional de~Sitter space, Mod. Phys.
Lett. A3 (1998) 257--265.

\bibitem{Vasiliev_inter}M.A. Vasiliev, Cubic
interactions of bosonic higher spin gauge fields in AdS(5), Nucl.
Phys. B616 (2001) 106--162 [Erratum-ibid. B 652 (2003) 407],
[arXiv:hep-th/0106200]; Algebraic aspects of the higher spin
problem, Phys. Lett. B257 (1991) 111--118; Nonlinear equations
for symmetric massless higher spin fields in (A)dS(d), Phys. Lett.
B 567 (2003) 139--151, [arXiv:hep-th/0304049].

\bibitem{Metsaev-0}R.R.Metsaev, Free totally (anti)symmetric massless fermionic
fields in d-dimensional anti-de Sitter space, Class. Quant. Grav. 14
(1997) L115--L121, [arXiv:hep-th/9707066]; Fermionic fields in the
d-dimensional anti-de Sitter spacetime, Phys. Lett. B419 (1998)
49--56, [arXiv:hep-th/9802097]; Arbitrary spin massless bosonic
fields in d-dimensional anti-de~Sitter space,
[arXiv:hep-th/9810231]; Light-cone form of field dynamics in
anti-de Sitter space-time and AdS/CFT correspondence, Nucl. Phys.
B563 (1999) 295--348, [arXiv:hep-th/9906217]; Massless arbitrary
spin fields in AdS(5) Phys. Lett. B531 (2002) 152--160,
[arXiv:hep-th/0201226]; Massive totally symmetric fields in
AdS(d), Phys. Lett. B590 (2004) 95--104, [arXiv:hep-th/0312297];
R.R. Metsaev, Mixed-symmetry massive fields in AdS(5),
Class. Quant. Grav. 22 (2005) 2777--2796, [arXiv:hep-th/0412311];
Cubic interaction vertices of massive and massless higher spin
fields, Nucl. Phys. B759  (2006) 147--201, [arXiv:hep-th/0512342];
R.R. Metsaev, Gauge invariant formulation of massive totally
symmetric fermionic fileds in (A)dS space, Phys. Lett. B643 (2006)
205--212, [arXiv:hep-th/0609029].

\bibitem{Mets-amb}R.R. Metsaev, Massless mixed symmetry bosonic free fields
in d-dimensional anti-de Sitter space-time, Phys. Lett. B354
(1995) 78--84; Fermionic fields in the d-dimensional anti-de
Sitter spacetime, Phys. Lett. B419 (1998) 49--56,
[arXiv:hep-th/9802097].

\bibitem{Deser}S. Deser, A. Waldron, Gauge invariances and phases of massive
higher spins in (A)dS, Phys. Rev. Lett. 87 (2001) 031601,
[arXiv:hep-th/0102166];  S. Deser, A. Waldron, Partial
Masslessness of Higher Spins in (A)dS, Nucl. Phys. B 607 (2001)
577--604, [arXiv:hep-th/0103198]; Null propagation of partially
massless higher spins in (A)dS and cosmological constant
speculations, Phys. Lett. B513 (2001) 137--141,
[arXiv:hep-th/0105181]; K. Hallowell, A. Waldron, Constant
curvature algebras and higher spin action generating functions,
Nucl. Phys. B724  (2005) 453--486, [arXiv:hep-th/0505255]; E.D.
Skvortsov,  M.A. Vasiliev, Geometric formulation for partially
massless fields, Nucl. Phys. B756  (2006) 117--147,
[arXiv:hep-th/0601095].

\bibitem{Zinoviev_m}Y.M. Zinoviev, On massive mixed symmetry tensor
fields in Minkowski space and (A)dS, [arXiv:hep-th/0211233]; First
order formalism for mixed symmetry tensor fields,
[arXiv:hep-th/0304067]; First order formalism for massive mixed
symmetry tensor fields in Minkowski and (A)dS spaces,
[arXiv:hep-th/0306292].

\bibitem{Sezgin}E. Sezgin,  P. Sundell, Doubletons and 5-D higher spin
gauge theory, JHEP 0109 (2001) 036, [arXiv:hep-th/0105001]; Towards
massless higher spin extension of D=5, N=8 gauged supergravity,
JHEP 0109 (2001) 025, [arXiv:hep-th/0107186]; Holography in 4D
(super) higher spin theories and a test via cubic scalar
couplings, JHEP 0507 (2005) 044, [arXiv:hep-th/0305040].

\bibitem{Sagnotti}A. Sagnotti, M.  Tsulaia, On higher spins
and the tensionless limit of string theory, Nucl. Phys.  B682
(2004) 83--116, [arXiv:hep-th/0311257].

\bibitem{Fotopoulos}A. Fotopoulos, K.L. Panigrahi, M. Tsulaia,
Lagrangian formulation of higher spin theories on AdS, Phys. Rev.
D74 (2006) 085029, [arXiv:hep-th/0607248]; I.L. Buchbinder, A.V.
Galajinsky, V.A. Krykhtin, Quartet unconstrained formulation for
massless higher spin fields, [arXiv: hep-th/0702161].

\bibitem{Barnich}G. Barnich, M.  Grigoriev, A.  Semikhatov, I.  Tipunin, Parent
field theory and unfolding in BRST first-quantized terms, Commun.
Math. Phys. 260 (2005) 147--181, [arXiv:hep-th/0406192]; G.
Barnich, M. Grigoriev, Parent form for higher spin fields on
anti-de Sitter space, [arXiv:hep-th/0602166]; M. Grigoriev,
off-shell gauge fields from BRST quantization,
[arXiv:hep-th/0605089].

\bibitem{AlkalaevVasiliev}
K.B. Alkalaev,
Free fermionic higher spin fields in AdS(5), Phys. Lett. B519
(2001) 121--128, [arXiv: hep-th/0107040];
K.B. Alkalaev, M.A. Vasiliev,
N=1 supersymmetric theory of higher spin gauge fields in AdS(5)
at the cubic level, Nucl. Phys. B655 (2003) 57--92, [arXiv:hep-th/0206068];
K.B. Alkalaev,
Two column higher spin massless fields in AdS(d), Theor. Math. Phys.
140 (2004) 1253--1263, [arXiv:hep-th/0311212];
K.B. Alkalaev, O.V. Shaynkman, M.A. Vasiliev,
On the frame-like formulation of mixed-symmetry massless fields in (A)dS(d),
Nucl. Phys. B692 (2004) 363--393, [arXiv:hep-th/0311164];  Lagrangian formulation
for free mixed-symmetry bosonic gauge fields in (A)dS(d), JHEP
0508 (2005) 069, [arXiv:hep-th/0501108]; Frame-like formulation for free
mixed-symmetry bosonic massless higher-spin fields in AdS(d),
[arXiv:hep-th/0601225];
K.B. Alkalaev,
Mixed-symmetry massless gauge fields
in AdS(5), Theor. Math. Phys. 149 (2006) 1338--1348.

\bibitem{reviews}M. Vasiliev, Higher spin gauge theories in various dimensions,
Fortsch. Phys. 52 (2004) 702--717, [arXiv:hep-th/0401177]; D.
Sorokin, Introduction to the classical theory of higher spins, AIP
Conf. Proc. 767 (2005) 172--202, [arXiv:hep-th/0405069]; N.
Bouatta, G. Comp\`ere,  A. Sagnotti, An introduction to free
higher-spin fields, [arXiv:hep-th/0409068]; A. Sagnotti, E.
Sezgin, P. Sundell, On higher spins with a strong Sp(2,R) sondition,
[arXiv:hep-th/0501156]; X. Bekaert, S. Cnockaert, C.
Iazeolla, M.A. Vasiliev, Nonlinear higher spin theories in various
dimensions, [arXiv:hep-th/0503128].

\bibitem{Heslop}N. Beisert, M. Bianchi, J.F. Morales, H. Samtleben, Higher spin
symmetries and N=4 SYM, JHEP 0407 (2004) 058,
[arXiv:hep-th/0405057]; A.C. Petkou, Holography, duality and
higher spin fields, [arXiv:hep-th/0410116]; M. Bianchi, P.J.
Heslop, F. Riccioni, More on la Grande Bouffe: towards higher spin
symmetry breaking in AdS, JHEP 0508 (2005) 088,
[arXiv:hep-th/0504156]; P.J. Heslop, F. Riccioni, On the fermionic
Grande Bouffe: more on higher spin symmetry breaking in AdS/CFT,
JHEP 0510 (2005) 060, [arXiv:hep-th/0508086]; M. Bianchi, V.
Didenko, Massive higher spin multiplets and holography,
[arXiv:hep-th/0502220].

\bibitem{Bonelli1}G. Bonelli,
On the boundary gauge dual of closed tensionless free strings in AdS,
JHEP 0411 (2004) 059, [arXiv:hep-th/0407144];
On the covariant quantization of tensionless bosonic strings in AdS spacetime,
JHEP 0311 (2003) 028, [arXiv:hep-th/0309222];
On the tensionless limit of bosonic strings, infinite symmetries and higher spins,
Nucl. Phys. B669 (2003) 159--172, [arXiv:hep-th/0305155]

\bibitem{Maldacena}J.M. Maldacena, The large N limit of
superconformal field theories and supergravity,
Adv. Theor. Math. Phys. 2 (1998) 231, [arXiv:hep-th/9711200].

\bibitem{Hull}C. Hull, Duality in gravity and higher spin gauge fields,
 JHEP 0109 (2001) 027, [arXiv:hep-th/0107149];
P. de Medeiros, C. Hull, Exotic tensor gauge theory and duality,
Commun. Math. Phys. 235 (2003) 255--273, [arXiv:hep-th/0208155];
H. Casini, R. Montemayor, L.F. Urrutia, Duality for symmetric
second rank tensors. 2. The Linearized gravitational field, Phys.
Rev. D 68 (2003) 065011, [arXiv:hep-th/0304228]; N. Boulanger, S.
Cnockaert, M. Henneaux, A note on spin s duality, JHEP 0306 (2003)
060, [arXiv:hep-th/0306023].

\bibitem{BFV}E.S. Fradkin,  G.A. Vilkovisky, Quantization of relativistic
systems with constraints, Phys. Lett. B55 (1975) 224--226; I.A.
Batalin, G.A. Vilkovisky, Relativistic S-matrix of dynamical
systems with boson and fermion constraints, Phys. Lett. B69 (1977)
309--312; I.A. Batalin, E.S. Fradkin, Operator quantization of
relativistic dynamical systems subject to first class constraints,
Phys. Lett. B128 (1983) 303.

\bibitem{bf}I.A. Batalin, E.S. Fradkin, Operator quantization method and
abelization of dynamical systems subject to first class
constraints, Riv. Nuovo Cimento, 9, No.~10 (1986) 1; I.A. Batalin,
E.S. Fradkin, Operator quantization of dynamical systems subject
to constraints.  A further study of the construction, Ann. Inst.
H. Poincare, A49 (1988) 145--214.

\bibitem{Henneaux}M. Henneaux, Hamiltonian form of the path integral for theories
with a gauge freedom, Phys. Rept. 126 (1985) 1--66; M. Henneaux,
C. Teitelboim, Quantization of gauge systems, Princeton Univ.
Press, 1992.

\bibitem{conversion}L.D. Faddeev, S.L. Shatashvili, Realization of
the Schwinger term in the Gauss low and the possibility of correct
quantization of a theory with anomalies, Phys.Lett. B167 (1986)
225--238; I.A. Batalin, E.S. Fradkin, T.E. Fradkina, Another
version for operatorial quantization of dynamical systems with
iireducible constraints, Nucl. Phys. B314 (1989) 158--174; I.A.
Batalin, I.V. Tyutin, Existence theorerm for the effective gauge
algebra in the generalized canonical formalism and Abelian
conversion of second class constraints, Int. J. Mod. Phys. A6
(1991) 3255--3282; E. Egorian, R. Manvelyan, Quantization of
dynamical systems with first and second class constraints, Theor.
Math. Phys. 94 (1993) 241--252.

\bibitem{BV-BFV}M. Grigoriev,  P.H. Damgaard,
Superfield BRST charge and the master action, Phys. Lett. B474
(2000) 323--330, [arXiv:hep-th/9911092]; G. Barnich,  M.
Grigoriev, Hamiltonian BRST and Batalin-Vilkovisky formalisms for
second quantization of gauge theories, Commun. Math. Phys. 254
(2005) 581Ц-601, [arXiv:hep-th/0310083];
 D.M. Gitman, P.Yu. Moshin, A.A. Reshetnyak, Local
superfield Lagrangian BRST quantization, J. Math. Phys. {46}
(2005) 072302-01--072302-24, [arXiv:hep-th/0507160]; An embedding
of the BV quantization into an N=1 local superfield formalism,
Phys. Lett. B 621 (2005) 295--308, [arXiv:hep-th/0507049].

\bibitem{Lyakh-Shar}S.L. Lyakhovich, A.A Sharapov, BRST theory without
Hamiltonian and Lagrangian, JHEP {03} (2005) 011,
[arXiv:hep-th/0312075];  Schwinger-Dyson equation for
non-Lagrangian field theory, JHEP 02 (2006) 007,
[arXiv:hep-th/0512119]; Quantizing non-Lagrangian gauge theories:
an augmentation method, JHEP 01 (2007) 047,
[arXiv:hep-th/0612086]; P.O. Kazinski, S.L. Lyakhovich, A.A.
Sharapov, Lagrange structure and quantization, JHEP {0507} (2005)
076, [arXiv:hep-th/0506093].

\bibitem{Tsulaia}A. Fotopoulos, M. Tsulaia, Interacting higher spins and the high energy
limit of the bosonic string, [arXiv:0705.2939].

\bibitem{Bizdadea}C. Bizdadea, C.C. Ciobirca, E.M. Cioroianu,
S.O. Saliu, Interactions between a massless tensor field with the
mixed symmetry of the Riemann tensor and a massless vector field,
[arXiv:0705.1054].

\bibitem{BarnichHenneaux1}G. Barnich, M. Henneaux,  Consistent couplings
between fields with a gauge freedom and deformations of the master
equation, Phys. Lett. B311 (1993) 123--129,
[arXiv:hep-th/9304057]; M. Henneaux, Consistent interactions
between gauge fields: the cohomological approach, Contemp. Math.
219 (1998) 93--105, [arXiv:hep-th/9712226], M. Dubois-Violette, M.
Henneaux, Generalized cohomology for irreducible tensor fields of
mixed Young symmetry type, Lett. Math. Phys. 49 (1999) 245--252,
[arXiv:math.qa/9907135].

\bibitem{0505092}I.L. Buchbinder, V.A. Krykhtin, Gauge invariant Lagrangian
construction for massive bosonic higher spin fields in D
dimensions, Nucl. Phys. B727 (2005) 536--563,
[arXiv:hep-th/0505092].

\bibitem{symint-ads}I.L. Buchbinder, A. Pashnev, M. Tsulaia, Lagrangian formulation of
the massless higher integer spin fields in the AdS background,
Phys. Lett. B523 (2001) 338--346, [arXiv:hep-th/0109067];
 X. Bekaert, I.L. Buchbinder, A.
Pashnev, M. Tsulaia, On higher spin theory: strings, BRST,
dimensional reductions, Class. Quant. Grav. 21 (2004) 1457--1464,
[arXiv:hep-th/0312252]; I.L. Buchbinder, V.A. Krykhtin, P.M.
Lavrov, Gauge invariant Lagrangian formulation of higher massive
bosonic field theory in AdS space, Nucl. Phys. B762 (2007)
344--376, [arXiv:hep-th/0608005].

\bibitem{symferm-flat}I.L. Buchbinder, V.A. Krykhtin, A. Pashnev, BRST approach to
Lagrangian construction for fermionic massless higher spin fields,
Nucl. Phys. B711 (2005) 367--391, [arXiv:hep-th/0410215]; I.L.
Buchbinder, V.A. Krykhtin, L.L. Ryskina, H. Takata, Gauge
invariant Lagrangian construction for massive higher spin
fermionic fields,  Phys. Lett. B641 (2006) 386--392,
[arXiv:hep-th/0603212].

\bibitem{symferm-ads}I.L. Buchbinder, V.A. Krykhtin, A.A.
Reshetnyak, BRST approach to Lagrangian construction for fermionic
higher spin fields in AdS space, Nucl. Phys. B787 (2007) 211,
[arXiv:hep-th/0703049].

\bibitem{0001195}C. Burdik, A.
Pashnev, M. Tsulaia, Auxiliary representations of Lie algebras and
the BRST constructions, Mod. Phys. Lett. A15 (2000) 281--291,
[arXiv:hep-th/0001195].

\bibitem {Dixmier}J. Dixmier, Algebres enveloppantes,
Gauthier-Villars, Paris (1974).

\bibitem{Burdik}C. Burdik, Realizations of the real simple Lie algebras:
the method of construction, J. Phys. A: Math. Gen. 18 (1985)
3101--3112.


\bibitem{DuffBuchbinder}M.J. Duff, J.K. Liu, H. Sati, Quantum
$\rm{M^2\to 2{\Lambda}/3}$
discontinuity for massive gravity with a Lambda
term, Phys. Lett. B515 (2001) 156--160, [arXiv:hep-th/0105008];
F.A. Dilkes, M.J. Duff, J.K. Liu, H. Sati, Quantum discontinuity
between zero and infinitesimal graviton mass with a Lambda term,
Phys. Rev. Lett. 87 (2001) 041301, [arXiv:hep-th/0102093]; I.L.
Buchbinder, G. de Berredo-Pexoto, I.L. Shapiro, Quantum effects in
softly broken gauge theories in curved space-times,
[arXiv:hep-th/0703189].
\bibitem {0902BKRT}I.L.~Buchbinder, V.A.~Krykhtin, L.L.~Ryskina,
Lagrangian formulation of massive fermionic totally antisymmetric
tensor field theory in AdS${}_d$ space, Nucl. Phys. B819 (2009)
453--477, [arXiv:0902.1471].

\bibitem {0903Zinoviev}Yu.M.~Zinoviev,
Note on antisymmetric spin-tensors, JHEP 0904 (2009) 035,
[arXiv:0903.0262].

\bibitem{ScherkShcwarz} J.~Scherk, J.H.~Schwarz, How to Get Masses from Extra Dimensions, Nucl.Phys. B153 (1979) 61--88

\bibitem{RINDANISAHDEV-dimred-ferm}S.D.~Rindani,  D.~Sahdev, M.~Sivakumar, Dimensional  reduction
of symmetric higher spin actions II: Fermions,    Mod. Phys. Lett. A, 04(03) (1989) 275--281.


\bibitem{ReshMosh}A.A. Reshetnyak, Constrained BRST-BFV lagrangian
formulations for higher spin fields in Minkowski spaces, JHEP 09 (2018)
104, [arXiv:1803.04678[hep-th]].
\end{thebibliography}
\end{document}